\documentstyle[12pt]{article}
\input epsf
\def \beq{\begin{equation}}

\def \eep{e^+ e^-}
\def \eeq{\end{equation}}
\def \es{E$_6$}
\def \eth{E_{\rm th}}
\def \evt{$e^+ e^- \gamma \gamma E\!\!\!/_T$}
\def \G{{\rm GeV}}
\def \Hc{{\rm H.c.}}
\def \ga{\gamma \gamma}

\def \ite{{\it et al.}}

\def \k{K^0}
\def \bk{\bar K^0}

\def \M{{\rm MeV}}
\def \met{E\!\!\!/_T}

\def \pipe{\pi^+ \pi^-}
\def \poop{\pi^0 \pi^0}

\def \s{\sqrt{2}}
\def \sso{\sigma/\sigma_0}
\def \st{\sqrt{3}}

\def \sui{SU(2)$_I$}
\def \sx{\sqrt{6}}
\begin{document}

\begin{titlepage}
\vspace{-3in}
\rightline{CERN-TH/96-245}
\rightline{EFI-96-34}
\rightline{hep-ph/9610222}
\begin{center}
{\large\bf Top Quark Mass
\footnote{Three lectures given at Carg\`ese Summer Institute on
Particle Physics, {\it Masses of Fundamental Particles}, August, 1996.
Proceeding to be published by Plenum.}} \\
\vspace{1.5cm}
{\large Jonathan L. Rosner\footnote{rosner@uchepa.uchicago.edu}} \\
\vspace{.5cm}
{\sl CERN, 1211-CH Geneva 23, Switzerland} \\
\vspace{.5cm}
{\sl Enrico Fermi Institute and Department of Physics}\\
{\sl University of Chicago, Chicago, IL 60637 USA}
\footnote{Permanent address.}\\
\vspace{1.5cm}
\begin{abstract}
The influence of the top quark mass on mixing processes and precise electroweak
measurements is described.  Experimental observation of the top quark in
proton-antiproton collisions is discussed, and some brief remarks are made
about electron-positron production.  Some speculations are noted about the
possible significance of the large top quark mass. 
\end{abstract}

\end{center}
\vfill
\leftline{CERN-TH/96-245}
\leftline{September 1996}
\end{titlepage}

\null
\vskip 140pt
\leftline{\bf TOP QUARK MASS}
\vskip 28pt

\parindent 1in

Jonathan L. Rosner

Enrico Fermi Institute and Department of Physics

University of Chicago

5640 South Ellis Avenue, Chicago, IL 60637
\vskip 28pt

\leftline{\bf 1.  INTRODUCTION}
\bigskip

\parindent 0.85cm

In the present article, based on a series of three lectures, we describe ways
in which the large top quark mass influences a number of processes, and some
ideas about how it might arise. After a brief introduction to the pattern of
masses and couplings of quarks and leptons and an overview of the top quark's
properties, we discuss in Section 2 the role of the top quark in mixing
processes, such as particle-antiparticle mixing in the neutral kaon and $B$
meson systems, and in other flavor-changing charge-preserving processes.
Section 3 is devoted to precision electroweak experiments and the impact upon
their interpretation of the precise measurements of the top quark's mass which
have recently been achieved at Fermilab (Section 4). Some brief remarks about
top quark production in electron-positron collisions occupy Section 5, while
Section 6 contains various speculations about the source of the top quark's
mass.  (See also Graham Ross' lectures at this Institute \cite{Ross}.)  Section
7 concludes. 
\bigskip

\leftline{\bf A.  Quark and lepton mass and coupling patterns}
\bigskip

The top quark is the heaviest known elementary particle, nearly 200 times as
heavy as a proton.  Its mass is compared with that of the other known quarks
and leptons in Figure 1. Also shown are the patterns of couplings in
charge-changing weak transitions. The main coupling of the top ($t$) appears to
be to the bottom ($b$) quark \cite{LeCompte}, while charm ($c$) couples mainly
to strange ($s$) and up to down.  Each charged lepton, in turn, appears to
couple to its own neutrino. However, the pattern for quarks known at present is
much richer, including weaker couplings of $t$ to $s$ and $d$, of $c$ to $b$
and $d$, and of $u$ to $b$ and $s$.  This pattern is encoded in a unitary $3
\times 3$ matrix $V$ known as the {\it Cabibbo-Kobayashi-Maskawa} \cite{Cab,KM}
matrix. 

In these lectures we shall explore a number of aspects of the pattern of masses
and couplings in Figure 1 and Table 1 \cite{JRCP}.  The large mass of the top
quark and its pattern of couplings to the $b,~s$, and $d$ quarks lead to a
number of consequences for processes in which the top quark participates as a
virtual particle.  Not the least of these is the ability of loop diagrams
involving the top quark to generate the observed CP violation in the neutral
kaon system \cite{KM}.  At the same time, the fact that the top quark is about
twice as heavy as the gauge bosons $W$ and $Z$ of the electroweak interactions
\cite{GWS} offers a hint that it may play a crucial role in the understanding
of all the quark and lepton masses. 

\begin{figure}
\centerline{\epsfysize = 3 in \epsffile {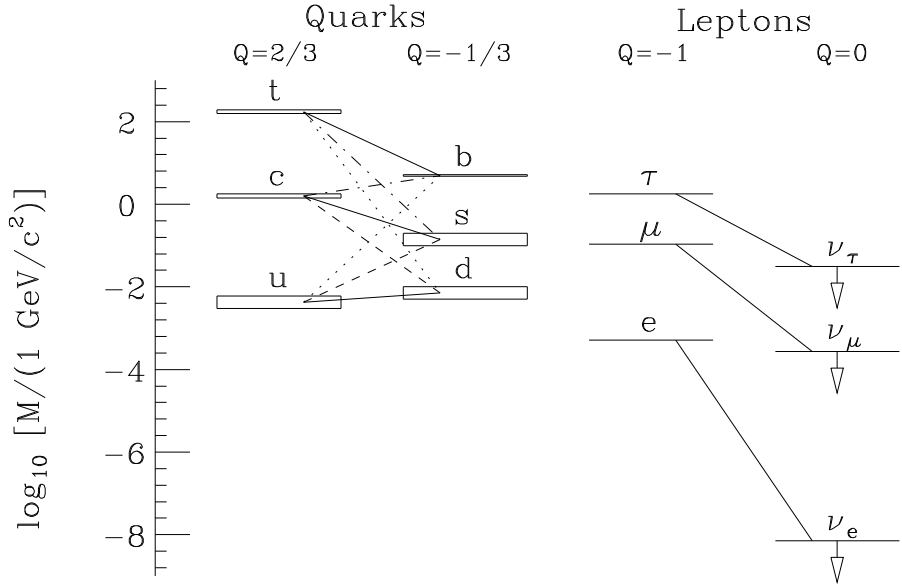}}
\caption{Patterns of charge-changing weak transitions among quarks and leptons.
Direct evidence for $\nu_\tau$ does not yet exist. The strongest inter-quark
transitions correspond to the solid lines, with dashed, dot-dashed, and dotted
lines corresponding to successively weaker transitions.} 
\end{figure}

\begin{table}
\caption{Relative strengths of charge-changing weak transitions.}
\begin{center}
\begin{tabular}{||c|c|l||} \hline
Relative & Transition & Source of information \\
amplitude & & ~~~~~~~~(example) \\ \hline
$\sim$ 1 & $u \leftrightarrow d$ & Nuclear $\beta$-decay \\
$\sim$ 1 & $c \leftrightarrow s$ & Charmed particle decays \\
$\sim 0.22$ & $u \leftrightarrow s$ & Strange particle decays \\
$\sim 0.22$ & $c \leftrightarrow d$ & Neutrino prod. of charm \\
$\sim 0.04$ & $c \leftrightarrow b$ & $b$ decays \\
$\sim 0.003$ & $u \leftrightarrow b$ & Charmless $b$ decays \\
$\sim$ 1 & $t \leftrightarrow b$ & Dominance of $t \to W b$ \\
$\sim 0.04$ & $t \leftrightarrow s$ & Only indirect evidence \\
$\sim 0.01$ & $t \leftrightarrow d$ & Only indirect evidence \\ \hline
\end{tabular}
\end{center}
\end{table}

A note of caution:  if its mass is viewed on a logarithmic scale, as in Figure
1, the top quark is not all that exceptional.  Although it is more than
100 times as heavy as the next lightest quark of charge 2/3, the charmed quark,
we have not yet explored similar logarithmic mass intervals above the bottom
quark or $\tau$ lepton.  But the $Z$ decays to just three pairs of light
neutrinos \cite{LEPEWWG}. so the pattern of any fermions heavier than those in
Figure 1 must at least change to a significant extent.
\newpage

\leftline{\bf B.  Quark and lepton families}
\bigskip

The case for a quark-lepton analogy is suggested by the doublets
\beq \label{eqn:firstfam}
\left( \begin{array}{c} \nu_e \\ e^- \end{array} \right): {\rm leptons}~~;~~~
\left( \begin{array}{c} u \\ d \end{array} \right): {\rm quarks}~~~.
\eeq
In both cases the states in the upper and lower rows are connected by emission
or absorption of a $W$ boson.  This pattern was extended to the second family
by several authors \cite{HMOBG}, who inferred the existence of a fourth quark
(charm) coupling mainly to the strange quark from the existence of the $\mu -
\nu_\mu$ doublet: 
\beq \label{eqn:secfam}
\left( \begin{array}{c} \nu_\mu \\ \mu^- \end{array} \right): {\rm leptons}
\Rightarrow \left( \begin{array}{c} c \\ s \end{array} \right): {\rm
quarks}~~~. 
\eeq

The suppression of processes involving weak decays of strange particles as
compared with strangeness-preserving weak decays could be made compatible with
the universal strength of the weak interactions if the $u$ quark coupled to an
appropriately normalized linear combination of $d$ and $s$: 
\beq \label{eqn:ucomb}
u \leftrightarrow d \cos \theta_c + s \sin \theta_c~~~.
\eeq
This universality property was proposed by Gell-Mann and L\'evy \cite{GL} and
used by Cabibbo \cite{Cab} (hence the subscript on the angle) to relate
successfully a number of beta-decay processes of non-strange and strange
particles.  The strength of the charge-changing weak current could be
normalized by considering it and its hermitian adjoint to be members of an
SU(2) triplet.  The neutral member of this triplet contained
strangeness-changing terms and thus had to be interpreted as unphysical, since
no strangeness-changing neutral-current processes had been seen.  However, if
one then considered the charmed quark to couple to the orthogonal combination, 
\beq \label{eqn:ccomb}
c \leftrightarrow - d \sin \theta_c + s \cos \theta_c~~~,
\eeq
the corresponding neutral current would couple to the combination $u \bar u + c
\bar c - d \bar d - s \bar s$, and would preserve flavor \cite{HMOBG,GIM}.
This property was preserved in higher orders of perturbation theory, and
neutral strangeness-changing processes were induced only to the extent that the
charm and $u$ quark masses differed from one another \cite{GIM}. 

A further motivation for the quark-lepton analogy was noted by Bouchiat,
Iliopoulos, and Meyer \cite{BIM} in 1972.  In a gauge theory of the electroweak
interactions, triangle anomalies associated with graphs of the type shown
in Figure 2 have to be avoided.  This cancellation requires the fermions $f$
in the theory to contribute a total of zero to the sum over $f$ of $Q_f^2
I^f_{3L}$.  Such a cancellation can be achieved by requiring quarks and leptons
to occur in complete families of the type mentioned above, so that the terms 
\beq
{\rm Leptons}:~~(0)^2 \left( \frac{1}{2} \right) + (-1)^2 \left( -\frac{1}{2}
\right) = - \frac{1}{2}
\eeq
\beq
{\rm Quarks}:~~3 \left[ \left( \frac{2}{3} \right)^2 \left( \frac{1}{2} \right)
+ \left( -\frac{1}{3} \right)^2 \left( -\frac{1}{2}\right) \right] =
\frac{1}{2} 
\eeq 
sum to zero for each family.

\begin{figure}
\centerline{\epsfysize = 1.3 in \epsffile {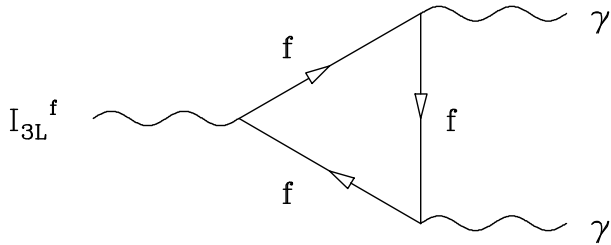}}
\caption{Example of triangle diagram for which leading behavior must cancel
in a renormalizable electroweak theory.}
\end{figure}

Some history of how the quark and lepton families were pieced together may
be of interest \cite{hist}.  In the family (\ref{eqn:firstfam}), the first
particle to be discovered was the electron, whose unique properties were
clinched by J. J. Thomson's measurement of its enormous charge-to-mass
ratio (nearly 2000 times that of the proton) in 1897.  The corresponding
neutrino was proposed by W. Pauli \cite{Pauli} in 1931.  As for the $u$ and $d$
quark, the need for two constituents of hadrons differing in charge by a
unit but with very similar behaviour under the strong interactions dates back
to the recognition of isotopic spin as a good symmetry \cite{isospin}, in the
1930's.

Evidence for the second family (\ref{eqn:secfam}) began with the discovery in
1937 by S. Neddermeyer and C. Anderson of the muon in cosmic radiation
\cite{NA}.  The corresponding neutrino $\nu_\mu$ was proved distinct from the
electron neutrino $\nu_e$ in 1962 by a group led by L. Lederman, M. Schwartz,
and J. Steinberger \cite{twonu}.  Meanwhile, evidence for particles containing
one or more strange ($s$) quarks, initially seen in cosmic radiation and later
produced at accelerators, began accumulating over the period 1946-53.  The
charmed ($c$) quark was discovered initially in the form of a $c \bar c$ bound
state, the $J$ \cite{Ting} or $\psi$ \cite{Richter} (the dual name $J/\psi$ has
survived).  Particles containing a single charmed quark (``bare charm'') were
identified a bit later \cite{charm}. 

At the same time evidence for the charmed quark was accumulating in
electron-positron annihilations, the first member of a third family:
\beq \label{eqn:thdfam}
\left( \begin{array}{c} \nu_\tau\\ \tau^- \end{array} \right): 
{\rm leptons}~~;~~~
\left( \begin{array}{c} t \\ b  \end{array} \right): {\rm quarks}~~~.
\eeq
was making its appearance.  This was the $\tau$ lepton, identified by M. Perl
and collaborators \cite{Perl}.  If this were to be a lepton like the $e$ and
$\mu$, having its own neutrino, the anomaly-cancellation mechanism mentioned
earlier \cite{BIM} then required there to be a third pair of quarks:  the
``top'' and ``bottom'' \cite{HH}, our present names for the quarks proposed by
Kobayashi and Maskawa \cite{KM} as candidates for explaining CP violation.

The bottom quark made its appearance in the form of $b \bar b$ bound states
discovered at Fermilab in 1977 by a group led by L. Lederman \cite{upsilons}
and in ``bare bottom'' (otherwise known as ``beauty'') mesons containing single
$b$ quarks seen at the Cornell Electron Storage Ring (CESR) somewhat later
\cite{beauty}. Eagerly awaited, the top quark finally made its appearance 17
years after the bottom, in proton-antiproton collisions at Fermilab at a
center-of-mass energy of 1.8 TeV \cite{top}. 

Even up to the present day, in the absence of direct evidence for the
$\nu_\tau$, we have never had a complete set of quark and lepton families
without hints of the next family.  By the time the $\nu_\tau$ is discovered,
perhaps at a forthcoming experiment at Fermilab \cite{nutau}, who knows what
other particles may appear? 
\bigskip

\leftline{\bf C.  Top quark properties}
\bigskip

The most recent measurements of the top quark mass, reported at the 1996 Warsaw
Conference, are \cite{Watop} 
\beq \label{eqn:topmass}
m_t = 176.8 \pm 4.4 \pm 4.8~\G/c^2~({\rm CDF})~;~~
169 \pm 8 \pm 8~\G/c^2~({\rm D0})~;~~
175 \pm 6~\G/c^2~({\rm avg.})~.
\eeq
The top is expected to decay essentially $100\% (\simeq |V_{tb}|^2)$ to $W +
b$, with branching ratios of only about $|V_{ts}|^2 \simeq 1.5 \times 10^{-3}$
to $W + s$ and $|V_{td}|^2 \simeq 10^{-4}$ to $W + d$.  We shall see in Sec.~4
that recent data \cite{LeCompte} support this expectation.

The predicted rate for the decay $t \to W + b$ is
\beq
\Gamma(t \to W + b) = \frac{G_F}{8 \pi \sqrt{2}} |V_{tb}|^2 m_t^3 \Phi_K
K_{\rm QCD}~~~,
\eeq
where, in the limit in which the $b$ quark mass can be neglected,
\beq
\Phi_K \equiv \left( 1 - \frac{M_W^2}{m_t^2} \right)^2 \left( 1 +
2 \frac{M_W^2}{m_t^2} \right) = 0.885
\eeq
and $K_{\rm QCD}$ is a QCD correction factor which is about 0.917 \cite{JKT}. 
For $m_t = 175$ GeV, one thus predicts $\Gamma = (1.76~\G)(0.885)(0.917)
\simeq 1.43~\G$, so that the top quark decays before it can form hadrons.
Thus, sad to say, we will not be able to study the properties of $t \bar t$
or $t \bar q$ states (where $q$ stands for one of the lighter quarks).
Nonetheless, some crude information about the $t \bar t$ interaction can
be recovered, as we shall mention in Sec.~5.

The spin of the top quark is expected to be 1/2, just like that of the other
quarks.  Direct evidence for this is lacking, but if we are prepared to admit
that $J(t) = J(b)$, we can use all of the bottom quark's rich spectroscopy to
conclude that $J(b) = 1/2$ and hence $J(t) = 1/2$. The spectra of $c \bar c$,
$b \bar b$, charm -- light quark  and bottom -- light quark bound states are
shown in Figs.~3, 4, 5, and 6 respectively.  We shall use some of the
terminology in these figures in what follows, so they are worth getting
acquainted with.

The charmonium ($c \bar c$) spectrum in Fig.~3 bears strong evidence that the
spin of the charmed quark is $1/2$, as one might have expected from its
partnership with the strange quark.  (Spins of strange particles have been
directly measured for a number of years and support the notion that $J(s) =
1/2$.)  The S-wave ($L=0$) levels have total angular momentum $J$, parity $P$,
and charge-conjugation eigenvalue $C$ equal to $J^{PC} = 0^{+-}$ and $1^{--}$
as one would expect for $^1S_0$ and $^3S_1$ states, respectively, of a quark
and antiquark.  The P-wave ($L=1$) levels have $J^{PC} = 1^{+-}$ for the
$^1P_1$, $0^{++}$ for the $^3P_0$, $1^{++}$ for the $^3P_1$, and $2^{++}$ for
the $^3P_2$. The $J^{PC} = 1^{--}$ levels are identified as such by their
copious production through single virtual photons in $\eep$ annihilations.  The
$0^{-+}$ level $\eta_c$ is produced via single-photon emission from the
$J/\psi$ (so its $C$ is positive) and has been directly measured to have
$J^{P}$ compatible with $0^-$ \cite{etacJP}.  Numerous studies have been made
of the electromagnetic (electric dipole) transitions between the S-wave and
$P$-wave levels \cite{EDT} and they, too, support the assignments shown. 

\begin{figure}
\centerline{\epsfysize=4in \epsffile{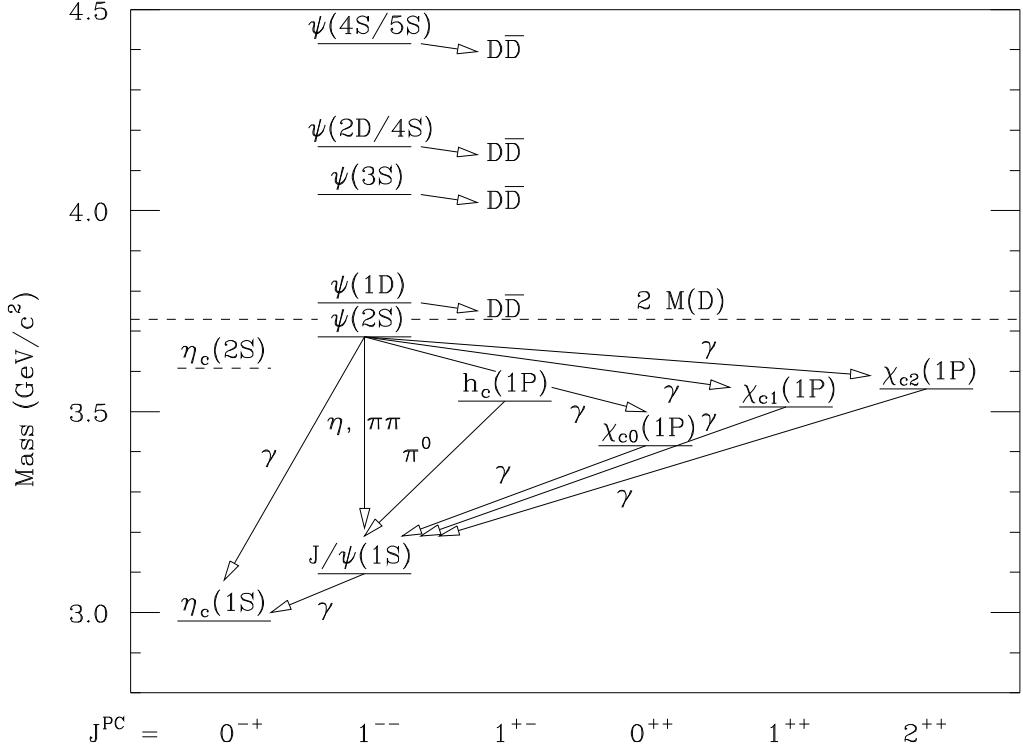}}
\caption{Charmonium ($c \bar c$) spectrum.  Observed and predicted levels are
denoted by solid and dashed horizontal lines, respectively.  Arrows denote
electromagnetic transitions (labeled by $\gamma$) and hadronic transitions
(labeled by emitted hadrons).}
\end{figure}

The $b \bar b$ (upsilon) levels shown in Fig.~4 have a very similar structure,
aside from an overall shift.  The similarity of the $c \bar c$ and $b \bar b$
spectra is in fact an accident of the fact that for the interquark distances
in question (roughly 0.2 to 1 fm), the interquark potential interpolates
between short-distance Coulomb-like and long-distance linear behavior in a
manner roughly compatible with $V \sim \log r$ \cite{QR} or with an effective
power near zero \cite{Martin}.  The copious production of $1^{--}$ candidiates
in $\eep$ annihilations and the pattern of electric dipole transitions between
S- and P-wave levels again supports the assignments shown. 

\begin{figure}
\centerline{\epsfysize=4in \epsffile{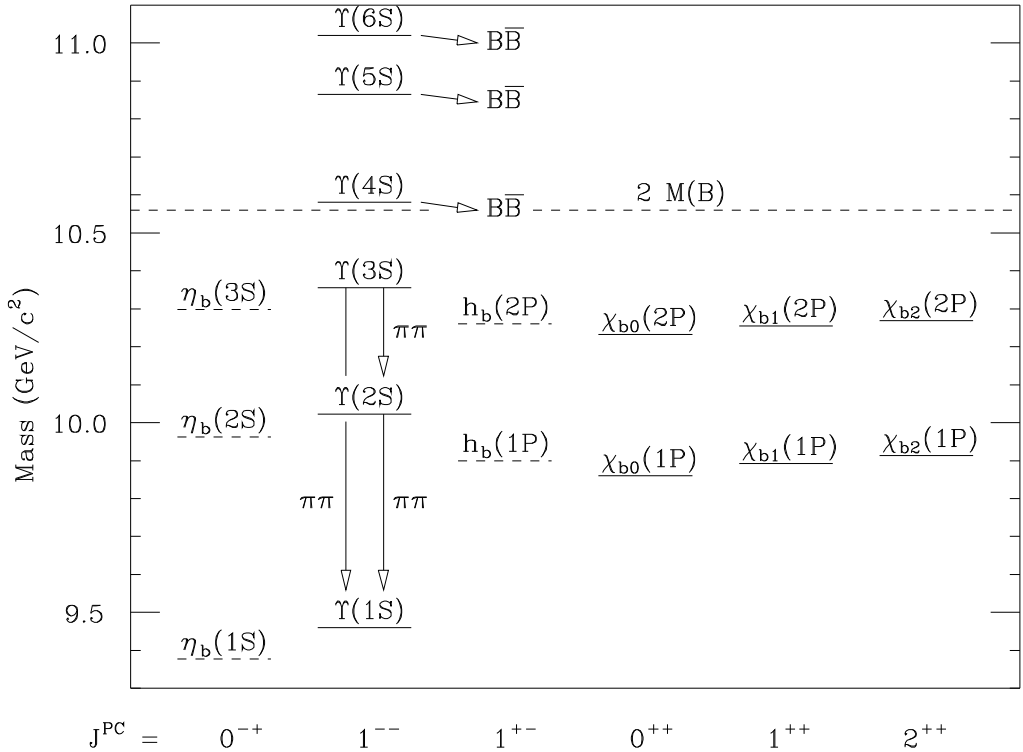}}
\caption{Spectrum of $b \bar b$ states.  Observed and predicted levels are
denoted by solid and dashed horizontal lines, respectively.  In addition to the
transitions labeled by arrows, numerous electric dipole transitions and decays
of states below $B \bar B$ threshold to hadrons containing light quarks have
been seen.} 
\end{figure}

\begin{figure}
\centerline{\epsfysize=4in \epsffile{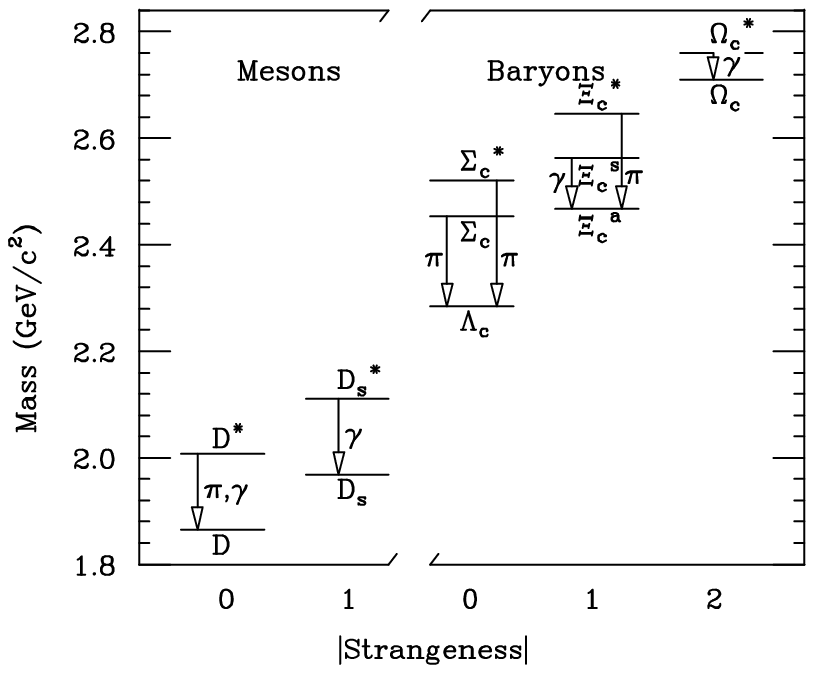}}
\caption{Spectrum of lowest-lying states containing one charmed and one light
quark. Observed and predicted levels are denoted by solid and broken horizontal
lines, respectively.} 
\end{figure}

\begin{figure}
\centerline{\epsfysize=4in \epsffile{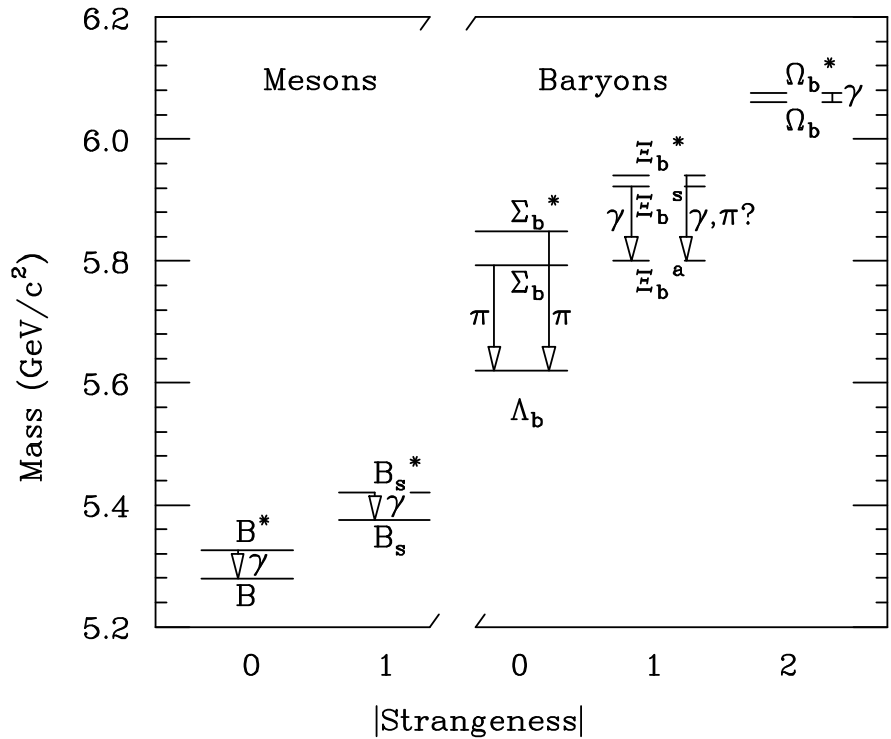}}
\caption{Spectrum of lowest-lying states containing one bottom and one light
quark. Observed and predicted levels are denoted by solid and broken horizontal
lines, respectively.} 
\end{figure}

States consisting of a single charmed quark and light ($u,~d$, or $s$) quarks
or antiquarks, shown in Fig.~5, again support $J(c) = 1/2$.  The lightest
mesons have $J^P = 0^-$ ($^1S_0$) and $1^-$ ($^3S_1$).  Two charmed baryons
discovered fairly recently, both strange \cite{Xicstar} and nonstrange
\cite{Sigmacstar}, are candidates for $J = 3/2$.  One expects such states as
bound states of three spin-1/2 quarks. 

Finally, the pattern of states containing a single $b$ quark (Fig.~6) is very
similar to that for singly-charmed states, though not as well fleshed-out.
In many cases the splittings between states containing a single $b$ quark
is less than that between the corresponding charmed states by roughly a
factor of $m_c/m_b \simeq 1/3$ as a result of the smaller chromomagnetic
moment of the $b$ quark.  (This feature has also been noted by Neubert at
this Institute \cite{MN}.)

The top quark's mass, just like every other mass in the standard electroweak
theory, arises as a result of Yukawa interactions of one or more Higgs
bosons with the fermions.  For the $t$ and $b$ quarks, the relevant part of
the Lagrangian can be written as
\beq
{\cal L}^{t,b}_{\rm Higgs} = - g_Y^b [ \bar Q_L \phi_1 b_R + \Hc]
- g_Y^t [ \bar Q_L \phi_2 t_R + \Hc]~~~
\eeq
where
\beq
Q_L \equiv \left( \begin{array}{c} t \\ b \end{array} \right)_L~,~~
\phi_1 \equiv \left( \begin{array}{c} \phi_1^+ \\ \phi_1^0 \end{array}
\right)~,~~
\phi_2 \equiv \left( \begin{array}{c} \phi_2^0 \\ \phi_2^- \end{array}
\right)
\eeq
if there are two Higgs doublets, or $\phi_2 = i \sigma_2 \phi_1^c$ if there is
only one. If the neutral member pf each Higgs doublet acquires a vacuum
expectation value $\langle \phi_i^0 \rangle = v_i/\s$, then $m_b = g_Y^b v_1
/\s$, $m_t = g_Y^t v_2/\s$.  Here 
\beq
2^{-1/4}G_F^{-1/2} = v_1^2 + v_2^2 = (246~\G)^2~~~,
\eeq
while the right-hand side is replaced by $v_1^2 \equiv v^2$ if $\phi_2 =
i \sigma_2 \phi_1^c$.

Suppose there were just one Higgs doublet, with $v \simeq 246~\G$.  Then
\beq
g_Y^t = \frac{246~\G}{\s m_t} = \frac{174~\G}{175 \pm 6~\G} \simeq 1~!
\eeq
The significance of this relation may be accidental, however.  Quark masses
``run'' as a result of their interactions with gauge fields (particularly
with gluons), and so their values depend on the scale at which the mass is
probed.  As one example, the top quark's ``pole'' mass (what one would measure
in a physical process of top quark production) and the value $\bar m_t(\mu)$
in the modified-minimal-subtraction scheme at a mass scale $\mu$ are related
\cite{RGtop} by
\beq
\frac{m_t^{\rm pole}}{\bar m_t(\mu)} = 1 + \frac{4}{3} \frac{\alpha_s(\mu)}
{\pi} + \ldots~~~.
\eeq
For example, with a pole mass of 175 \G, and $\mu = M_W$ (which will frequently
be appropriate for the loop calculations we shall perform), so that $\alpha_s
(M_W) \simeq 0.12$, one has $m_t^{\rm pole}/\bar m_t(M_W) \simeq 1.05$.  Taking
into account higher-order effects as well, we shall take $\bar m_t(M_W) = 165
\pm 6~\G/c^2$ \cite{AL} in a number of subsequent calculations.
\bigskip

\leftline{\bf 2.  THE TOP QUARK IN MIXING PROCESSES}
\bigskip

\leftline{\bf A.  Cabibbo-Kobayashi-Maskawa matrix parameters}
\bigskip

The electroweak Lagrangian, before electroweak symmetry breaking, may be
written in flavor-diagonal form as 
\beq
{\cal L}_{\rm int} = - \frac{g}{\sqrt{2} }[ \overline{U'}_L 
\gamma^\mu W_\mu^{(+)} {D'}_L + \Hc]~~~,
\eeq
where $U' \equiv (u',c',t')$ and $D' \equiv (d',s',b')$ are column vectors
decribing {\em weak eigenstates}. Here $g$ is the weak $SU(2)_L$ coupling
constant, and $\psi_L \equiv (1 - \gamma_5) \psi /2$ is the left-handed
projection of the fermion field $\psi = U$ or $D$. 

Quark mixings arise because mass terms in the Lagrangian are permitted to
connect weak eigenstates with one another. Thus, the matrices ${\cal M}_{U,~D}$
in 
\beq
{\cal L}_m = - [\overline{U'}_R {\cal M}_U {U'}_L + \overline {D '}_R {\cal
M}_D {D'}_L + \Hc]
\eeq
may contain off-diagonal terms. One may diagonalize these matrices by separate
unitary transformations on left-handed and right-handed quark fields: 
\beq
R_{Q}^+ {\cal M}_{Q} L_{Q} = L_{Q}^+ {\cal M}_{Q}^+ R_Q = \Lambda_Q ~~~.
\eeq
where
\beq
{Q'}_L = L_Q Q_L ; ~~ {Q'}_R = R_Q Q_R ~~~ (Q = U, D)~~~.
\eeq
Using the relation between weak eigenstates and mass eigenstates: 
${U'}_L = L_U U_L , ~ {D'}_L = L_D D_L$, we find 
\beq
{\cal L}_{\rm int} = - \frac{g}{\sqrt{2}} [ \overline{U}_L \gamma^\mu W_\mu
V D_L + \Hc] ~~~,
\eeq
where $U \equiv (u,c,t)$ and $D \equiv (d,s,b)$ are the mass eigenstates, and
$V \equiv L_U^{\dag} L_D$. The matrix $V$ is just the Cabibbo-Kobayashi-Maskawa
matrix. By construction, it is unitary: $V^{\dag}V = VV^{\dag} = 1$. It carries
no information about $R_U$ or $R_D$. More information would be forthcoming from
interactions sensitive to right\--handed quarks or from a genuine theory of
quark masses.  Because of the unitarity of the matrix, the neutral currents to
which the $Z^0$ couples are flavor-diagonal:  $\overline{Q'} \Gamma Q' =
\overline{Q} \Gamma Q$, where $\Gamma$ is any combination of $\gamma^\mu$ and
$\gamma^\mu \gamma_5$. 

For $n~u$-type quarks and $n~d$-type quarks, $V$ is $n \times n$ and
unitary. An arbitrary $n \times n$ matrix has $2n^2$ real parameters, but
unitarity $(V^{\dag} V = 1)$ provides $n^2$ constraints, so only $n^2$ real
parameters remain. We may remove $2n-1$ of these by appropriate redefinitions
of relative quark phases. The number of remaining parameters is then $n^2 -
(2n-1) = (n-1)^2$. Of these, $n (n - 1)/2$ (the number of independent rotations
in $n$ dimensions) correspond to angles, while the rest, $(n-1)(n-2)/2$,
correspond to phases. We summarize these results in Table 2. 

\begin{table}
\caption{Parameters of KM matrices for $n$ doublets of quarks.}
\begin{center}
\renewcommand{\arraystretch}{1.2}
\begin{tabular}{| c | c c c c |} \hline
 & $n =$  & 2~~~ &~~~ 3~~~ &~~~ 4 \\ \hline
Number of parameters & $(n-1)^2 $ & 1~~~ &~~~ 4~~~ &~~~ 9 \\
Number of angles & $n(n-1)/2$  & 1~~~ &~~~ 3~~~  &~~~ 6 \\
Number of  phases& $ (n-1)(n-2)/2$ & 0~~~ &~~~ 1~~~ &~~~ 3 \\ \hline
\end{tabular}
\end{center}
\renewcommand{\arraystretch}{1.0}
\end{table}

For $n=2$, we have one angle and no phases. The matrix $V$ then can always be
chosen as orthogonal \cite{Cab,HMOBG,GIM}. For $n=3$, we have three angles and
one phase, which in general cannot be eliminated by arbitrary choices of phases
in the quark fields. It was this phase that motivated Kobayashi and
Maskawa \cite{KM} to introduce a third quark doublet.  It provides a potential
source of CP violation, serving as the leading contender for the observed
CP-violating effects in the kaon system and suggesting substantial CP
asymmetries in the decays of mesons containing $b$ quarks. 

The CKM matrix $V$ is then, explicitly,
\beq \label{eqn:CKM}
V = \left( \begin{array}{c c c}
V_{ud} & V_{us} & V_{ub} \\
V_{cd} & V_{cs} & V_{cb} \\
V_{td} & V_{ts} & V_{tb}
\end{array} \right) ~~~.
\eeq
We now parametrize its elements.

It is convenient to choose quark phases \cite{QP} so that the $n$ diagonal
elements and the $n-1$ elements just above the diagonal are real and positive.
The parametrization we shall employ is one suggested by Wolfenstein \cite{WP}. 

The diagonal elements of $V$ are nearly $1$, while the dominant off-diagonal
elements are $V_{us} \simeq - V_{cd} \equiv \lambda \simeq 0.22$. Thus to order
$\lambda^2$, the upper $2 \times 2$ submatrix of $V$ is already known from the
four-quark pattern.  One may identify $\lambda = \sin \theta_c$, where
$\theta_c$ is the Gell-Mann -- L\'evy -- Cabibbo angle mentioned earlier.  It
is the single parameter needed to describe the mixing in the four-quark ($n=2$)
system.  More precisely, analyses of strange particle decays \cite{strange}
have yielded $\sin \theta_c = 0.2205 \pm 0.0018$.

The matrix element $|V_{ud}|^2$ may be obtained by comparing the strengths of
certain beta-decay transitions involving vector transitions with that of muon
decay. One can also measure the neutron decay rate (which involves both vector
and axial vector transitions), and extract the vector coupling strength by
finding $g_A$ from decay asymmetries. This vector coupling strength may be
compared with that obtained in muon decay to learn $|V_{ud}|^2$. Finally, one
can study the decay $\pi^+ \to \pi^0 e^+ \nu_e$.  Overall, one finds
\cite{JRCP} $|V_{ud}| \simeq 0.975 \pm 0.001$, so that $|V_{ud}|^2 +
|V_{us}|^2 = 0.999 \pm 0.002$.  One cannot deduce the need for the additional
contribution of $|V_{ub}|^2$ in the relation $|V_{ud}|^2 + |V_{us}|^2 +
|V_{ub}|^2 = 1$ required by the unitarity of the CKM matrix.  Measurements
of $|V_{cd}|$ and $|V_{cs}|$ (see Ref.~\cite{JRCP}) are also consistent with
the predictions of unitarity, but with larger errors.

The long $b$ quark lifetime (about 1.5 to 1.6 ps \cite{BH}) and the
predominance of charmed quarks among $b$ decay products implies that $V_{cb}
\simeq 0.04$, allowing one to express it as $A \lambda^2$, where $A = {\cal
O}(1)$.  A recent compilation of results \cite{AL,Gib} on semileptonic $b \to
c$ decays yields $V_{cb} = A \lambda^2 = 0.0393 \pm 0.0028$, or $A = 0.808 \pm
0.058$.  [M. Neubert \cite{MN} quotes $V_{cb} = 0.0388 \pm 0.0020_{\rm exp} \pm
0.0012_{\rm th}$.]  Unitarity then requires $V_{ts} \simeq - A \lambda^2$ as
long as $V_{td}$ and $V_{ub}$ are small enough (which they are). 

The magnitude of the element $V_{ub}$ is learned by comparing charmless $b$
decays to those with charm.  One thus finds that $|V_{ub}|$ appears to be of
order $A \lambda^3$. Here one must allow for a phase, so one must introduce two
new parameters $\rho$ and $\eta$:  $V_{ub} = A \lambda^3 (\rho - i \eta )$. 
The measured ratio \cite{AL} $|V_{ub}/V_{cb}| = 0.08 \pm 0.016$, whose main
error is dominated by theoretical extrapolation from a small part of the lepton
spectrum in semileptonic $b \to u$ decays, implies
\beq \label{eqn:Vubcon}
(\rho^2 + \eta^2)^{1/2} = 0.363 \pm 0.073~~~.
\eeq
This constraint can be plotted in the $(\rho,\eta)$ plane as a band bounded by
circles centered at (0,0). 

Finally, unitarity specifies uniquely the form $V_{td} = A \lambda^3 (1 - \rho
- i \eta )$. To summarize, the CKM matrix may be written in terms of the
4 parameters $\lambda$, $A$, $\rho$, and $\eta$ as
\beq \label{eqn:CKMwp}
V \approx \left[ \matrix{1 - \lambda^2/2 & \lambda & A \lambda^3 (\rho -
i \eta) \cr
- \lambda & 1 - \lambda^2 /2 & A \lambda^2 \cr
A \lambda^3 (1 - \rho - i \eta) & - A \lambda^2 & 1 \cr } \right]~~~.
\eeq
The form (\ref{eqn:CKMwp}) is only correct to order $\lambda^3$ in the matrix
elements. For certain purposes it may be necessary to exhibit corrections of
higher order to the elements. This can be done using the unitarity of the
matrix.

The unitarity of the CKM matrix implies that the scalar product of one row (or
column) with the complex conjugate of any other row (or column) must vanish: 
for example, 
\beq \label{eqn:ur}
V_{ud}^* V_{td} + V_{us}^* V_{ts} + V_{ub}^* V_{tb} = 0 ~~~.
\eeq
Since $V_{ud}^* \approx 1,~V_{us}^* \approx \lambda,~V_{ts} \approx - A
\lambda^2$, and $V_{tb} \approx 1$ we have $V_{td} + V_{ub}^* = A \lambda^3$, a
useful relation expressing the least-known CKM elements in terms of relatively
well-known parameters.  This result can be visualized as a triangle \cite{UT}
in the complex plane [Fig.~7(a)]. In this figure the angles $\alpha, \beta$,
and $\gamma$ are defined as in the review by Nir and Quinn \cite{NQ}. 

\begin{figure}
\centerline{\epsfysize = 1.2 in \epsffile {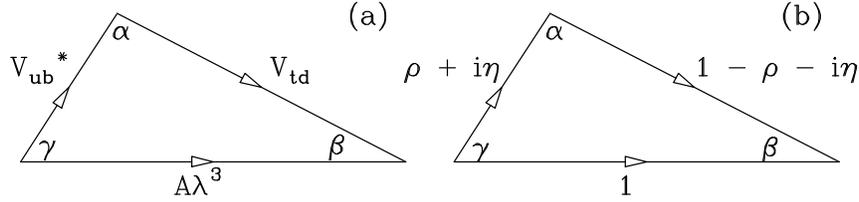}}
\caption{Unitarity triangle for CKM elements.
(a) The relation (\ref{eqn:ur}) in the complex plane;
(b) Eq.~(\ref{eqn:ur}) divided by the normalizing factor $A \lambda^3$.}
\end{figure}

Dividing (\ref{eqn:ur}) by $A \lambda^3$, since $V_{ub}^*/ A \lambda^3 = \rho +
i \eta, ~~ V_{td} / A \lambda^3 = 1 - \rho - i \eta$, one obtains a triangle of
the form shown in Fig.~7(b). The value of $V_{ub}^* / A \lambda^3$ may then be
depicted as a point in the $(\rho,\eta)$ plane. The major ambiguity which still
remains in the determination of the CKM matrix elements concerns the shape of
the unitarity triangle. The answer depends on the magnitude of $V_{td}$. As we
shall see, decays alone will not provide the answer.  In order to learn about
the elements $V_{td}$ and $V_{ts}$ one must resort to indirect means, which
involve loop diagrams. 
\bigskip

\leftline{\bf B.  Mixing of neutral $B$ mesons}
\bigskip

The box diagrams of Fig.~8 dominate the mixing between $B^0$ and $\bar B^0$,
leading to a splitting $\Delta m$ between the mass eigenstates.  (We present
an abbreviated account of mixing; for more details see \cite{JRCP,CL,TASI,IL}.)
The leading contributions in each of these diagrams cancel one another when one
sums over all the intermediate quarks of charge 2/3, since $V_{ub}V^*_{ud} +
V_{cb}V^*_{cd} + V_{tb}V^*_{td} = 0$.  The remaining contributions are
dominated by the top quark since the corresponding CKM products of pairs of CKM
elements $V_{ib}V^*_{id}$ are each of order $\lambda^3$ and $m_t \gg m_c,m_u$. 
One then finds \cite{IL}

\begin{figure}
\centerline{\epsfysize = 1.3 in \epsffile {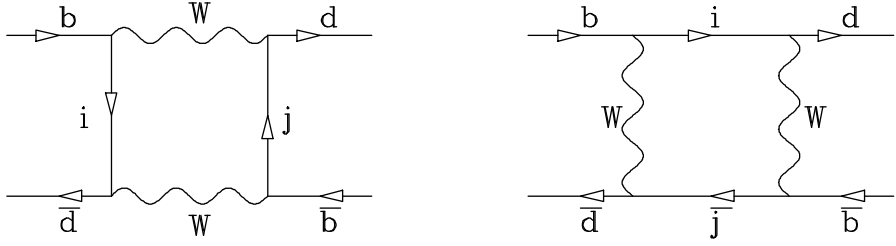}}
\caption{Box diagrams for mixing of $B^0$ and $\bar B^0$.}
\end{figure}

\beq \label{eqn:dmd}
\Delta m = {G_F^2 \over 6 \pi^2} |V_{td}|^2 M_W^2 m_B f_B^2 B_B \eta_B S \left(
{m_t^2 \over M_W^2} \right)~~~,
\eeq
where
\beq \label{eqn:sdef}
S(x) \equiv {x \over 4} \left[ 1 + {3-9x \over (x-1)^2} + {6x^2\ln x \over
(x-1)^3} \right]~~~.
\eeq
This factor behaves as $x$ for small $x$ and $x/4$ for large $x$ and equals 3/4
for $x = 1$.

We take $m_t = 175 \pm 6~\G/c^2$ as noted in Sec.~1 C, $M_W = 80.34 \pm
0.10~\G/c^2$ (see Sec.~3), $m_B = 5.279~\G/c^2$ (see \cite{PDG}), and $f_B
\sqrt{B_B} = 200 \pm 40$ MeV \cite{MN,AL}.  Here $f_B$ is the $B$ meson decay
constant, defined in such a way that the matrix element of the weak
axial-vector current $A_\mu \equiv \bar b \gamma_\mu \gamma_5 d$ between a
$B^0$ meson and the vacuum is $\langle 0 | A_\mu | B^0(p) \rangle = i p_\mu
f_B$.  With this normalization, the decay constants of the light pseudoscalar
mesons are $f_\pi = 131$ MeV and $f_K = 160$ MeV.  The meson decay constants
express the amplitude for the corresponding quark and antiquark (e.g., $b$ and
$\bar d$) to be found at a point, as they must in order to participate in the
short-distance processes expressed by Figs.~8. 

The factor $B_B$ expresses the degree to which the diagrams of Fig.~8 actually
provide the contribution to $B - \bar B$ mixing.  It is sometimes known as the
``vacuum saturation parameter'' or, more obscurely, as the ``bag parameter''
since an early estimate of the corresponding quantity for kaons was performed
in a ``bag'' model of quarks confined in hadrons.  It has been estimated in
lattice gauge theories \cite{latBB} to be $1.16 \pm 0.08$.  Finally, $\eta_B =
0.55$ is a QCD correction.  All quantities are quoted in the same consistent
renormalization scheme \cite{QCDB} and often appear in the literature with
hats. 

The first hints of $B - \bar B$ mixing were obtained by the UA1 Collaboration
in 1986 \cite{UA1mix}.  In proton-antiproton collisions, an excess was observed
of like-sign muons above background, which could be interpreted as the effects
of $B - \bar B$ pair production, followed by oscillation of one of the
(probably strange) $B$'s to its antiparticle and semileptonic decay of both
$B$'s.  (As we shall see, the mixing amplitude for strange $B$'s is expected to
be much larger than that for nonstrange ones.) 

The first evidence for mixing of nonstrange $B$'s was obtained by the ARGUS
Collaboration in 1987 \cite{ARmix}.  The reaction $e^+ e^- \to B^0 \bar B^0$
was studied at threshold; the observation of like-sign lepton pairs from the
semileptonic decays of $b$ quarks (e.g., $b \to c \ell^- \nu_\ell$, where $\ell
= e$ or $\mu$) indicated that one of the neutral $B$ mesons had oscillated to
its antiparticle.  The large mixing amplitude, $\Delta m /\Gamma \simeq 0.7$
(where $\Delta m$ is the mass difference between mass eigenstates and $\Gamma$
is the $B$ meson decay rate), was one early indication of a very heavy top
quark. 

With the rapidly moving $B$ mesons and the fine vertex information now
available at LEP \cite{LEPmix}, CDF \cite{CDFmix}, and SLD \cite{SLDmix} (see
also the review by \cite{BH}), it has become possible to directly observe
time-dependent $B^0 - \bar B^0$ oscillations with a modulating factor
$\sin(\Delta m t)$ (where $t$ is the proper decay time).  The current world
average \cite{Dmavg} is $\Delta m_d = 0.470 \pm 0.017~{\rm ps}^{-1}$, where the
subscript refers to the mixing between $B^0 \equiv \bar b d$ and $\bar B^0
\equiv b \bar d$.  Using the expression (\ref{eqn:dmd}) and the parameters
mentioned above, we can then obtain an estimate of $|V_{td}|$, which leads,
once we factor out a term $A \lambda^3$, to the constraint 
\beq \label{eqn:Bmixcon}
|1 - \rho - i \eta | = 1.01 \pm 0.22~~~.
\eeq  
This result can be plotted in the $(\rho,\eta)$ plane as a band bounded by
circles with centers at (1,0).

Now that the top quark mass is known so precisely, the dominant source of error
in Eq.~(\ref{eqn:Bmixcon}) is uncertainty in the $B$ meson decay constant
$f_B$.  Pseudoscalar meson decay constants, through the axial-current matrix
element mentioned earlier, govern purely leptonic processes such as $\pi \to
\mu \nu$, $K \to \mu \nu$, and the recently observed $D_s \to \mu \nu$ and $D_s
\to \tau \nu$. (The $D_s = c \bar s$ is the lowest-lying charmed-strange meson;
see Fig.~5.)  Information on heavy meson decay constants can be improved in
several ways \cite{JRCP}. 

Direct measurements of $D_s$ leptonic decays have been reported by the WA75,
CLEO, BES, and E653 Collaborations \cite{Dsmeas}, and one can also estimate
$D_s$ by assuming factorization in certain $B$ decays in which the charged weak
current produces a $D_s$ \cite{Dsfact}.  A recent compilation of experimental
values \cite{Mart} yields $f_{D_s} = (241 \pm 21 \pm 30)$ MeV. 

Estimates based on flavor SU(3), whereby one relates $f_{D_s}$ to the
corresponding decay constant $f_D$ of the nonstrange charmed meson, yield
ratios $f_D/f_{D_s}$ in the range 0.8 -- 0.9.  Thus, one expects a value of
$f_D$ not far below the current experimental upper bound \cite{MkIII} $f_D <
290$ MeV (90\% c.l.).  This limit was placed by studying the reaction $e^+ e^-
\to D^+ D^-$ just above threshold at the SPEAR storage ring, and looking for
the distinctive decay $D \to \mu \nu$ opposite an identified $D$.  The Beijing
Electron-Positron Collider collaboration \cite{BEPC} has identified one
$D \to \mu \nu$ event, leading to $f_D = 300^{+180+80}_{-150-40}$ MeV.

Direct measurements of $f_B$ are possible once $|V_{ub}|$ is fairly well known,
since the partial width $\Gamma(B \to \ell \nu)$ is proportional to $(f_B
|V_{ub}|)^2$.  The expected branching ratios are about $(1/2) \times 10^{-4}$
for $\tau \nu$ and $2 \times 10^{-7}$ for $\mu \nu$ \cite{JRCP}.  Theoretical
estimates of $f_B$ take many forms, including quark models \cite{FBQ} and
lattice gauge theory calculations \cite{FBL}, leading to a rough range
$f_B = 180 \pm 40$ MeV.

The same SU(3) estimates \cite{FBQ,FBL} for the ratio of $f_D/f_{D_s}$ also
give a very similar ratio of $f_B/f_{B_s}$.  It has been shown that the
equality of these two ratios is to be expected within a few percent
\cite{Grin}. 
\bigskip

\leftline{\bf C.  CP-violating $\k - \bk$ mixing}
\bigskip

The top quark (and its charge $-1/3$ partner, the bottom) were invented
\cite{KM} to explain CP violation in the neutral kaon system.  More than thirty
years after its discovery \cite{CCFT}, this remarkable phenomenon has only a
candidate theory to explain it, with confirmation of the explanation still to
come.  (For a compendium of literature on CP violation, see \cite{CJ}.)

The $\k$ and $\bk$ are strong-interaction eigenstates of opposite strangeness,
but the weak interactions do not conserve strangeness.  Hence the weak
interactions may, and do, pick out linear combinations of $\k$ and $\bk$ in
decay processes \cite{GP}.  As of 1957, when the weak interactions were
understood to violate charge-conjugation invariance C and spatial reflection P
but to preserve their product CP, one expected \cite{KCP} the linear
combination $K_1^0 \equiv (\k + \bk)/\s$, with even CP, to have a much more
rapid decay rate since it could decay to the CP-even final state of two
pions.  The orthogonal linear combination $K_2^0 \equiv (\k - \bk)/\s$, with
odd CP, would live much longer since it was forbidden by CP invariance to decay
to two pions and would have to decay to three pions or a pion and a
lepton-neutrino pair.  In fact, a long-lived neutral kaon does exist \cite{KL}, 
with a lifetime about 600 times that of the short-lived variety.

In 1964 J. Christenson, J. Cronin, V. Fitch, and R. Turlay reported that in
fact the long-lived neutral kaon {\it did} decay to two pions, with an
amplitude whose magnitude is about $2 \times 10^{-3}$ that for the short-lived
$K \to 2 \pi$ decay.  To reach this conclusion they constructed a beam of
neutral long-lived kaons and observed their decays in a spark chamber
\cite{CCFT}. 

One then can parametrize the mass eigenstates as
\beq
K_S~({\rm ``short''}) \simeq K_1 + \epsilon K_2~,~~
K_L~({\rm ``long''}) \simeq K_2 + \epsilon K_1~~~,
\eeq
where $|\epsilon| \simeq 2 \times 10^{-3}$ and the phase of $\epsilon$ turns
out to be about $\pi/4$.  The parameter $\epsilon$ encodes all that is
currently known about CP violation in the neutral kaon system.  But where
does it come from?

One possibility, proposed \cite{sw} immediately after the discovery and still
not excluded, is a ``superweak'' CP-violating interaction which directly mixes
$\k = d \bar s$ and $\bk = s \bar d$.  This interaction would have no other
observable consequences since the $\k - \bk$ system is so sensitive to it!

\begin{figure}
\centerline{\epsfysize = 1.3 in \epsffile {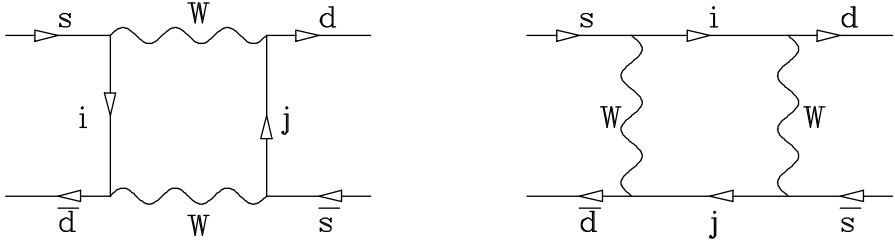}}
\caption{Box diagrams for mixing of a neutral kaon and its antiparticle.}
\end{figure}

The presence of three quark families \cite{KM} poses another opportunity for
explaining CP violation through the mixing diagram shown in Fig.~9.  With three
quark families, phases in complex coupling coefficients cannot be removed by
redefinition of quark phases.  Within some approximations \cite{JRCP}, the
parameter $\epsilon$ is directly proportional to the imaginary part of the
mixing amplitude described in Fig.~9.  Its magnitude (see \cite{CL} or
\cite{TASI} for a calculation in the limit of $m_t \ll M_W$) is \cite{IL} 
\beq \label{eqn:eps}
|\epsilon| \simeq \frac{G_F^2 m_K f_K^2 B_K M_W^2}{\s (12 \pi^2)
\Delta m_K} [\eta_1 S(x_c) I_{cc} + \eta_2 S(x_t) I_{tt} + 2 \eta_3 S(x_c,x_t)
I_{ct}]~~~, 
\eeq
where $I_{ij} \equiv {\rm Im}(V_{id}^* V_{is} V_{jd}^* V_{js})$.
In order to evaluate these expressions we need to work to sufficiently high
order in small parameters in $V$. The application of the unitarity relation to
the first and second rows tells us, in fact, that a more precise expression for
$V_{cd}$ is $V_{cd} = - \lambda - A^2 \lambda^5 (\rho + i \eta).$  We then find
$I_{cc} = - 2 A^2 \lambda^6 \eta$, $I_{ct} = A^2 \lambda^6 \eta$, and $I_{tt} =
2 A^2 \lambda^6 \eta [A^2 \lambda^4 (1- \rho)]$. The factors $\eta_1 =
1.38,~\eta_2=0.57,~\eta_3=0.47$ are QCD corrections \cite{QCDK}, while $x_i
\equiv m_i^2/M_W^2.$  The function $S(x)$ was defined in Eq.~(\ref{eqn:sdef}),
while 
\beq
S(x,y) \equiv xy\left\{ \left[ {1 \over 4} + {3 \over 2(1-y)} - {3 \over
4(1-y)^2} \right] {\ln y \over y-x} + (y \leftrightarrow x) - {3 \over
4(1-x)(1-y)} \right\}~~. 
\eeq

Eq.~(\ref{eqn:eps}) may then be rewritten (updating \cite{HRS}) as
\beq
|\epsilon| = 4.39 A^2 B_K \eta [\eta_3 S(x_c,x_t) - \eta_1 S(x_c) + \eta_2
A^2 \lambda^4 (1-\rho) S(x_t) ]~~~.
\eeq
Using the experimental values \cite{PDG} $|\epsilon| = (2.28 \pm 0.02) \times
10^{-3}$, $f_K = 160~\M$, $\Delta m_K = 3.49 \times 10^{-15}~\G$, and $m_K =
0.4977~\G$, the value $B_K = 0.75 \pm 0.15$ \cite{BKlat}, and the top quark mass
$\bar m_t (M_W) = 165 \pm 6$ GeV/$c^2$, we find that CP-violating $K - \bar K$
mixing leads to the constraint 
\beq \label{eqn:Kmixcon}
\eta(1 - \rho + 0.44) = 0.51 \pm 0.18~~~,
\eeq
where the term $1 - \rho$ in parentheses corresponds to the loop diagram with
two top quarks, and the term 0.44 corresponds to the additional contribution of
charmed quarks.  The major source of error on the right-hand side is the
uncertainty in the parameter $A \equiv V_{cb}/\lambda^2$.
Eq.~(\ref{eqn:Kmixcon}) can be plotted in the $(\rho,\eta)$ plane as a band
bounded by hyperbolae with foci at (1.44,0). 
\bigskip

\leftline{\bf D.  Summary of parameter space}
\bigskip

The constraints (\ref{eqn:Vubcon}), (\ref{eqn:Bmixcon}), and
(\ref{eqn:Kmixcon}) define the allowed region of parameters shown in Fig.~10.
The boundaries shown are $1 \sigma$ errors, but are dominated by theoretical
uncertainties in each case.

\begin{figure}
\centerline{\epsfysize = 3 in \epsffile {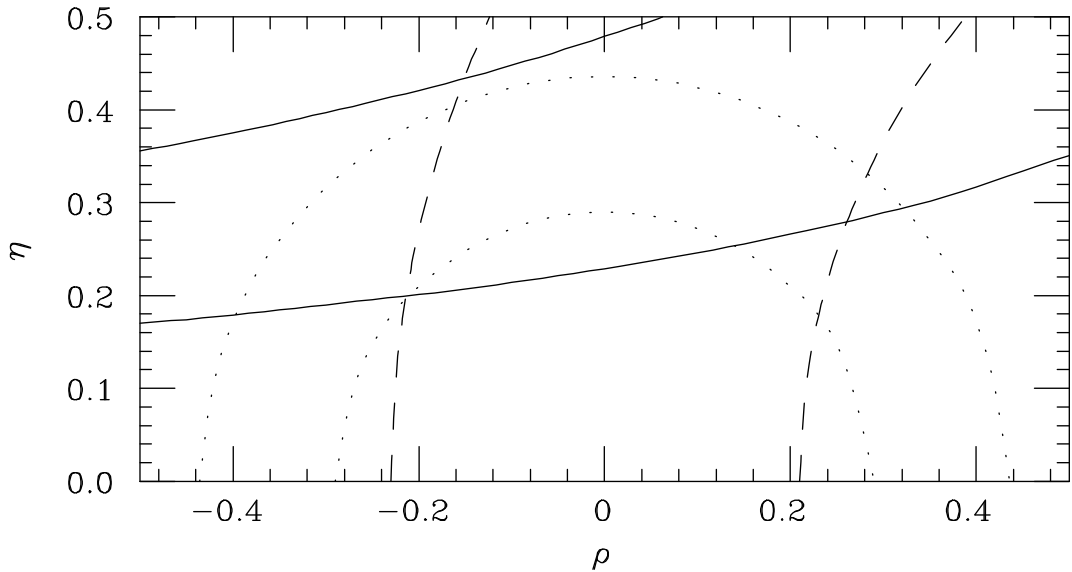}}
\caption{Region in the $(\rho,\eta)$ plane allowed by constraints on
$|V_{ub}/V_{cb}|$ (dotted semicircles), $B^0 - \bar B^0$ mixing (dashed
semicircles), and CP-violating $K - \bar K$ mixing (solid hyperbolae).}
\end{figure}
 
A large region centered about $\rho \simeq 0$, $\eta \simeq 0.35$ is permitted.
Nonetheless, it could be that the CP violation seen in kaons is due to an
entirely different source, perhaps a superweak mixing of $K^0$ and $\bar K^0$
\cite{sw}. In that case one could probably still accommodate $\eta = 0$, and
hence a real CKM matrix, by going slightly outside the $1 \sigma$ bounds based
on $|V_{ub}/V_{cb}|$ or $B - \bar B$ mixing.  In order to confirm the
predicted nonzero value of $\eta$, we turn to other experimental possibilities.
\bigskip

\leftline{\bf E.  CP violation in $B$ meson decays}
\bigskip

We have already mentioned the information provided by $B^0 - \bar B^0$ mixing.
The large value of $\Delta m/\Gamma \simeq 0.7$ was an early hint, through
graphs of the form of Fig.~8, that the top quark was very heavy.  This large
value also has another very important benefit:  It makes possible a class of
incisive studies of CP violation in neutral $B$ meson decays \cite{BCP}.

By comparing rates decays for a state which is produced as $B^0$ and one which
is produced as a $\bar B^0$ to final states which are eigenstates of CP, one
can directly measure angles in the unitarity triangle of Fig.~7.  Because of
the interference between direct decays (e.g., $B^0 \to J/\psi K_S$) and those
which proceed via mixing (e.g., $B^0 \to \bar B^0 \to J/\psi K_S$), these processes are
described by time-dependent functions whose difference when integrated over all
time is responsible for the rate asymmetry. Thus, if we define 
\beq
C_f \equiv \frac{\Gamma(B_{t=0} \to f) - \Gamma(\bar B_{t=0} \to f)}
{\Gamma(B_{t=0} \to f) + \Gamma(\bar B_{t=0} \to f)}~~~,
\eeq
we have, in the limit of a single direct contribution to decay amplitudes,
\beq \label{eqn:asymms}
A(J/\psi K_S, \pi^+ \pi^-) = - \frac{x_d}{1+x_d^2} \sin(2\beta,2\alpha)~~~,
\eeq
where $x_d \equiv \Delta m(B^0)/\Gamma(B^0)$.  This limit is expected to be
very good for $J/\psi K_S$, but some correction for penguin contributions (see
Sec.~2 G) is probably needed for $\pi^+ \pi^-$.  The value $x_d \simeq 0.7$ is
nearly optimum to maximize the coefficient of $\sin(2\beta,2\alpha)$. 

To see this behavior in more detail \cite{Isi}, we note that the time-dependent
partial rates for a state which is initially $B^0~(\bar B^0)$ to decay to a
final state $f$ may be written as 
\beq \label{eqn:dgdt}
d \Gamma [B^0 (\bar B^0) \to f]/dt  \sim e^{- \Gamma t}
[1 \mp {\rm Im} \lambda_0 \sin (\Delta m t)]~~~,
\eeq
where in order to obtain this simple result we neglect $\Delta \Gamma / \Gamma$
in comparison with $\Delta m / \Gamma$. This step is justified for $B$'s, in
contrast to the situation for $K$'s.  The final states to which both $B$ and
$\bar{B}$ can decay are only a small fraction of those to which $B$ or $\bar B$
normally decay, and so one should expect quite similar lifetimes for the two
mass eigenstates.  Integration of (\ref{eqn:dgdt}) gives 
\beq \label{eqn:asy}
C_f = \frac{-x_d}{1 + x_d^2} {\rm Im}\lambda_0(f)
\eeq
for the total asymmetry.  For the CP-eigenstate final states mentioned,
$\lambda_0(J/\psi K_S) = - e^{- 2 i \beta}$ and $\lambda_0(\pi^+ \pi^-) = e^{2
i \alpha}$.  The extra minus sign in the first relation is due to the odd CP of
the $J/\psi K_S$ final state. 

The asymmetry (\ref{eqn:asy}) is suppressed both when $\Delta m/\Gamma$ is very
small and when it is very large (e.g., as is expected for $B_s$). For $B_s$, in
order to see an asymmetry, one must not integrate with respect to time.
Experiments planned with detection of $B_s$ as their focus will require precise
vertex detection to measure mixing as a function of proper time.  For $B^0$, on
the other hand, the value of $x/(1+x^2)$ for $x=0.7$ is 0.47, very close to its
maximum possible value of 1/2 for $x=1$. 

When more than one eigenchannel contributes to a decay, there can appear terms
of the form $\cos (\Delta m t)$ as well as $\sin (\Delta m t)$ in results
analogous to Eqs.~(\ref{eqn:dgdt}) \cite{PP}. These complicate the analysis
somewhat, but information can be obtained from them \cite{pipi} on the relative
contributions of various channels to decays.
\bigskip

\leftline{\bf F.  Lifetime and mass differences for strange $B$'s}
\bigskip

The mixing between strange $B$'s due to diagrams like those in Fig.~8 is
considerably enhanced relative to that between nonstrange $B$'s: 
\beq
\frac{\Delta m_s}{\Delta m_d} = \frac{f_{B_s}^2 B_{B_s}}{f_B^2 B_B}
\left| \frac{V_{ts}}{V_{td}} \right|^2 \simeq 17 - 52~~~,
\eeq
where we have taken the expected ranges of decay constant and CKM element
ratios, and $\Delta m_s$ refers to mixing between the $B_s \equiv \bar b s$
and $\bar B_s \equiv b \bar s$.  Alternatively, we may retrace the steps
of Sec.~2 B, replacing appropriate quantities in Eq.~(\ref{eqn:dmd}), to
derive an analogous expression for $\Delta m_s$, which we then evaluate
directly.  For $V_{ts} = 0.040 \pm 0.004$, $m_{B_s} = 5.37~\G/c^2$,
$f_{B_s} \sqrt{B_{B_s}} = 225$ MeV, $\eta_{B_s} = 0.6 \pm 0.1$, and
\cite{BH} $\tau_{B_s} \equiv 1/\Gamma_s = 1.55 \pm 0.10$ ps, we find
$\Delta m_s/\Gamma_s = 22 \pm 6$, with an additional 40\% error associated
with $f_{B_s}^2 B_{B_s}$.  This result implies many particle-antiparticle
oscillations in a decay lifetime, requiring good vertex resolution and highly
time-dilated $B_s$'s for a measurement.  The present experimental bound
$\Delta m_s > 9.2$ ps$^{-1}$ based on combining ALEPH and DELPHI results
\cite{dms} begins to restrict the parameter space in an interesting manner.

The large value of $\Delta m_s$ entails a value of $\Delta \Gamma_s$ between
mass eigenstates of strange $B$'s which may be detectable.  After all, the
short-lived and long-lived neutral kaons differ in lifetime by a factor of 600.
 Strong interactions and the presence of key channels (e.g., $\pi \pi$) are a
crucial effect in strange particle (e.g., $\k$ and $\bk$) decays.  While the
$b$ quark decays as if it is almost free, so that strong interactions here are
much less important, it turns out that a corresponding difference in lifetimes
for strange $B$'s of the order of 20\% is not unlikely \cite{Blifes}. 

In the ratio $\Delta m_s/\Delta \Gamma_s$, uncertainties associated with the
meson decay constants cancel out, and in lowest order (before QCD corrections
are applied) one finds \cite{IsiBs,BP} $\Delta m_s/\Delta \Gamma_s \simeq {\cal
O}(-[1/\pi] [m_t^2/m_b^2]) \simeq - 200$.  The sign indicates that the heavier
state is expected to be the longer-lived one, as in the neutral kaon system.
The top quark does not contribute to the width difference associated with the
imaginary part of the graphs in Fig.~8, since no $t \bar t$ pairs are produced
in $B_s$ decays. 

Aside from small CP-violating effects, the mass eigenstates of strange $B$'s
correspond to those $B_s^{(\pm)}$ of even and odd CP. The decay of a $\bar B_s$
meson via the quark subprocess $b (\bar s) \to c \bar c s (\bar s)$ gives rise
to predominantly CP-even final states \cite{CPeven}, so the CP-even eigenstate
should have a greater decay rate. One calculation \cite{Blifes} gives 
\beq \label{eqn:widthdiff}
\frac{\Gamma(B_s^{(+)}) - \Gamma(B_s^{(-)}) }{\overline \Gamma} \simeq 0.18
\frac{f_{B_s}^2}{(200~\M)^2}~~~,
\eeq
while a more recent estimate \cite{Beneke} is $0.16^{+0.11}_{-0.09}$.
The lifetime difference between CP-even and CP-odd strange $B$'s thus can
provide useful information on $f_{B_s}$, and hence indirectly on the weak
interactions at short distances.  One way to learn this lifetime difference
\cite{DDLR} is to study angular distributions in $B_s \to J/\psi + \phi \to e^+
e^- K^+ K^-$ (or $\mu^+ \mu^- K^+ K^-$).  The $J/\psi$ and $\phi$ are both
spin-1 particles and hence can be produced in states of orbital angular momenta
$L = 0,~1$, and 2 from the spinless $B_s$ decay.  $L=1$ corresponds to $P = CP
= -$, while $L = 0$ or 2 corresponds to $P = CP = +$.  A simple transversity
analysis \cite{Tr} permits one to separate the two cases. 

In the rest frame of the $J/\psi$, let the $x$ axis be defined by the direction
of the $\phi$, the $x-y$ plane be defined by the kaons which are its decay
products, and the $z$ axis be the normal to that plane.  Let the $e^+$ (or
$\mu^+$) make an angle $\theta$ with the $z$ axis.  Then the CP-even final
states give rise to an angular distribution $1 + \cos^2 \theta$, while
the CP-odd state gives rise to $\sin^2 \theta$.  In case both CP eigenstates
are present in $B_s \to J/\psi \phi$, one will see a gradual increase of the
$\sin^2 \theta$ component relative to the $1 + \cos^2 \theta$ component.
More likely (if predictions \cite{CPeven} are correct), the CP-even state
will dominate, so that one will be measuring mainly the lifetime of this
eigenstate when following the time-dependence of the decay.  The average
decay rate $\bar \Gamma \equiv (\Gamma_+ + \Gamma_-)/2$ (the subscripts
denote CP eigenvalues) is measured in flavor-tagged decays of $B_s = \bar b s
\to \bar c + \ldots$ or $\bar B_s = b \bar s \to c + \ldots$.

A recent analysis \cite{CLEOTr} of $B \to J/\psi K^*$ has some bearing on
the $B_s \to J/\psi \phi$ partial-wave structure.  The two processes are
related by flavor SU(3), involving a substitution $s \leftrightarrow d$
of the spectator quark.  Thus, one expects the same partial waves in the
two decays.  The CLEO Collaboration has studied 146 $B \to J/\psi K^*$ decays
in $3.36 \times 10^6~B \bar B$ pairs produced at the Cornell Electron
Storage Ring (CESR).

One can decompose the amplitude $A$ for a spinless meson to decay to two vector
mesons into three independent components \cite{FMJR}, corresponding to linear
polarization states of the vector mesons which are either longitudinal (0), or
transverse to their directions of motion and parallel ($\parallel$) or
perpendicular ($\perp$) to one another. The states $0$ and $\parallel$ are
P-even, while the state $\perp$ is P-odd. Aside from the case of $A_0$, which
is special for massive vector mesons, these arguments were advanced some time
ago \cite{Yang} to determine the parity of the neutral pion in its decays to
two photons. Since $J/\psi$ and $\phi$ are both C-odd eigenstates, the
properties under P are the same as those under CP.

A suitably normalized amplitude $A_\perp$ thus can be expressed in terms of
partial-wave amplitudes $S,~P$, and $D$ as $|A_\perp|^2 = |P|^2 /
(|S|^2+|P|^2+|D|^2)$.  In $B_s \to J/\psi \phi,~|P|^2$ would correspond to a
CP-odd final state and hence to the decay of the odd-CP eigenstate.  The CLEO
results \cite{CLEOTr} for the related process $B \to J/\psi K^*$ are
$|A_\perp|^2 = 0.21 \pm 0.14$ from a fit to the transversity angle, and
$|A_\perp|^2 = 0.16 \pm 0.08 \pm 0.04$ from a fit to the full angular
distribution.  This implies (via flavor SU(3)) that $B_s \to J/\psi \phi$ is
dominated by the CP-even final state, and thus the lifetime in this state
measures approximately $\tau(B_s^{(+)})$.  It will be interesting to see if
any evidence for non-zero $|P|^2$ can be gathered in $B \to J/\psi K^*$, in
which case $B_s \to J/\psi \phi$ should exhibit the time-variation in the
transversity-angle distribution mentioned above \cite{DDLR}. 
\bigskip

\leftline{\bf G.  Processes dominated by penguin diagrams}
\bigskip

Although the unitarity of the CKM matrix implies flavor conservation for
charge-preserving electroweak interactions in lowest order, we have seen
that loop diagrams can induce flavor-changing charge-preserving interactions
in higher order.  Another example of this phenomenon is provided by the
``penguin'' diagram \cite{pen} illustrated in Fig.~11.  Although the penguin's
``leg'' is a gluon in this illustration, it can also be a photon or $Z$. When
the external quarks $x$ and $y$ have charge $-1/3$, the intermediate quarks
have charge 2/3 and can include the top quark.  Because of the top quark's
large mass, such penguin diagrams can be very important. 
 
\begin{figure}
\centerline{\epsfysize = 2 in \epsffile {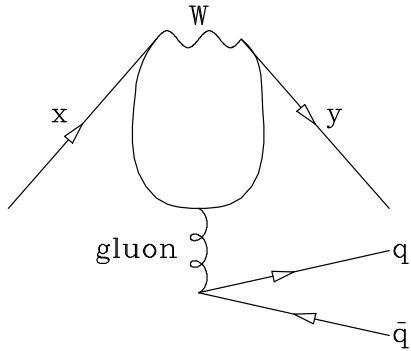}}
\caption{``Penguin'' diagram describing transition of a quark $x$ to another
quark $y$ with the same charge.  The intermediate quarks have charge
differing from $Q(x) = Q(y)$ by one unit.  Here $q = (u,d,s)$.}
\end{figure}

An example of a predicted penguin effect in $s \to d$ transitions is a phase
arising in the decays of neutral kaons to $\pi \pi$.  This phase can lead to a
``direct'' contribution to the ratios for CP-violating and CP-conserving
decays, in addition to that provided by the mixing parameter $\epsilon$
measured earlier. 

One may define
\beq
\eta_{+-} \equiv \frac{A(K_L \to \pipe)}{A(K_S \to \pipe)}~;~~
\eta_{00} \equiv \frac{A(K_L \to \poop)}{A(K_S \to \poop)}~~;
\eeq
the effect of ``direct'' decays then shows up in a parameter $\epsilon'$
which causes $\eta_{+-}$ and $\eta_{00}$ to differ from one another:
\beq
\eta_{+-} = \epsilon + \epsilon'~~~;~~~\eta_{00} = \epsilon - 2 \epsilon'~~~.
\eeq
Since $\epsilon'$ and $\epsilon$ are expected to have approximately the same
phase (see, e.g., \cite{JRCP}), one expects
$$
| \eta_{+-} | \simeq | \epsilon | [ 1 + {\rm Re} (\epsilon '/\epsilon) ] ~~~,
$$
\beq
| \eta_{00} | \simeq | \epsilon | [ 1 - 2 {\rm Re} (\epsilon'/\epsilon ] ~~~,
\eeq
and hence
\beq \label{eqn:ratrat}
\left | \frac{\eta_{00}}{\eta_{+-}} \right |^2 =
\frac{\Gamma (K_L \to 2 \pi^0)}{\Gamma ( K_S \to 2 \pi^0 )} /
\frac{\Gamma (K_L \to \pipe )}{\Gamma (K_S \to \pipe )} = 1 - 6~{\rm Re}
\frac{\epsilon '}{\epsilon} ~~~.
\eeq
Present expectations \cite{epspth} are that $\epsilon'/\epsilon$ could be a few
parts in $10^4$ (but in any case $\epsilon'/\epsilon \le 10^{-3}$), requiring
the ratio of ratios (\ref{eqn:ratrat}) to be measured to about one part in
$10^3$.  Experiments now in progress at Fermilab and CERN should have the
required sensitivity.  The previous results of these experiments are:
\beq
{\rm E731}~\cite{E731}: ~~ {\rm Re} (\epsilon '/ \epsilon) = (7.4 \pm 6.0)
\times 10^{-4} ~~~,
\eeq
\beq
{\rm NA31}~\cite{NA31}: ~~ {\rm Re} (\epsilon '/\epsilon) = (23.0 \pm 6.5)
\times 10^{-4} ~~~,
\eeq
leading to some question about whether a non-zero value has been observed.
Because of the cancelling effects of gluonic and ``electroweak'' penguins
(in which the curly line in Fig.~11 is a photon or $Z$), the actual magnitude
of $\epsilon'/\epsilon$ is difficult to estimate, so that one's best hope is
for a non-zero value within the rather large theoretical range, thereby
disproving the superweak model \cite{sw} of CP violation.

The contribution of the (gluonic) penguin diagram to the effective weak
Hamiltonian may be written (for $x,y$ equal to quarks of charge $-1/3$)
$$
{\cal H}_W^{\rm penguin} \simeq \frac{G_F}{\s}\frac{\alpha_s}{6 \pi}
\left[ \xi_c \ln \frac{m_t^2}{m_u^2} + \xi_t \ln \frac{m_t^2}{m_u^2} \right]
$$
\beq \label{eqn:pen}
\left[ (\bar y_L \gamma^\mu \lambda^a x_L)(\bar u \gamma_\mu \lambda^a u
+ \bar d \gamma_\mu \lambda^a d + \ldots) + \Hc \right]~~~,
\eeq
where $\xi_i \equiv V_{ix} V^*_{iy}$, and $\lambda^a$ are color SU(3) matrices
[$a = (1,\ldots,8)$] normalized so that Tr$(\lambda^a \lambda^b) = 2 
\delta^{ab}$.  The top quark is dominant in the flavor-changing processes
$b \to d$ and $b \to s$ (with corrections due to charm which can be important
in some cases \cite{BuFl}), while the charmed quark dominates the $s \to d$
process.  (The top quark plays a key role, however, in the electroweak penguin
contribution to this process \cite{epspth}.)  One may imitate the effect of
an infrared cutoff for the gluonic penguin graph by using a constituent-quark
mass $m_u \sim 0.3~\G/c^2$.

One can estimate the effect of Eq.~(\ref{eqn:pen}) for the $b \to s q \bar q$
penguin graph; one finds it is comparable to that of the $b \to u d \bar u$
``tree'' contribution $(G_F/\s)V_{ub}V^*_{ud} [\bar u \gamma_\mu (1 - \gamma_5)
b][\bar d \gamma^\mu (1 - \gamma_5) u]$.  The penguin contribution to $b \to d
q \bar q$ and the $b \to u s \bar u$ tree contribution are both expected to be
suppressed by approximately one power of the Wolfenstein parameter $\lambda
\sim 0.2$, as one can see by comparing CKM elements.  Thus, $B^0 \to \pi^+
\pi^-$ is expected to be dominated by the tree amplitude; $B^0 \to K^+ \pi^-$
is expected to be dominated by the penguin amplitude; and the rates of the two
processes should be similar.

In Fig.~12 we show a contour plot of the significance of detection by the
CLEO Collaboration \cite{Battle} of the decays $B^0 \to \pipe$ and $B^0 \to
K^+ \pi^-$.  Evidence exists for a combination of $B^0 \to K^+ \pi^-$ and
$\pi^+ \pi^-$ decays, generically known as $B^0 \to h^+ \pi^-$. On the basis of
2.4 fb$^{-1}$ of data, the most recent published result \cite{Wurt} is $B(B^0
\to h^+ \pi^-) = (1.8^{~+0.6~+0.2} _{~-0.5~-0.3} \pm 0.2) \times 10^{-5}$. 
Although one still cannot conclude that either decay mode is nonzero at the $3
\sigma$ level, the most likely solution is roughly equal branching ratios
(i.e., about $10^{-5}$) for each mode.  Only upper limits exist for other modes
of two pseudoscalars \cite{CLEOGlas}, but these are consistent with predictions
\cite{BPP}. 

\begin{figure}
\centerline{\epsfysize = 4 in \epsffile {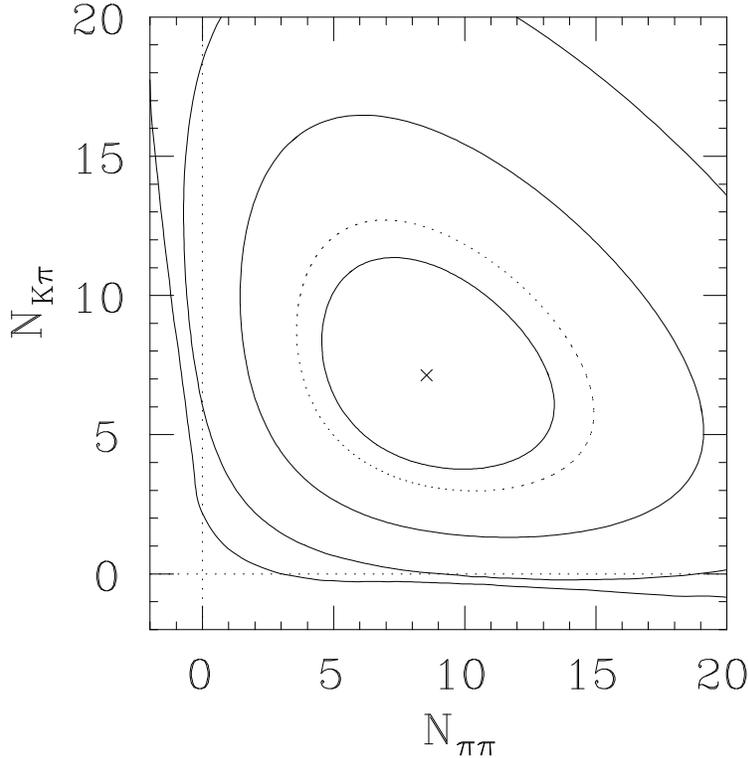}}
\caption{Contours of significance of detection of $B^0 \to \pipe$ and $B^0 \to
K^+ \pi^-$ by the CLEO Collaboration.}
\end{figure}

Penguin diagrams play a number of roles in $B$ decays.  We enumerate several of
them; others are mentioned in \cite{JRCP}. 

1.  The process $B^+ \to K^0 \pi^+$ is expected to be almost completely due to
the penguin graph; one cannot write a corresponding tree graph for it.  By
comparison with the $B^0 \to K^+ \pi^-$ decay, where the penguin graph is
expected to be the main contribution, one expects $B(B^+ \to K^0 \pi^+) \simeq
10^{-5}$.  The weak phase of the process (which changes sign under
charge-conjugation) thus is expected to be Arg($V_{tb}^* V_{ts}) = \pi$, so
that the charge-conjugate process has the same weak phase. As a result, one
turns out to be able to separate strong final-state interaction phases from
weak phases and obtain estimates of quantities like the angle $\gamma = {\rm
Arg}(V_{ub}^*)$ in Fig.~7 by comparing rates for $B^+ \to (K^0 \pi^+,~K^+
\pi^0,~K^+ \eta,~K^+ \eta')$ with the corresponding $B^-$ rates \cite{GReta}. 
One can also obtain this information by measuring the time-dependence in $B^0
(\bar B^0) \to \pipe$ and the rates for $B^0 \to K^+ \pi^-$, $B^+ \to K^0
\pi^+$, and the charge-conjugate processes \cite{DGR}.  (The rates for $B^+ \to
\k \pi^+$ and $B^- \to \bk \pi^-$ are expected to be equal because of the
penguin dominance mentioned above.)  The weak phases of the major amplitudes
contributing to these decays are summarized in Table 3.  The relative weak
phase of tree and penguin amplitudes for strangeness-preserving decays is
$\gamma + \beta = \pi - \alpha$ (assuming the unitarity triangle to be valid),
while the corresponding relative phase for strangeness-changing decays (aside
from a sign) is just $\gamma$.  As a result, one can measure both $\alpha$
and $\gamma$.

\begin{table}
\caption{Phases of amplitudes contributing to decays of $B$ mesons to
$\pi \pi$ and $K \pi$.  Here $\Delta S$ refers to the change of strangeness
in the process.}
\begin{center}
\begin{tabular}{c|c c|c c} \hline
  & \multicolumn{2}{c|}{``Tree''} & \multicolumn{2}{c}{``Penguin''} \\
\cline{2-3} \cline{4-5}
 $|\Delta S|$ & CKM elements     &   Phase  & CKM elements & Phase \\
       0     & $V^*_{ub}V_{ud}$ & $\gamma$ & $V^*_{tb} V_{td}$ & $-\beta$ \\
       1     & $V^*_{ub}V_{us}$ & $\gamma$ & $V^*_{tb} V_{ts}$ & $\pi$ \\
\hline 
\end{tabular}
\end{center}
\end{table}

2.  A number of processes (in addition to the decay $B^+ \to K^0 \pi^+$
mentioned above) are dominated by penguin graphs.  By comparing the rates for
strangeness-preserving and strangeness-changing processes, one can measure the
ratio $|V_{td}/V_{ts}|$ \cite{GRP}.  Examples of useful ratios are $B(B^+ \to
\bar K^{*0} K^+)/B(B^+ \to \phi K^+)$ and $B(B^+ \to \bar K^{*0} K^{*+})/B(B^+
\to \phi K^{*+})$. 

3.  We mentioned in Sec.~2 E that the time-integrated rate asymmetry in $B \to
\pipe$ could provide information on the angle $\alpha$ of the unitarity
triangle.  The most direct test is based on the assumption that the tree
process $\bar b \to \bar u u \bar d$ is the only direct contribution to this
decay.  However, ``penguin pollution'' \cite{PP} (due to the $b \to d$
transition) makes the analysis less straightforward, even thought the penguin
amplitude is expected to be only about 0.2 of the tree amplitude.  Ways to
circumvent this difficulty include the detailed study of the isospin structure
of the $\pi \pi$ final state \cite{pipi}, and the use of flavor SU(3) to
estimate penguin effects using $B \to K \pi$, where they are expected to be
dominant \cite{BPP,DGR,SilWo}. 
\bigskip

\leftline{\bf H.  Rare kaon decays}
\bigskip

We give a sample of processes influenced by loop diagrams in which the
top quark plays a role.  The present discussion is based on a recent
discussion of $K \to \pi \nu \bar \nu$ decays \cite{BB}.  Details on other
processes may be found in Refs. \cite{JRCP,TASI,QCDK,Dib,RVW}.

We are concerned with the subprocess $s \to d \nu \bar \nu$, which can
proceed via loop diagrams involving exchange of a pair of $W$'s (a box
diagram) or one $W$ and one $Z$ (an electroweak penguin).  In each case
results will be quoted for a sum over the three neutrino species.

1.  The decay $K^+ \to \pi^+ \nu \bar \nu$ is predicted to have a branching
ratio
\beq \label{eqn:kpinunu}
B(K^+ \to \pi^+ \nu \bar \nu) \simeq 10^{-11} A^4 |T(x_t)(1 - \rho - i \eta
+ [0.41 \pm 0.09])|^2~~~,
\eeq
where $x_t \equiv m_t^2/M_W^2$, and
\beq
T(x) \equiv \frac{x}{4} \left[ \left( \frac{3x-6}{(x-1)^2} \right)
\ln x + \frac{x+2}{x-1} \right]~~~.
\eeq
The term $1 - \rho - i \eta$ in (\ref{eqn:kpinunu}) is the contribution of
the top quark, while the term in square brackets represents the contribution
of charm, uncertain primarily because the charmed quark mass is not all that
well known.  Uncertainty in the Wolfenstein parameter $A$ contributes a
good deal to the error on the predicted branching ratio.  The top quark's
mass used to be a major source of error in this estimate, but is no longer.
For the present value of $m_t$, $T(x_t) \simeq 3$.

An experimentally measured branching ratio will define a band in the
$(\rho,\eta)$ plane whose boundaries are circles with centers at $(1.41 \pm
0.09,0)$.  Given the region of parameters already allowed in Fig.~10, such a
measurement essentially specifies $\rho$.  The predicted branching ratio is
$10^{-10}$, give or take a factor of 2.  The present experimental limit from an
experiment (E787) at Brookhaven National Laboratory \cite{E787} is $B(K^+ \to
\pi^+ \nu \bar \nu) < 2.4 \times 10^{-9}$ (90\% c.l.).

2.  The decay $K_L \to \pi^0 \nu \bar \nu$ is expected to be purely
CP-violating \cite{GaL}.  The predicted branching ratio is
\beq
B(K_L \to \pi^0 \nu \bar \nu) \simeq 4.4 \times 10^{-11} \eta^2 A^4
|T(x_t)|^2~~~. 
\eeq
For the allowed range of $A$ and $\eta$, one expects a branching ratio of $2
\times 10^{-11}$, give or take a factor of 2.  While the present upper limit 
\cite{pinunu} of $5.8 \times 10^{-5} (90\%$ c.l.) is quite far from this, it
represents a considerable advance from previous limits, and much improvement
is expected in the next few years.
\newpage

\leftline{\bf 3.  PRECISION ELECTROWEAK EXPERIMENTS}
\bigskip

Tests of the electroweak theory have reached the precision that they are
sensitive to the mass of the top quark.  Indeed, the large top quark mass had
been anticipated to some extent by these experiments.  In this section we
review the lowest-order theory, show how it is affected by the top quark in
loop diagrams, and explore the sensitivity of precision electroweak
measurements to further types of heavy particles which could shed light on the
top quark's mass.  We draw heavily on the discussion in Ref.~\cite{APV}, for
which the present discussion serves in part as an update.
\bigskip

\leftline{\bf A.  Lowest-order formalism}
\bigskip

In order to construct a renormalizable theory of the weak interactions, it was
proposed quite early \cite{Wref} to regard the four-fermion interaction as the
limit at zero momentum transfer of a process in which an intermediate vector
boson (now called $W$) was exchanged, so that one identifies $G_F/\s =
g^2/8M_W^2$.  The Fermi coupling constant is well known: $G_F = 1.16639(2)
\times 10^{-5}~\G^{-2}$.  However, neither the coupling constant $g$ nor the
$W$ mass $M_W$ were known when this proposal was made.  Searches for $W$'s as
light as a couple of $\G/c^2$ were undertaken in the early 1960's. 

If $W^\pm$ exchange is to be described by a gauge interaction, the simplest
group containing charged $W$'s is SU(2), which also contains a $W^0$.  But
one cannot identify the $W^0$ with a photon.  Aside from the fact that the
symmetry must be badly broken, so that the photon remains massless while the
charged $W$'s become very heavy, there are two other difficulties.  First, the
charged $W$'s couple to matter with a $V-A$ interaction \cite{VA}, i.e., to
left-handed particles, while the photon couples via a $V$ interaction, i.e., to
both left-handed and right-handed particles.  Second, if the photon really were
identified with $W^0$ all the charges of particles would have to be
half-integer, just like values of $J_3$, the third component of angular
momentum. 

The solution \cite{GWS} is to add an SU(2) singlet $B$ to the theory, with
another coupling constant $g'$.  The photon can then be one mixture of the
$B$ and the $W^0$, while there will be another neutral particle which is
the orthogonal mixture:
$$
{\rm Photon:}~~ A =  B \cos \theta + W^0 \sin \theta~~~(m = 0)~~~;
$$
\beq
{\rm ~~New~~~:}~~Z = -B \sin \theta + W^0 \cos \theta~~~(m \ne 0)~~.
\eeq
The $W^0$ couples to the third component $I_{3L}$ of the SU(2). The subscript
``$L$'' stands for ``left-handed'' and refers to the fact that all three
components of the SU(2), both the charged $W$'s and the $W^0$, couple only
to left-handed particles or right-handed antiparticles.  The $B$ couples
to a new quantum number ``weak hypercharge,'' invented to account for
the difference between electromagnetic charge and $I_{3L}$:
\beq
Q_{\rm em} = I_{3L} + \frac{Y_W}{2}~~~.
\eeq
Once the interaction Lagrangian is expressed in terms of the photon and $Z$,
one finds that the electron charge $e$ is related to the other parameters in
the theory by $e = g \sin \theta = g' \cos \theta$, so that
\beq \label{eqn:lomw}
\frac{1}{e^2} = \frac{1}{g^2} + \frac{1}{g'^2}~~;~~~
M_W = \left( \frac{\pi \alpha}{\s G_F} \right)^{1/2} / \sin \theta~~~.
\eeq
Moreover, the $Z$ mass is related to the $W$ mass:
\beq \label{eqn:lomz}
M_Z = \frac{M_W}{\cos \theta}~~;~~~
\frac{G_F}{\s} = \frac{g^2+g'^2}{8 M_Z^2}~~~.
\eeq
\bigskip

\leftline{\bf B.  Effects in loop diagrams}
\bigskip

The effects of fermions in loop diagrams, as shown in Fig.~13, are seen
in several ways in the electroweak interactions.

\begin{figure}
\centerline{\epsfysize = 1.2 in \epsffile {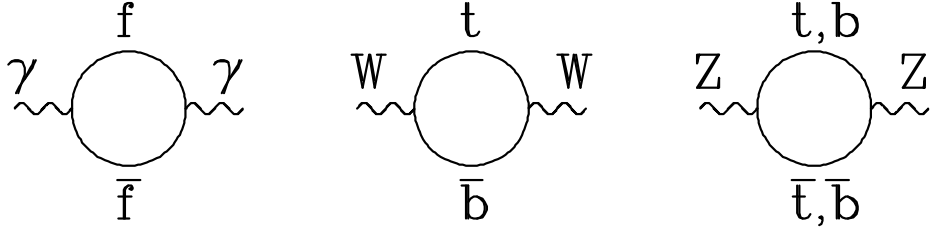}}
\caption{Loop diagrams with virtual fermions leading to important vacuum
polarization effects.  (a) Photons (all fermions $f$ below the top quark
are important); (b) $W$; (c) $Z$.}
\end{figure}

The photon vacuum polarization effects shown in Fig.~13(a) lead to an
effective fine structure constant $\alpha(q^2)$ at an invariant momentum
scale $q^2$ which is related to the usual $\alpha \simeq 1/137.036$ at
$q^2 = 0$ by
\beq
\alpha^{-1}(q) = \alpha^{-1} [1 - \Pi_{\ga}(q^2)]~~;~~~
\Pi_{\ga} = \frac{\alpha}{3 \pi} \sum_f Q_f^2 \left[ - \frac{5}{3}
+ \ln \frac{q^2}{m_f^2} \right]~~~.
\eeq
The quantity $\Pi_{\ga}$ can be evaluated in part using data on $\eep \to {\rm
hadrons}$; several recent precise determinations \cite{alpha} lead to
$\alpha^{-1}(M_Z) \simeq 128.9 \pm 0.1$.  This quantity is scheme-dependent;
one also sees it quoted in as $127.9 \pm 0.1$ in the $\overline{{\rm MS}}$
scheme \cite{Sirlinalpha}.

If we had used $\alpha^{-1} = 137.036$, the expressions we have written for the
$W$ and $Z$ masses would imply, using the latest value \cite{LEPEWWG} $\sin^2
\theta = 0.23165$, that $M_W = 77.5~\G/c^2$ and $M_Z = 88.4~\G^2$.  The use of
the correct value of $\alpha^{-1}(M_Z)$ changes these predictions to $M_W =
79.86~\G/c^2$, $M_Z = 91.11~\G/c^2$.  These values are to be compared with the
experimental ones $M_W = 80.34 \pm 0.10~\G/c^2$ (see below) and $M_Z = 91.1863
\pm 0.0020~\G/c^2$ (see \cite{LEPEWWG}, \cite{Pich} and \cite{Treille}). The
use of the correct value of $\alpha$ considerably improves the predictions, but
the experimental ratio $M_W/M_Z$ is higher than predicted.  The major part of
the reason may be traced to the effect of the very heavy top quark in the loop
diagrams of Figs.~13(b) and 13(c). 

Both vector and axial-vector currents enter in the couplings of $W$ and
$Z$ to fermions; the axial-vector currents are not conserved when the
fermions have mass.  Moreover, even the vector currents are not conserved
when the fermions have unequal masses [as in Fig.~13(b)].  The result is
the appearance of contributions to $\Pi$ for the $W$ and $Z$ which are
quadratic in fermion masses.  No such contributions appear in $\Pi_{\ga}$,
as a result of electromagnetic gauge invariance and current conservation.

The low-energy limits of $W$ and $Z$ exchange are now described by
\beq \label{eqn:low}
\frac{G_F}{\sqrt{2}} = \frac{g^2}{8M_W^2} ~~~,~~~
\frac{G_F}{\sqrt{2}} \rho = \frac{g^2 +{g'}^2}{8M_Z^2} ~~~,
\eeq
where the parameter $\rho$, which receives contributions from quark loops to
$W$ and $Z$ self-energies, is dominated by the top \cite{Tini}: 
\beq \label{eqn:rho}
\rho \simeq 1 + \frac{3G_F m_t^2}{8 \pi^2 \sqrt{2}} ~~~,
\eeq
Consequently, if we define $\theta$ by means of the precise measurement
at LEP of $M_Z$,
\beq
M_Z^2 = \frac{\pi \alpha}{\sqrt{2} G_F \rho \sin^2 \theta \cos^2 \theta}
~~~,
\eeq
then $\theta$ will depend on $m_t$, and so will $M_W$ in Eq.~(\ref{eqn:lomw}).
The ratio $M_W/M_Z = \sqrt{\rho} \cos \theta$ then may be used to extract a
value $\sqrt{\rho} = 1.0050 \pm 0.0013$, implying $m_t = 179 (1 \pm 0.13)
~\G/c^2$.  We shall see that the inclusion of other electroweak data does not
greatly alter this simple result. 

In order to display dependence of electroweak observables on such quantities as
the top quark and Higgs boson masses $m_t$ and $M_H$, we choose to expand the
observables about ``nominal'' values calculated for specific $m_t$ and $M_H$.
We thereby bypass a discussion of ``direct'' radiative corrections which are
independent of $m_t,~M_H$, and new particles.  We isolate the dependence on
$m_t,~M_H$, and new physics arising from ``oblique'' corrections \cite{PT},
associated with loops in the $W$ and $Z$ propagators like those shown in
Figs.~13(b,c). 

For $m_t = 175$ GeV, $M_H = 300$ GeV, the measured value of $M_Z$ leads to a
nominal expected value of $\sin^2 \theta_{\rm eff} = 0.2315.$  In what follows
we shall interpret the effective value of $\sin^2 \theta$ as that measured via
leptonic vector and axial-vector couplings: $\sin^2 \theta_{\rm eff} \equiv
(1/4)(1 - [g_V^{\ell}/g_A^{\ell}])$.  We have corrected the nominal value of
$\sin^2 \theta_{\rm \overline{MS}} \equiv \hat s^2$ as quoted by DeGrassi,
Kniehl, and Sirlin \cite{DKS} for the difference \cite{GS} $\sin^2 \theta_{\rm
eff} - \hat s^2 = 0.0003$ and for the recent change in the evaluation of
$\alpha(M_Z)$ \cite{alpha}. 

Defining the parameter $T$ by $\Delta \rho \equiv \alpha T$, we find 
\begin{equation} \label{eqn:Teq}
T \simeq \frac{3}{16 \pi \sin^2 \theta} \left[ \frac{m_t^2 - (175
~{\rm GeV})^2}{M_W^2} \right] - \frac{3}{8 \pi \cos^2 \theta}
\ln \frac{M_H}{300~{\rm GeV}} ~~~.
\end{equation}
The weak mixing angle $\theta$, the $W$ mass, and other electroweak observables
depend on $m_t$ and $M_H$. 

The weak charge-changing and neutral-current interactions are probed under a
number of different conditions, corresponding to different values of momentum
transfer.  For example, muon decay occurs at momentum transfers small with
respect to $M_W$, while the decay of a $Z$ into fermion-antifermion pairs
imparts a momentum of nearly $M_Z/2$ to each member of the pair. Small
``oblique'' corrections \cite{PT}, logarithmic in $m_t$ and $M_H$, arise from
contributions of new particles to the photon, $W$, and $Z$ propagators. Other
(smaller) ``direct'' radiative corrections are important in calcuating actual
values of observables. 

We may then replace (\ref{eqn:low}) by
\beq
\frac{G_F}{\sqrt{2}} = \frac{g^2}{8 M_W^2} \left( 1 + \frac{\alpha S_W}{4
\sin^2 \theta} \right)~~~,~~~
\frac{G_F \rho}{\sqrt{2}} = \frac{g^2 + {g'}^2}{8M_Z^2} \left( 1 + \frac{\alpha
S_Z}{4 \sin^2 \theta \cos^2 \theta} \right)~~~, 
\eeq
where $S_W$ and $S_Z$ are coefficients representing variation with momentum
transfer. Together with $T$, they express a wide variety of electroweak
observables in terms of quantities sensitive to new physics.  (The presence of
such corrections was noted quite early by Veltman \cite{MVS}.)  The
Peskin-Takeuchi \cite{PT} variable $U$ is equal to $S_W - S_Z$, while $S \equiv
S_Z$.

Expressing the ``new physics'' effects in terms of deviations from nominal
values of top quark and Higgs boson masses, we have the expression
(\ref{eqn:Teq}) for $T$, while contributions of Higgs bosons and of possible
new fermions $U$ and $D$ with electromagnetic charges $Q_U$ and $Q_D$ to $S_W$
and $S_Z$, in a leading-logarithm approximation, are \cite{KenL} 
\beq \label{eqn:sz}
S_Z = \frac{1}{6 \pi} \left [
\ln \frac{M_H}{300~\G/c^2} + \sum N_C \left ( 1 - 4 \overline Q \ln
\frac{m_U}{m_D} \right ) \right ] ~~~,
\eeq
\beq \label{eqn:sw}
S_W = \frac{1}{6 \pi} \left [
\ln \frac{M_H}{300 ~\G/c^2} + \sum N_C \left ( 1 - 4 Q_D \ln \frac{m_U}{m_D}
\right ) \right ]~~.
\eeq
The expressions for $S_W$ and $S_Z$ are written for doublets of fermions with
$N_C$ colors and $m_U \geq m_D \gg m_Z$, while $\overline Q \equiv (Q_U + Q_D )
/2$. The sums are taken over all doublets of new fermions. In the limit $m_U =
m_D$, one has equal contributions to $S_W$ and $S_Z$. For a single Higgs boson
and a single heavy top quark, Eqs.~(\ref{eqn:sz}) and (\ref{eqn:sw}) become 
$$
S_Z = \frac{1}{6 \pi} \left [ \ln \frac{M_H}{300~\G/c^2} - 2 \ln
\frac{m_t}{175~\G/c^2} \right ] ~,~~
$$
\beq
S_W = \frac{1}{6 \pi} \left [ \ln \frac{M_H}{300~\G/c^2} + 4 \ln
\frac{m_t}{175~\G/c^2} \right ] ~,
\eeq
where the leading-logarithm expressions are of limited validity for $M_H$
and $m_t$ far from their nominal values.  A degenerate heavy fermion doublet
with $N_c$ colors thus contributes $\Delta S_Z = \Delta S_W = N_c/(6 \pi)$.
For example, in a minimal dynamical symmetry-breaking (``technicolor'')
scheme about which we shall have more to say presently, with a single
doublet of $N_c = 4$ fermions, one will have $\Delta S = 2/(3 \pi) \simeq
0.2$.  This will turn out to be marginally acceptable, while many non-minimal
schemes, with large numbers of doublets, will be seen to be ruled out.
\bigskip

\leftline{\bf C.  Analysis of present data}
\bigskip

We shall discuss a number of observables, including neutrino deep inelastic
scattering, direct $W$ mass measurements, direct measurements of a number of
$Z$ properties, and parity violation in atoms.

Among the first pieces of evidence for neutral currents was the presence of
muonless events in neutrino deep inelastic scattering \cite{Hasert}. The ratios
of neutrino and antineutrino neutral-current (NC) to charged-current (CC) cross
sections, $R_\nu$ and $R_{\bar \nu}$, are predicted to be \cite{Lls} 
$$
R_\nu \equiv
\frac{\sigma_{NC} ( \nu N)}{\sigma_{CC} ( \nu N)}
= \rho^2 \left [ \frac{1}{2} - x + \frac{5}{9} x^2 (1+r) \right ]~~~,
$$
\beq
R_{\bar \nu} \equiv
\frac{\sigma_{NC} (\bar  \nu N)}{\sigma_{CC} (\bar  \nu N)}
= \rho^2 \left [ \frac{1}{2} - x + \frac{5}{9} x^2
(1+ \frac{1}{r}) \right ] ~~,
\eeq
where $x \equiv \sin^2 \theta$ and $r \equiv \sigma_{CC} ( \bar \nu N) /
\sigma_{CC} ( \nu N )$.  These expressions are proportional to $\rho^2$ since
they involve neutral current {\em cross sections} (squares of amplitudes). 

\begin{figure}
\centerline{\epsfysize = 3in \epsffile {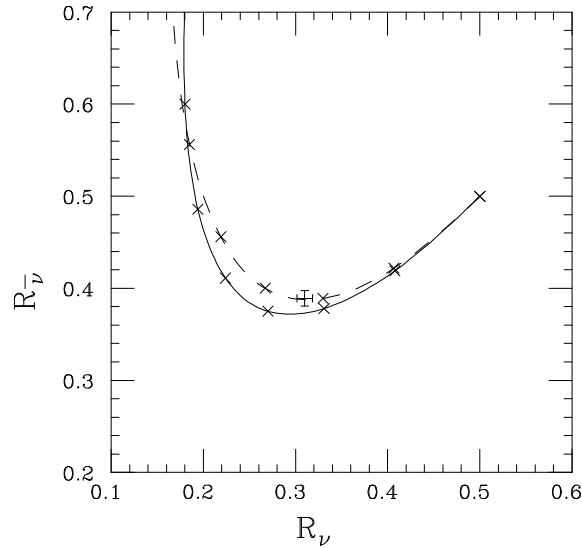}}
\caption{Ratios $R_\nu$ and $R_{\bar \nu}$ plotted against one another for
varying values of $\sin^2 \theta$ and $\rho = 1$.  The $\times$ marks denote
units of 0.1 in $\sin^2 \theta$, starting at the point (0.5,0.5) for $\sin^2
\theta = 0$. The dashed line corresponds to $r = 1/3$, which would be
appropriate if the nucleon had no antiquarks but were composed only of valence
quarks.  The solid line corresponds to $r = 0.4$ (approximately the measured
value).  Plotted point illustrates measured values.} 
\end{figure}

The resulting plot of $R_\nu$ vs. $R_{\bar \nu}$ as a parametric function
of $\sin^2 \theta$ is shown in Fig.~14.  One gets a ``nose-like'' curve. The
experimental rate for $\bar \nu N \to \bar \nu + \ldots$ (the plotted point is
based on Ref.~\cite{CCFR}) is about as low as it can be, while the rate for
$\nu N \to \nu + \ldots$ is very sensitive to $\sin^2 \theta$ and provides a
good measure of it.  It turns out that the $x$ and $\rho$ dependences
combine in such a way that $R_\nu$ actually depends on $m_t$ and $M_H$ in very
much the same way as does $M_W$.  As a result, measurements of $R_\nu$ are
often quoted as effective measurements of $M_W$, which is what we shall do
here. 

Averaging a new measurement \cite{CCFR} with previous determinations
\cite{CDHS,CHARM}, one finds a result equivalent to \cite{LEPEWWG,Blondel} $M_W
= 80.306 \pm 0.218~\G/c^2$ when one takes into account the measured value of
$M_Z$. [A somewhat different average was quoted in \cite{CCFR}, implying $M_W =
80.220 \pm 0.208~\G/c^2$.] 

Direct measurements of the $W$ mass have been presented by the CDF and D0
Collaborations \cite{CDFW,D0W}.  These use kinematic fitting to the sharp
Jacobian peak in $\bar p p \to W + \ldots \to \ell \nu_\ell + \ldots$, where
$\ell = (e,\mu)$.  Now, for the first time, it is possible as a result of the
increased energy at LEP to directly measure a point on the excitation curve in
$\eep \to W^+ W^-$ \cite{LEPW}, yielding a surprisingly accurate value for only
a few dozen events.  These results are compared with an earlier determination
by the UA2 Collaboration \cite{UA2W} in Table 4.  (See also
Ref.~\cite{Treille}.) 

\begin{table}
\caption{Recent direct measurements of the $W$ mass.}
\begin{center}
\begin{tabular}{c c c c c} \hline
Collaboration & $M_W~(\G/c^2)$ \\ \hline
CDF (Run 1A) & $80.41 \pm 0.18$\\
D0 (Run 1A) & $80.35 \pm 0.27$\\
D0 (Run 1B) & $80.38 \pm 0.17$\\
LEP & $80.3 \pm 0.4 \pm 0.1$ \\
UA2 & $80.36 \pm 0.37$ \\ \hline
Average & $80.356 \pm 0.125$ \\ \hline
\end{tabular}
\end{center}
\end{table}

The $W$ {\it width} $\Gamma_W$ is an interesting quantity (see also
Ref.~\cite{Treille}). A popular way of measuring it which is sometimes quoted
as giving the total width is to measure the {\it ratio} of signals $\sigma(\bar
p p \to W + \ldots \to \ell \nu_\ell + \ldots)/\sigma(\bar p p \to Z + \ldots
\to \ell^+ \ell^- + \ldots)$.  In fact, however, since one needs to combine the
standard model prediction for $\Gamma(W \to \ell \nu)$ with this ratio in
order to extract $\Gamma_W$, what one actually ends up measuring is the
{\it branching ratio} $B(W \to \ell \nu_\ell)$.

The standard model predicts \cite{Wwidth}
\beq
B(W \to \ell \nu_\ell) \simeq \frac{1}{3 + 6[1 + (\alpha_s(M_W)/\pi)]} = 0.1084
\pm 0.0002
\eeq
if the $W \to t \bar b$ channel is closed and there are no unanticipated
channels other than the three species of $\ell \nu_\ell$ and the lighter-quark
channels.  The world average \cite{Treille} is $0.110 \pm 0.003$.  On the other
hand, a cruder but less theory-dependent measurement of the total width
\cite{CDFWwidth} makes use of the different dependence on $\Gamma_W$ of charged
lepton pair production on and off the $W$ peak, and leads to $\Gamma_W = 2.11
\pm 0.28 \pm 0.26$ GeV in accord with the standard model prediction
\cite{Treille} of $2.077 \pm 0.014$ GeV.  The ``oblique'' corrections mentioned
above turn out not to affect the $W$ width \cite{Wwidth}. 

A number of measurements of $Z$ properties have become possible as a result
of the huge statistics amassed at the LEP $\eep$ Collider in the reaction
$\eep \to Z \to \ldots$.  The ability of the SLC Collider at Stanford
to polarize electron beams longitudinally has permitted useful information
to be obtained at that machine despite a much lower collision rate.  The
numbers we shall quote are those contained in the latest version of the
LEP Electroweak Working Group document \cite{LEPEWWG}, which should be
consulted for further references.

The mass of the $Z$, now measured to be $M_Z = 91.1863 \pm 0.0020~\G/c^2$, can
serve an input to the electroweak theory when combined with $G_F$ and
$\alpha(M_Z)$ to predict all other observables as functions of $m_t$ and $M_H$.
Deviations of observables from ``nominal'' values predicted for specific
choices of $m_t$ and $M_H$ can be described as linear functions of the
parameters $S,~T$, and $U$ mentioned above.  Every observable is a homogeneous
function of degree 0, 1, or 2 of the factor $\rho = \rho^{\rm nominal}(1 +
\alpha T)$ and of $\sin^2 \theta(S,T)$.  Thus, every observable will define a
band in the $S-T$ plane (assuming for present purposes that $S_W = S_Z$, i.e.,
$U=0$). 

Other measured $Z$ parameters \cite{LEPEWWG} include the total width $\Gamma_Z
= 2.4946 \pm 0.0027~{\rm GeV}$, the hadron production cross section
$\sigma_h^0 = 41.508 \pm 0.056$ nb, and $R_\ell \equiv \Gamma_{\rm
hadrons}/\Gamma_{\rm leptons} = 20.778 \pm 0.029$, which may be combined to
obtain the $Z$ leptonic width $\Gamma_{\ell\ell}(Z) = 83.91 \pm 0.11$ MeV.

A number of determinations of $\sin^2 \theta$ are obtained from asymmetries
measured at LEP.  These are summarized in Table 5.

\begin{table}
\caption{Determinations at LEP of $\sin^2 \theta_{\rm eff}$}
\begin{center}
\begin{tabular}{c l c c} \hline
Observable & Name & Value & Error \\ \hline
$A_{FB}^{\ell}$ & Forward-backward asymmetry & 0.23085 & 0.00056 \\
$A_\tau$ & See notes (a,b) & 0.23240 & 0.00085 \\
$A_e$ & See notes (a,b) & 0.23264 & 0.00096 \\
$\langle Q_{FB} \rangle$ & F-B charge asymm., light quarks & 0.23200 &
0.00100 \\
$A_{FB}^c$ & F-B charm asymmetry & 0.23155 & 0.00112 \\
$A_{FB}^b$ & F-B bottom asymmetry & 0.23246 & 0.00041 \\ \hline
$\sin^2 \theta_{\rm eff}^{\rm LEP}$ & LEP Average & 0.23200 & 0.00027 \\
\hline
\end{tabular}
\end{center}
\leftline{(a) $A_f \equiv 2 g_V^f g_A^f/[(g_V^f)^2 + (g_A^f)^2]$}
\leftline{(b) $A_\tau,A_e$ extracted from angular dependence of $\tau$
polarization}
\end{table}

The LEP average, $\sin^2 \theta_{\rm eff}^{\rm LEP} = 0.23200 \pm 0.00027$, is
to be compared with that based on the left-right asymmetry parameter $A_{LR}$
measured with polarized electrons at SLC \cite{SLC}: $\sin^2 \theta_{\rm
eff}^{\rm SLC} = 0.23061 \pm 0.00047$.  The $\chi^2$ for the average in Table 5
is 6.3 for 5 degrees of freedom, while it rises to 12.8 for 6 degrees of
freedom when the SLC value is added. 

Parity violation in atoms, stemming from the interference of $Z$ and photon
exchanges between the electrons and the nucleus, provides further information
on electroweak couplings.  The most precise constraint at present arises from
the measurement of the {\it weak charge} (the coherent vector coupling of the
$Z$ to the nucleus), $Q_W = \rho(Z - N - 4 Z \sin^2 \theta)$, in atomic cesium
\cite{CW}, with the result $Q_W({\rm Cs}) = -71.04 \pm 1.58 \pm 0.88$.  The
first error is experimental, while the second is theoretical \cite{Csth}.  The
prediction \cite{MR} $Q_W({\rm Cs}) = -73.20 \pm 0.13$ is insensitive to
standard-model parameters \cite{MR,JRRC,PS} once $M_Z$ is specified;
discrepancies are good indications of new physics.  Recently the weak charge
has also been measured in atomic thallium.  Averaging determinations by the
Seattle \cite{TlS} and Oxford \cite{TlO} in a manner described in more detail
in \cite{APV}, we find $Q_W({\rm Tl}) = -115.0 \pm 4.5$, to be compared with a
prediction \cite{PSBL,Tlth,Tlthr} of $-116.8$. 

We have performed a fit to the electroweak observables listed in Table 6.  The
``nominal'' values (including \cite{DKS} $\sin^2 \theta_{\rm eff} = 0.2315$)
are calculated for $m_t = 175$ GeV and $M_H = 300$ GeV.  We use $\Gamma_{\ell
\ell}(Z)$, even though it is a derived quantity, because it has little
correlation with other variables in our fit.  It is mainly sensitive to the
axial-vector coupling $g_A^\ell$, while asymmetries are mainly sensitive to
$g_V^\ell$.

In order to focus on electroweak parameters and avoid the use of highly
correlated measurements, we omit several quantities from the fit which are
normally included in complete analyses. (1) The total width $\Gamma_{\rm
tot}(Z)$ is omitted since it is highly correlated with $\Gamma_{\ell \ell}(Z)$
and mainly provides information on the value of the strong fine-structure
constant $\alpha_s$. With $\alpha_s = 0.12 \pm 0.01$, the observed total $Z$
width is consistent with predictions.  For the same reason, we omit the
quantity $R_\ell$.  (2) The invisible width of the $Z$ is compatible with three
species of neutrinos.  It would provide information on $\rho/\rho^{\rm nominal}
 = 1 + \alpha T$ but is less useful as a source of that information than the
leptonic width, and is highly correlated with it.  (3) The experimental
situation regarding the partial width $\Gamma(Z \to b \bar b)$ has changed
recently \cite{AZbb}.  We omit it from the fit while discussing it separately
below. 

\begin{table}
\caption{Electroweak observables described in fit.}
\begin{center}
\begin{tabular}{c c c} \hline
Quantity        &   Experimental   &   Theoretical \\
                &      value       &    value      \\ \hline
$Q_W$ (Cs)      & $-71.0 \pm 1.8^{~a)} $  &  $ -73.2^{~b)} - 0.80S - 0.005T$\\
$Q_W$ (Tl)      & $-115.0 \pm 4.5^{~c)} $ &  $ -116.8^{~d)} -1.17S - 0.06T$ \\
$M_W$ (GeV)     & $80.341 \pm 0.104^{~e)}$  & $80.35^{~f)} -0.29S + 0.45T$ \\
$\Gamma_{\ell\ell}(Z)$ (MeV) & $83.91 \pm 0.11^{~g)}$ & $83.90 -0.18S
+ 0.78T$ \\
$\sin^2 \theta_{\rm eff}$ & $0.23200 \pm 0.00027^{~h)}$ & $0.2315^{~i)}
 + 0.0036S - 0.0026T$ \\
$\sin^2 \theta_{\rm eff}$ & $0.23061 \pm 0.00047^{~j)}$ & $0.2315^{~i)} +
0.0036S - 0.0026T$ \\ \hline 
\end{tabular}
\end{center}
\leftline{$^{a)}$ {\small Weak charge in cesium \cite{CW}}}
\leftline{$^{b)}$ {\small Calculation \cite{MR} incorporating 
atomic physics corrections \cite{Csth}}}
\leftline{$^{c)}$ {\small Weak charge in thallium \cite{TlS,TlO} (see text)}}
\leftline{$^{d)}$ {\small Calculation \cite{PSBL} incorporating
atomic physics corrections \cite{Tlth}}}
\leftline{$^{e)}$ {\small Average of direct measurements
and indirect information}}
\leftline{{\small \quad from neutral/charged current ratio in
deep inelastic neutrino scattering \cite{CCFR,CDHS,CHARM}}}
\leftline{$^{f)}$ {\small Including perturbative QCD corrections \cite{DKS}}}
\leftline{$^{g)}$ {\small LEP average \cite{LEPEWWG}}}
\leftline{$^{h)}$ {\small From leptonic asymmetries at LEP \cite{LEPEWWG}}}
\leftline{$^{i)}$ {\small As calculated \cite{DKS} with correction for
relation between $\sin^2 \theta_{\rm eff}$ and $\hat s^2$ \cite{GS}}}
\leftline{$^{j)}$ {\small From left-right asymmetry in annihilations at
SLC \cite{SLC}}}
\end{table}

Each observable in Table 6 specifies a band in the $S - T$ plane with different
slope, as seen from the ratios of coefficients of $S$ and $T$.  Parity
violation in atomic cesium and thallium is sensitive almost entirely to
$S$ \cite{MR,PS}. The impact of $\sin^2 \theta_{\rm eff}$ determinations on $S$
is considerable. The leptonic width of the $Z$ is sensitive primarily to $T$.
The $W$ mass specifies a band of intermediate slope in the $S-T$ plane; here we
assume $S_W = S_Z$.  Strictly speaking, the ratio $R_\nu$ specifies a band
with slightly more $T$ and less $S$ dependence than $M_W$ \cite{PT,MR};
we have ignored this difference here.

The resulting constraints on $S$ and $T$ are shown in Fig.~15(a).  A top quark
mass of $175 \pm 6~\G/c^2$ (the CDF and D0 average), corresponding as we noted
in Sec.~2 to $\bar m_t(M_W) = 165 \pm 6~\G/c^2$, is compatible with all Higgs
boson masses between 100 and 1000 $\G/c^2$, as seen by the curved lines
intersecting the error ellipses.  The centers of the ellipses lie at $S =
0.05$, $T = 0.01$, indicating that the fit is very comfortable with the nominal
values chosen for $m_t$ and $M_H$. 

Independently of the standard model predictions, values of $S$ between $-0.3$
and $0.3$ are permitted at the 90\% confidence level.  The values from atomic
parity violation, $S = -2.7 \pm 2.3$ \cite{CW} based on cesium and $S = -1.5
\pm 3.8$ based on thallium \cite{TlS,TlO}, have an average $S = -2.4 \pm 2.0$.
The value of $S$ is now known much more precisely than specified by these
experiments. On the other hand, since the weak charge $Q_W$ provides unique
information on $S$, its determination with a factor of four better accuracy
than present levels could have a noticeable effect on global fits, as shown in
Fig.~15(b). 

\begin{figure}
\centerline{\epsfysize = 3in \epsffile {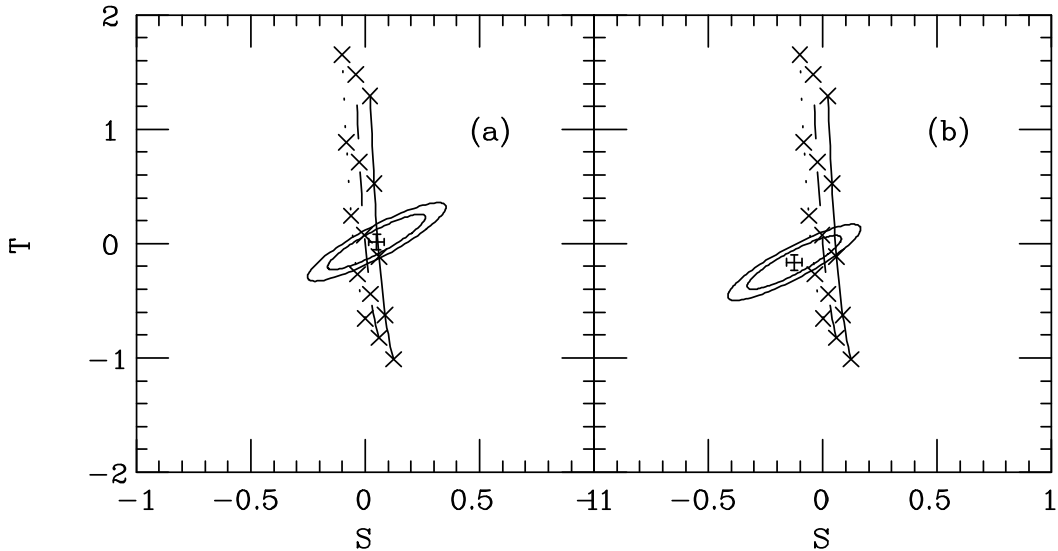}}
\caption{Allowed ranges of $S$ and $T$ at 68\% (inner ellipses) and 90\%
(outer ellipses) confidence levels, corresponding to $\chi^2 = 2.3$ and 4.6
above the minimum (crosses at center of ellipses).  Dotted, dashed, and solid
lines correspond to standard model predictions for $M_H = 100$, 300, 1000 GeV. 
Symbols $\times$, from bottom to top, denote predictions for $m_t = 100$, 140,
180, 220, and 260 GeV. (a) Fit including APV experiments with present errors;
(b) errors on APV experiments reduced by a factor of 4, with present central
values of $Q_W$ retained.}
\end{figure}

In contrast to many fits in the literature (see, e.g., \cite{LLM,EFL,deB}),
ours does not exhibit a preference for any particular Higgs boson mass in the
absence of separate information about $m_t$.  We show curves of $\chi^2$
as a function of $m_t$ for several values of $M_H$ in Fig.~16.

\begin{figure}
\centerline{\epsfysize = 4in \epsffile {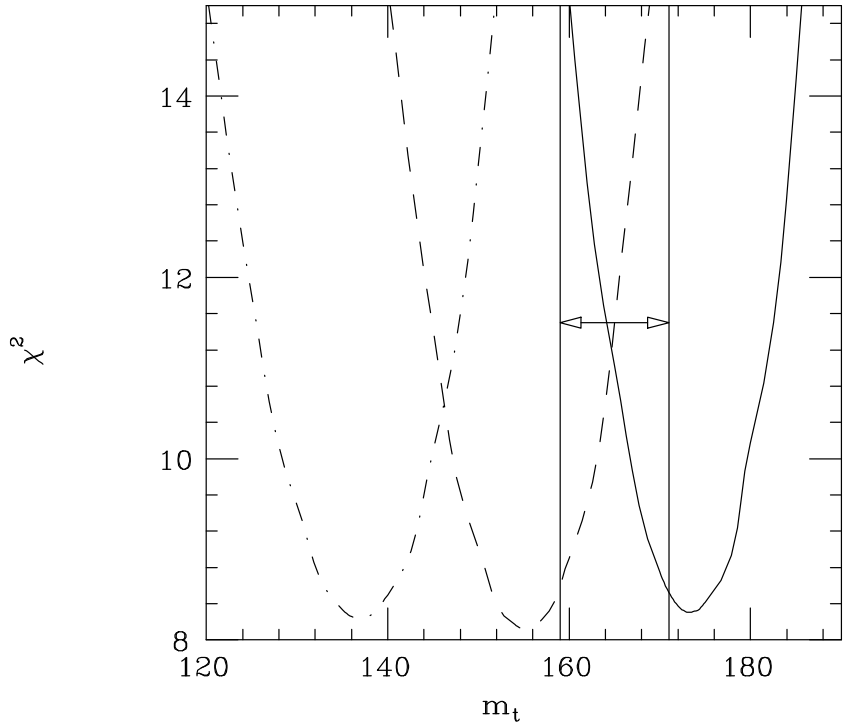}}
\caption{Values of $\chi^2$ for fits to the data in Table 6, as function of
$m_t$ for several values of $M_H$.  Dot-dashed, dashed, and solid curves
correspond to $M_H = 100,~300$, and $1000~\G/c^2$, respectively.  Horizontal
bands denote the $1 \sigma$ limits for $\bar m_t(M_W) = 165 \pm 6~\G/c^2$.}
\end{figure}

The minima of the $\chi^2$ curves are nearly identical (slightly above
8 for the six data points in Table 6, i.e., for 5 degrees of freedom) for
all three choices of $M_H$.  When the value $\bar m_t(M_W) = 165 \pm 6~\G/c^2$
(delimited by the horizontal band) is selected, some preference arises for a
Higgs boson mass in the range between 300 and 1000 $\G/c^2$.  To see this more
clearly, we include $m_t$ in the fit in Fig.~17. 

\begin{figure}
\centerline{\epsfysize = 3in \epsffile {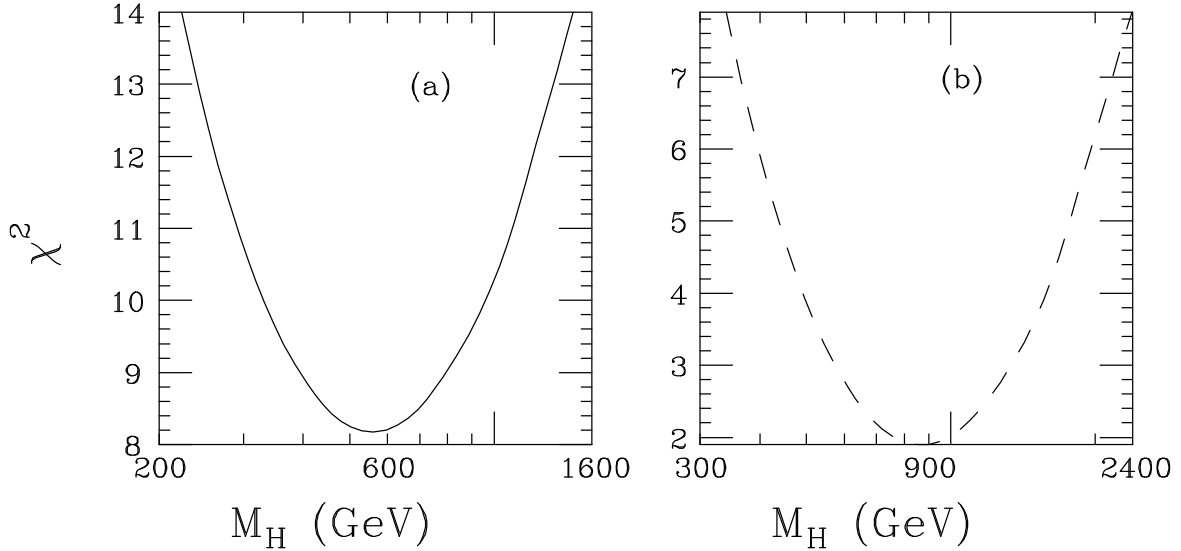}}
\caption{Values of $\chi^2$ for fits to the data in Table 6 and the value $\bar
m_t(M_W) = 165 \pm 6~\G/c^2$, as function of $M_H$.  (a) SLC data (last row
of Table 6) included (6 d.o.f.); (b) SLC data omitted (5 d.o.f.).}
\end{figure}

The favored value of $M_H$ in the fit based on the data in Table 6 and the top
quark mass mentioned above is $M_H = (560 \times 1.5^{\pm 1})~\G/c^2$. (See
Fig.~17(a).) This is considerably higher than most other determinations (see,
e.g., \cite{LLM,EFL,deB}).  The experimental average of the two last rows in
Table 6, $\sin^2 \theta_{\rm eff} = 0.23165 \pm 0.00024$ is slightly above our
``nominal'' value of $\sin^2 \theta_{\rm eff} = 0.2315$; a discrepancy in this
direction tends to push the Higgs boson mass up from its nominal value of 300
$\G/c^2$.  To accentuate this effect we explore the effect of omitting the SLC
data on $A_{LR}$ (the last row in Table 6).  The small upward change in $\sin^2
\theta_{\rm eff}$, by $+0.00035$, leads the favored $M_H$ to increase by a
factor of 1.5, to $(820 \times 1.7^{\pm 1})~\G/c^2$.  (See Fig.~17(b).)  The
$\chi^2$, incidentally, drops by more than 6. 

An effect which leads us to obtain a larger $M_H$ than some other fits is that
our ``nominal'' $\sin^2 \theta_{\rm eff} = 0.2315$ is slightly below theirs.
For example \cite{LEPEWWG}, the fitted value $\sin^2 \theta_{\rm eff}= 0.2318
\pm 0.0002$ is quoted for $m_t = 177 \pm 7~\G/c^2$ when $M_H = 300 ~\G/c^2$.

A further possible reason for the discrepancy between our fits and others is
that the others incorporate $R_\ell \equiv \Gamma(Z \to {\rm hadrons})/\Gamma(Z
\to \ell^+ \ell^-)$, whose experimental value of $20.778 \pm 0.029$ is somewhat
above the nominal value (20.757 in Ref.~\cite{LEPEWWG}).  We have chosen to
omit this quantity since it depends on $\alpha_s$.  Experimental determinations
of $\alpha_s(M_Z)$ may be converging rapidly enough that one can trust them;
the value \cite{Schmelling} $\alpha_s(M_Z) = 0.118 \pm 0.003$ was quoted
recently. The quantity $\rho$ cancels in $R_\ell$.  Aside from its dependence
on $\alpha_s$ (we find $\partial R_\ell/\partial \alpha_s = 6.7$), $R_\ell$
provides a measure of $x \equiv \sin^2 \theta$; we find $\partial R/\partial x
= -17.5$.  Thus a high $R_\ell$ implies a low $\sin^2 \theta$.  Low
experimental values of $\sin^2 \theta$ push $M_H$ down. 

Another possibility is that some fits allow a separate parameter to describe
$A_{FB}^b$.  This quantity, as shown by the last row of Table 5, has a small
quoted error.  If it is omitted, the LEP average becomes $\sin^2 \theta_{\rm
eff} = 0.23163 \pm 0.00037$ with $\chi^2 = 4.0$ for 4 d.o.f., while if the SLC
result is added to this sample, one finds $\sin^2 \theta_{\rm eff} = 0.23124
\pm 0.00029$ with $\chi^2 = 6.9$ for 5 d.o.f.  This low value again favors
low $M_H$.

In any event the difference between Figs.~17(a) and (b) shows the pitfalls of
drawing a conclusion about $M_H$ when its value is so sensitive to a
single datum.  One is really only determining $\log M_H$, whose value is
affected by less than $1 \sigma$ by including or dropping the SLC data. 
\bigskip

\leftline{\bf D.  The decay $Z \to b \bar b$}
\bigskip

We now return to the question of the $Z \to b \bar b$ branching ratio, which
has engaged considerable theoretical attention in the past couple of years
\cite{Zbb}.  We first discuss the calculation of $\Gamma (Z \to f \bar f)$ for
any fermion $f$. 

The interaction Lagrangian for a $Z$ and a fermion described by the Dirac field
$\psi$ is
\beq
{\cal L}_{\rm int} = - \sqrt{g^2 + g'^2} \bar \psi \gamma^\mu Z_\mu
(I_{3L} {\cal P}_L - Q \sin^2 \theta) \psi~~~,
\eeq
where ${\cal P}_L \equiv (1 - \gamma_5)/2$. As an example, for a $b$ quark with
$I_{3L} = -(1/2)$, the axial and vector coupling constants (in a convenient
normalization) are $g_A = 1/4$, $g_V = -(1/4) + (1/3)\sin^2 \theta$.  The $Z$
partial width is then 
$$
\Gamma(Z \to f \bar f) = \frac{\sqrt{2} G_F M_Z^3 \rho}{3 \pi} N_c
\left( 1 - \frac{4 m_f^2}{M_Z^2} \right)^{1/2} 
$$
\beq
\times \left[ g_A^2 \left( 1 - \frac{4 m_f^2}{M_Z^2} \right)
+ g_V^2 \left( 1 + \frac{2 m_f^2}{M_Z^2} \right) \right]~~~.
\eeq
For a massless neutrino $g_V = -g_A = 1/4$, so
\beq
\Gamma(Z \to \nu \bar \nu) = \frac{G_F M_Z^3 \rho}{12 \pi \s} = 166~\M
\times \rho~~~.
\eeq
As mentioned, the fact that the invisible width of the $Z$ is about 1/2
GeV leads one to conclude that there are three species of light neutrinos.

A calculation of $\Gamma(Z \to d \bar d) \equiv \Gamma_d$, for which $g_A^2 =
1/16 = 0.0625$ and $g_V^2 = 0.0299$, gives $(368~\M)(\rho)(1 +
[\alpha_s/\pi])$, or, with $\rho \simeq 1.01$ and $1 + (\alpha_s/\pi) = 1.04$,
about 387 MeV.  When compared with \cite{LEPEWWG} $\Gamma_h \equiv \Gamma(Z \to
{\rm hadrons}) = 1743.6 \pm 2.5~\M$, this gives $R_d^0 \equiv \Gamma(Z \to d
\bar d) /\Gamma(Z \to {\rm hadrons} \simeq 0.222$.  More precise calculations
(see, e.g., \cite{Zbb}) give $R_d^0 \simeq 0.221$.  We distinguish between
$R_q^0 \equiv \Gamma(Z \to q \bar q)/\Gamma(Z \to {\rm hadrons})$ and what is
actually measured at the $Z$ peak, $R_q \equiv \sigma(e^+ e^- \to q \bar q)/
\sigma(e^+ e^- \to {\rm hadrons})$.  As a result of a small photon
contribution, $R_b^0 = R_b + 0.0003$ \cite{heavy}. 

The non-zero $b$ mass must be taken into account in $Z \to b \bar b$.  The
predominance of $g_A^2$ over $g_V^2$ for the $Z b \bar b$ coupling (in a
ratio of about 2 to 1) means that the kinematic factor $\beta^3 \simeq 0.980$
for axial couplings, where $\beta = [1 - (4m_b^2/M_Z^2)]^{1/2}$, has an
appreciable effect.  The corresponding kinematic factor for vector couplings,
$\beta(3 - \beta^2)/2$, is nearly 1.  Thus there is an overall reduction in
$R_b/R_d$ by about 1.3\% just due to $m_b$, leading to the kinematically
corrected prediction \cite{JS} $R^0_{b~\rm kin} = 0.2183 \pm 0.0001$.

The top quark affects $\Gamma(Z \to b \bar b)$ through graphs like those shown
in Fig.~18.  (In 't Hooft -- Feynman gauge one must also include contributions
of charged Higgs bosons, replacing $W$ in all possible ways.) Contributions
quadratic in $m_t$ appear \cite{Zbbcorrs}, as in the $\rho$ parameter, so that
$R_b^0/R^0_{b~\rm kin} = 1 - a G_F m_t^2$, where $a$ is a dimensionless
constant.  For the observed value of the top quark mass the vertex-corrected
standard-model prediction is $R_b^0 = 0.2158$ \cite{LEPEWWG}, amounting to an
additional correction of $-1.2\%$. 

\begin{figure}
\centerline{\epsfysize = 2in \epsffile {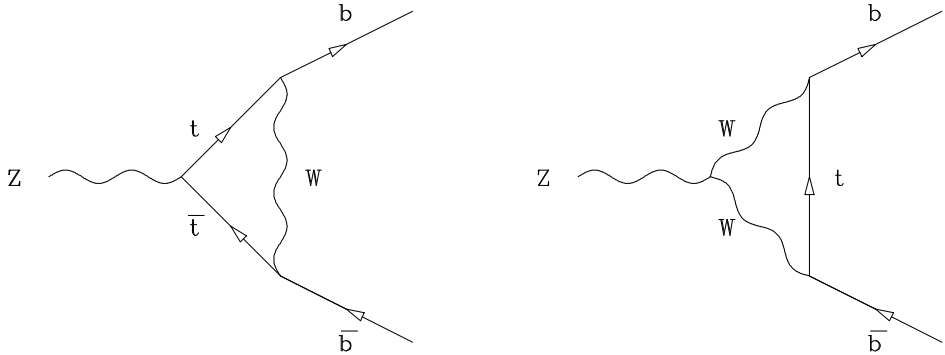}}
\caption{Examples of graphs contributing to $\Gamma(Z \to b \bar b)$.}
\end{figure}

Until this past summer, results from the LEP and SLD groups seemed to indicate
a value of $R_b$ noticeably above the standard-model prediction.  More recently
\cite{AZbb}, the ALEPH Collaboration has analyzed the full LEP I data set using
two different methods.  Using a lifetime tag, they find $R_b = 0.2169 \pm
0.0011 \pm 0.0017$, while using five mutually exclusive hemisphere tags they
find $R_b = 0.2158 \pm 0.0009 \pm 0.0011$.  The current LEP average
\cite{LEPEWWG}, incorporating some new results from other LEP groups and from
SLD \cite{SLDRb}, is $R_b^0 = 0.2178 \pm 0.0011$. These results are in accord
with standard model predictions.  They do not yet conclusively confirm the
$-1.2\%$ effect due to top-quark graphs like those in Fig.~18, but at least
indicate that $R_b < R_d$. 
\newpage

\leftline{\bf 4.  TOP QUARK OBSERVATION AT THE TEVATRON}
\bigskip

\leftline{\bf A.  Production and decay}
\bigskip

Quite some time ago \cite{EHLQ,JRtop} it was realized that the Fermilab
Tevatron would be able to produce top quarks heavy enough to decay to $W + b$.
An estimate \cite{JRtop} in 1984 predicted $\sigma(\bar p p \to t \bar t +
\ldots) \sim 10~{\rm pb}(m_t/150~\G/c^2)^{-5}$ at c.m. energy $\sqrt{s} = 2$
TeV. The major subprocesses contributing to $t \bar t$ production ($q$ and $g$
stand for quark and gluon) are $q \bar q \to g^* \to t \bar t$ and $g g \to t
\bar t$ (via $t$ exchange or a gluon in the direct channel).  The $q \bar q$
subprocess is dominant for the actual situation, so one has to know the
distribution of quarks and antiquarks inside the nucleon.  This information is
provided, for example, by deep inelastic scattering \cite{Feltesse}.  We shall
quote more precise predictions presently. 

The decay of the top quark for $m_t > M_W + m_b$ is essentially 100\% to
$W + b$ unless other channels (like $H^+ + b$, where $H^+$ is a charged
Higgs boson) exist.  We have already mentioned that this expectation is
confirmed \cite{LeCompte}, though not yet with overwhelming accuracy.
Assuming that only the $W + b$ channel is significant, the decays of a $t
\bar t$ pair then may be decomposed according to Table 7.  We have used the
shorthand $d'$ and $s'$ for the linear combinations (\ref{eqn:ucomb}) and
(\ref{eqn:ccomb}) of $d$ and $s$ which couple to $u$ and $c$ via $W$.

\begin{table}
\caption{Final states of a $t \bar t$ pair as function of decays of the $W$
bosons in $t \to W + b$. Here $\met$ stands for ``missing transverse energy,''
while $j$ stands for ``jet''.} 
\begin{center}
\begin{tabular}{|c|c|c|c|c|} \hline
$~~~~W^+ \to$ &   1/9   &    1/9    &    1/9     &      2/3            \\
$W^- \to~~~~$ & $e \nu$ & $\mu \nu$ & $\tau \nu$ & $u \bar d'+c \bar s'$ \\
\hline
$1/9~e \nu$   & $ee \met$ & $e \mu \met$ & $e \tau \met$ & $e \met$ \\
              &  $+2j$    &    $+2j$     &    $+2j$      &   $+4j$  \\ \hline
$1/9~\mu \nu$ &$e\mu\met$ & $\mu\mu\met$ & $\mu\tau\met$ & $\mu\met$ \\
              &  $+2j$    &    $+2j$     &    $+2j$      &   $+4j$  \\ \hline
$1/9~\tau\nu$ &$e\tau\met$&$\mu\tau\met$ &$\tau\tau\met$ & $\tau\met$ \\
              &  $+2j$    &    $+2j$     &    $+2j$      &   $+4j$  \\ \hline
    $2/3$     & $e \met$  & $\mu \met$   &  $\tau \met$  & \\
$\bar u d'+\bar c s'$&$+4j$&   $+4j$     &    $+4j$      &   $6j$   \\ \hline
\end{tabular}
\end{center}
\end{table}

Energy balance in Tevatron collisions can only be checked in the transverse
direction since particles can escape undetected at small angles relative to
the beam.  Large missing transverse energy serves as a signature of neutrino
production in $W$ leptonic decay.  One always expects at least two jets of
hadrons associated with the two $b$ quarks in $t \bar t$ decays.  If one
or two $W$'s decays hadronically a total of 4 or 6 jets should then be
produced.

All the channels in Table 7 except for those with one or two $\tau$ leptons
have now been studied.  These include dileptons (about 5\%), leptons $+$
jets (about 30\%), and jets (about 45\%).  A start is even being made on
channels with one $\tau$ and one $e$ or $\mu$.  Thus a large fraction of the $t
\bar t$ channels are seen.
\bigskip

\leftline{\bf B.  Recent CDF and D0 analyses}
\bigskip

We summarize in this subsection some results which were presented at the 1996
Warsaw Conference by the CDF and D0 Collaborations \cite{Watop}. For CDF
results, in addition to those presented at Warsaw, one may consult
Refs.~\cite{CDFMor} for details. As an example of how one exploits the
possibilities of Table 7, we now give some details of the CDF analyses. 

1.  In the {\it dilepton} sample, where $\ell = e$ or $\mu$, one asks that the
leptons have opposite charges, demands missing transverse energy ($\met$) as a
signature of one or more neutrinos, and requires two jets (signatures of the
two $b$ quarks).  As a result, 7 events with $e \mu$, two with $\mu \mu$, and
one with $ee$ were found.  Six ``tagged'' jets in a total of 4 events were
identified as $b$'s after the fact, but $b$ identification was not required {\it
a priori}. 

2.  In the {\it lepton + jets} sample, with $\ell = e$ or $\mu$, one again
asks for $\met$, at least three observed jets, and at least one identified
$b$ quark.  The $b$ quark can be identified by its displaced vertex, since
the average $b$ lifetime is 1.5 to 1.6 ps, using the silicon vertex detector
(``SVX'').  Alternatively, its semileptonic decay, with a branching ratio
of about 10.5\% to each of $e + \ldots$ and $\mu + \ldots$, gives rise to
a soft lepton tag (``SLT'').  One sees 34 SVX events and 40 SLT events.

3.  In the {\it 6 jets} sample, one asks to see at least 5 of them, requiring
a $b$ to be tagged via the SVX.  Here 192 events are seen.

\begin{table}
\caption{CDF top quark events in different channels based on an integrated
luminosity of 110 pb$^{-1}$.  Only the statistical error is shown for the
channels with $\tau$.  Acceptances were calculated for $m_t = 175~\G/c^2$.}
\begin{center}
\begin{tabular}{c c c c} \\ \hline
Channel & $N_{\rm data}$ & $N_{\rm background}$ & $\sigma(t \bar t)$ (pb) \\
\hline
Dilepton & 10 & $2.1 \pm 0.4$ & $9.3^{+4.4}_{-3.4}$ \\
$\ell + j~({\rm SVX})$ & 34 & $7.96 \pm 1.37$ & $6.8^{+2.3}_{-1.8}$ \\
$\ell + j~({\rm SLT})$ & 40 & $24.3 \pm 3.5$ & $8.0^{+4.4}_{-3.6}$ \\
Hadronic & 192 & $137 \pm 11$ & $10.7^{+7.6}_{-4.0}$ \\
$\tau + (e,\mu)$ & 4 & $2.0 \pm 0.35$ & $15.6^{+18.6}_{-13.2}$ \\ \hline
Total & & & $7.5^{+1.9}_{-1.6}$ \\ \hline
\end{tabular}
\end{center}
\end{table}

In Table 8 we summarize the CDF measurements in different channels. The CDF
cross sections in various channels average to a value $\sigma_{\rm CDF~av.}(p
\bar p \to t \bar t + \ldots) = 7.5^{+1.9}_{-1.6}$ pb at $\sqrt{s} = 1.8$ TeV. 

The corresponding D0 results make use of dilepton events, lepton + jets events
with kinematic discrimination, and lepton + jets events with a soft muon
``tag'' associated with the semileptonic decay $b \to c \mu^- \bar \nu_\mu$.
The results, based on the full 1992--5 sample with 100 pb$^{-1}$ of integrated
luminosity, are shown in Table 9.  The column labeled ``expected'' refers to
the number of events anticipated for a top quark with mass $m_t = 180~\G/c^2$
based on cross sections of Ref.~\cite{Laenen}.  The D0 top quark cross sections
are $4.7 \pm 3.2$ pb in the dilepton channel, $4.6 \pm 5.1$ pb in the hadron
channel, $4.2 \pm 2.0$ pb in the lepton + jets channel with kinematic
discrimination, and $7.0 \pm 3.3$ pb in the lepton + jets/tag channel.

\begin{table}
\caption{D0 top quark events in different channels based on an integrated
luminosity of 100 pb$^{-1}$.}
\begin{center}
\begin{tabular}{c c c c} \\ \hline
Channel & $N_{\rm expected}$ & $N_{\rm background}$ & $N_{\rm data}$ \\
\hline
 $e \mu$  & $1.7 \pm 0.3$ & $0.4 \pm 0.1$ & 3 \\
   $ee$   & $0.9 \pm 0.1$ & $0.7 \pm 0.2$ & 1 \\
$\mu \mu$ & $0.5 \pm 0.1$ & $0.5 \pm 0.3$ & 1 \\ \hline
$e + {\rm jets}$ & $6.5 \pm 1.4$ & $3.8 \pm 1.4$ & 10 \\
$\mu + {\rm jets}$ & $6.4 \pm 1.5$ & $5.4 \pm 2.0$ & 11 \\
$e + {\rm jets/tag}$ & $2.4 \pm 0.4$ & $1.4 \pm 0.4$ & 5 \\
$\mu + {\rm jets/tag}$ & $2.8 \pm 0.9$ & $1.1 \pm 0.2$ & 6 \\ \hline
Total & $21.2 \pm 3.8$ & $13.4 \pm 3.0$ & 37 \\ \hline
\end{tabular}
\end{center}
\end{table}

Combining all CDF and D0 cross section determinations, one finds
\beq
\sigma(p \bar p \to t \bar t + \ldots) = 6.4^{+1.3}_{-1.2}~{\rm pb}~~~
(\sqrt{s} = 1.8~{\rm TeV})~~~.
\eeq
This is to be compared with theoretical estimates such as 
\beq
\sigma_{\rm Theory}(p \bar p \to t \bar t + \ldots) =
\left\{ \begin{array}{c r} 4.84^{+0.73}_{-0.69}~{\rm pb} & \cite{ESW} \\
   4.75^{+0.73}_{-0.62}~{\rm pb} & \cite{CMNT} \\ \end{array} \right\}~~~.
\eeq
\newpage

\leftline{\bf C.  Mass}
\bigskip

The most precise CDF value of the top quark mass \cite{Watop} comes from a
combined analysis including (a) 34 events with a lepton, jets, and at least one
SVX tag (leading to $m_t = 177 \pm 7~\G/c^2$), (b) a more refined analysis of
the same sample, weighting events in terms of the quality of information they
provide, leading to the value $174 \pm 8~\G/c^2$, (c) the dilepton sample,
leading to $160 \pm 28~\G/c^2$ and $162 \pm 22~\G/c^2$ in two different
analyses (the second using the average $M^2_{\ell b}$ of the lepton and $b$),
and (d) the multijet sample, leading to $187 \pm 15~\G/c^2$.  The resulting
mass plot is shown in Fig.~19. 

\begin{figure}
\centerline{\epsfysize = 4in \epsffile {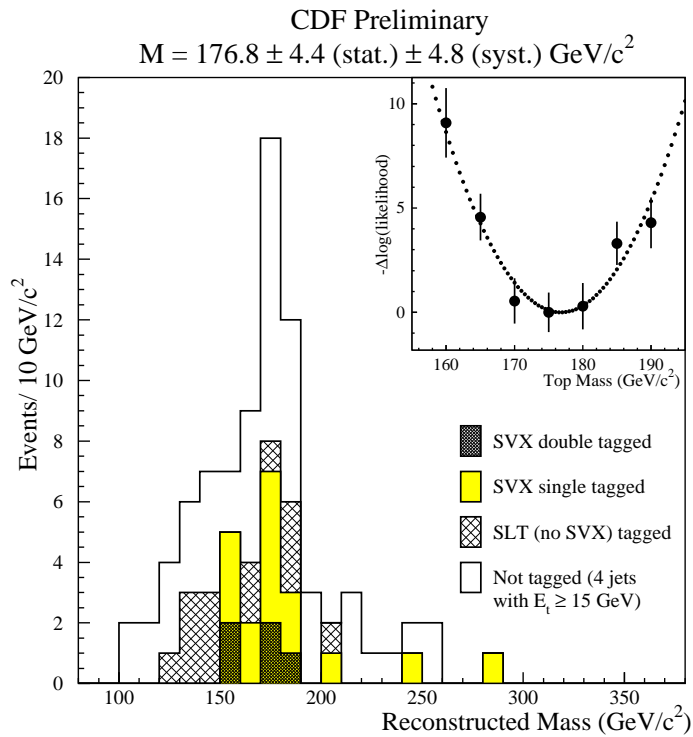}}
\caption{CDF top quark mass distributions as of July 1996.}
\end{figure}

\begin{figure}
\centerline{\epsfysize = 4in \epsffile {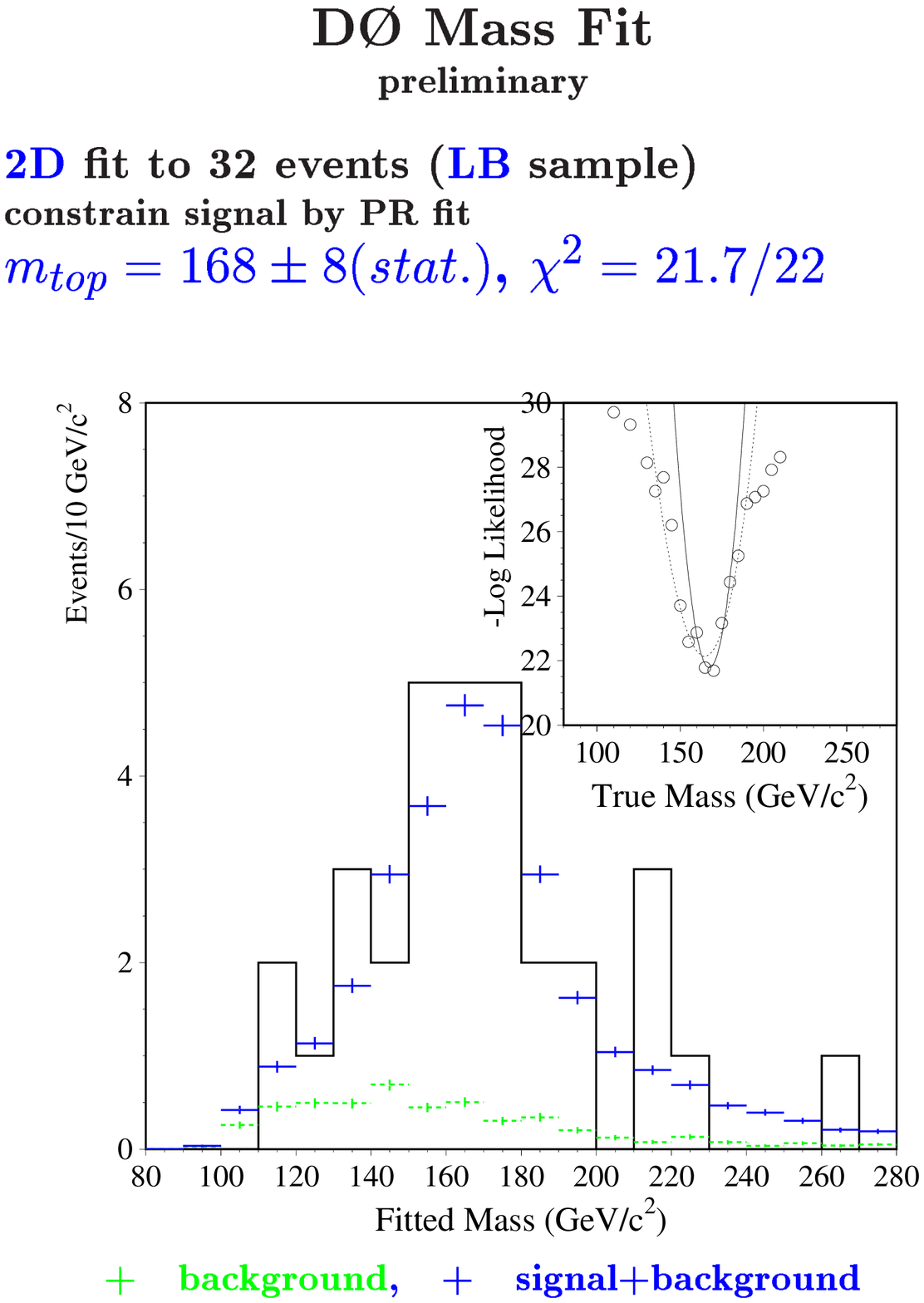}}
\caption{D0 top quark mass distribution from lepton + jets sample as of July
1996. A correction of $+1~\G/c^2$ must be applied for the energy scale.} 
\end{figure}

The CDF Collaboration has also presented distributions in top quark transverse
momentum and in $m_{t \bar t}$, the effective mass of the $t \bar t$ system
\cite{CDFMor}.  These data are in accord with standard model expectations.

The D0 top quark mass obtained from the lepton + jets sample is $169 \pm
11~\G/c^2$, while $158 \pm 26~\G/c^2$ is obtained from the dilepton sample.
The mass distribution from the lepton + jets sample (before a 1 $\G/c^2$ jet
energy correction has been applied) is shown in Fig.~20. The world average of
$175 \pm 6~\G/c^2$ is based on all the CDF and D0 determinations. 

In Fig.~21 we plot the expected top quark cross section \cite{ESW} as a
function of $m_t$ at the c.m. energies $\sqrt{s} = 1.8$ and 2 TeV.  Also
shown are the individual CDF and D0 mass and cross section points.  It is
clear there is satisfactory agreement with the standard model, though there
would be room, if one so chose, for an additional 50\% extra contribution
to the cross section from some non-standard process.

\begin{figure}
\centerline{\epsfysize = 4in \epsffile {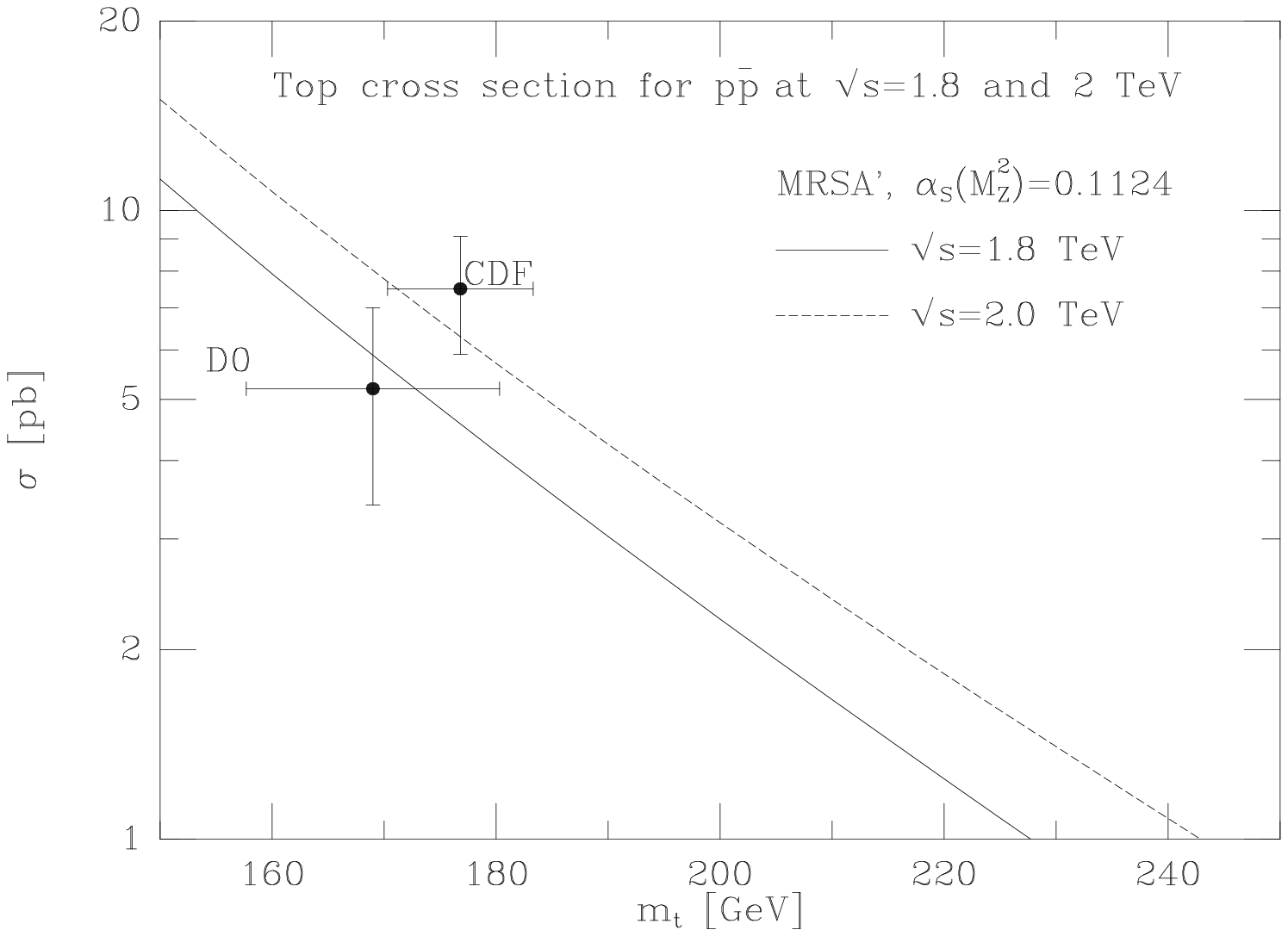}}
\caption{Top quark cross section prediction as a function of $m_t$.}
\end{figure}
\bigskip

\leftline{\bf D.  Decays}
\bigskip

The branching ratio of the top quark to $W + b$ can be measured relative to all
$W + q$ decays by comparing double $b$ tags with single $b$ tags
\cite{LeCompte}.  The result is 
\beq
\frac{B(t \to W + b)}{B(t \to W + q)} = 0.94 \pm 0.27 \pm 0.13~~~,
\eeq
implying $V_{tb} = 0.97 \pm 0.15 \pm 0.07$ in accord with standard-model
expectations.
\newpage

\leftline{\bf 5.  TOP QUARK PRODUCTION IN ELECTRON-POSITRON COLLISIONS}
\bigskip

\leftline{\bf A.  Excitation curve}
\bigskip

Eventually it will become possible to produce top quarks in pairs in
high-energy electron-positron annihilations.  Accordingly, we discuss the
production of pairs of arbitrary fermions $f$ via virtual photons ($\gamma^*$)
and $Z$'s ($Z^*$) in the direct channel.  These results will be of use in
Sec.~6 when we discuss exotic quarks and leptons. 

Define $x \equiv \sin^2 \theta$, $s \equiv E_{\rm c.m.}^2$, and $r \equiv
[s/(s-m_Z^2)x(1-x)]$.  Then far from the $Z$ pole, where the $Z$ width can be
neglected, the contribution of a virtual photon and $Z$ in the direct channel
to the cross section for production of a fermion with electric charge $Q_f$ and
axial and vector $Z$ couplings $g_A = -(1/2)I_{3L}^f$ and $g_V = (1/2)I_{3L}^f
- Q_f x$ is 
\beq
\sigma(e^+ e^- \to f \bar f) = \sigma_\gamma \left\{
Q_f^2 - 2 r Q_f g_V^e g_V^f + r^2 [ (g_V^e)^2 + (g_A^e)^2] [(g_V^f)^2 +
\frac{\beta^2}{K_V}(g_A^f)^2] \right\}~~,
\eeq
where
\beq
\sigma_\gamma \equiv \frac{4 \pi \alpha^2}{3 s} N_c \beta K_V~,~~~
\beta \equiv \left( 1 - \frac{4 m_f^2}{s} \right)^{1/2}~,~~~
K_V \equiv \frac{3 - \beta^2}{2}~~,
\eeq
and $N_c$ is the number of colors of fermions.  For quarks ($N_c = 3$) the
cross section should be multiplied by an additional correction factor of $1 +
(\alpha_s/\pi) \approx 1.04$.  The values of $\sso$ far above pair production
threshold, where $\sigma_0 \equiv \sigma(e^+ e^- \to \gamma^* \to \mu^+
\mu^-$), are compared in Table 10 for various fermion species $f$ when the
energy is far below the $Z$ pole (where only the virtual photon dominates) and
when it is far above the $Z$ (where the interference in vector contributions of
the photon and $Z$ is possible).  All the cross sections far above the $Z$ are
enhanced relative to the values they would have due to the virtual photon
alone.

\begin{table}
\caption{Cross sections $\sigma$ [in units of $\sigma_0 \equiv \sigma(e^+ e^-
\to \gamma^* \to \mu^+ \mu^-$)] for $e^+ e^-$ production of pairs of fermions
$f \bar f$ via virtual photons and $Z$'s in the direct channel.  Here
$t$-channel exchanges are neglected for $e$ and $\nu_e$. Values of $g_V^f$ are
quoted for $x = 0.2315$.  QCD corrections to quark production have been
neglected.} 
\begin{center}
\begin{tabular}{c c c c c c} \hline
Fermion & $Q_f$  & $g_V^f$   & $g_A^f$ &$\sso$ far &$\sso$ far \\
  $f$   &        &           &         & below $Z$ & above $Z$ \\ \hline
  $u$   & ~~2/3  & ~~0.0957  &  $-1/4$ &   4/3     &   1.80    \\
  $d$   & $-1/3$ & $-0.1728$ &   1/4   &   1/3     &   0.92    \\
  $e^-$ &  $-1$  & $-0.0185$ &   1/4   &    1      &   1.13    \\
$\nu_e$ &    0   &    1/4    &  $-1/4$ &    0      &   0.25    \\ \hline
\end{tabular}
\end{center}
\end{table}

The predicted top quark cross section, shown in Fig.~22, is composed of
incoherent contributions corresponding to vector and axial-vector couplings
to the top.  The vector part involves $\gamma^*-Z^*$ interference.  These
contributions have rather different energy dependence as a result of
different kinematic factors.  The vector contribution peaks around
$E_{\rm c.m.} = 410~\G$, while the axial vector contribution peaks around
550 GeV.  At 410 GeV, the axial contribution is only 5\% of the total,
rising to about 25\% asymptotically.  The sum of the two contributions peaks
around 420 GeV at a value of about 0.7 pb.

\begin{figure}
\centerline{\epsfysize = 4in \epsffile {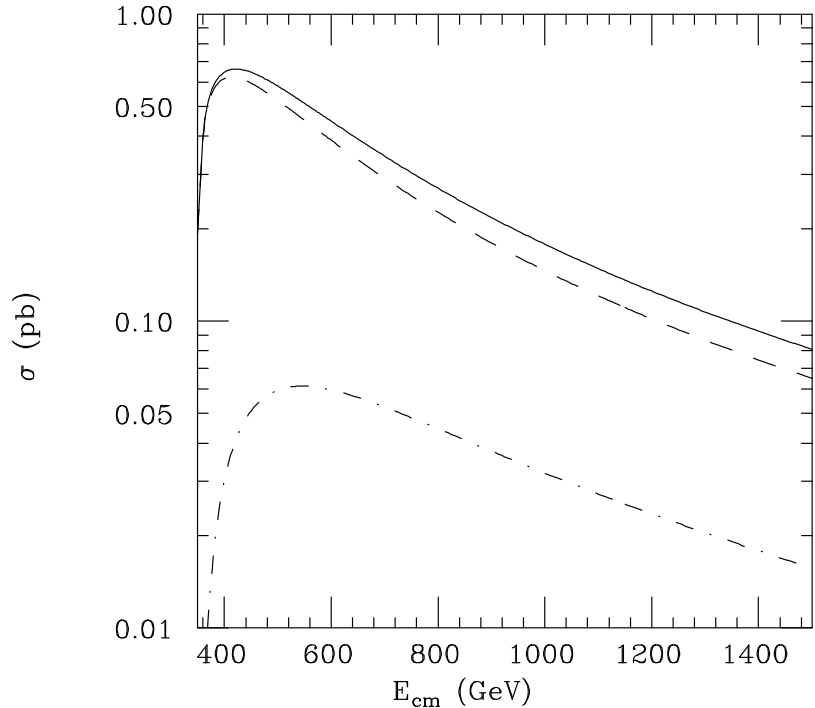}}
\caption{Cross section prediction for $\eep \to (\gamma^*,Z^*) \to t \bar t$
as a function of $E_{\rm c.m.}$.  Solid line: full cross section; dashed
line:  contribution of vector couplings of $t$; dot-dashed line:  contribution
of axial coupling of $t$.)}
\end{figure}
\bigskip

\leftline{\bf B.  Sensitivity to interactions}
\bigskip

The top quark mass is too large to permit the formation of $t \bar t$ bound
states or top mesons ($t$ quarks bound to light antiquarks).  The top quark
decays to $W + b$ with a partial width of about 1.4 GeV, as noted in the
Introduction.  This is considerably larger than the spacings between quarkonium
levels shown in Figs.~3 and 4.  Nonetheless one can learn some things about the
top-antitop interaction by studying the shape of the threshold function and how
it is distorted in comparison with the lowest-order prediction in Fig.~22
\cite{Kwong,PeS}.  One probably sees a vestige of the $1S$ level as a slight
threshold enhancement with respect to the free-quark behavior, while Higgs
interactions \cite{PeS} can alter the threshold shape considerably.
\bigskip

\leftline{\bf 6.  SPECULATIONS ON SOURCE OF TOP QUARK MASS}
\bigskip

\leftline{\bf A.  Mass matrix lore}
\bigskip

In Sec.~2 A we saw that the $3 \times 3$ quark mass matrices ${\cal M}_{U,D}$
were diagonalized by separate unitary transformations $L_{U,D}$ and $R_{U,D}$
on left- and right-handed quarks.  The CKM matrix $V = L_U^{\dag} L_D$ carries
no information about the transformations $R_{U,D}$.  As a result, separate
unitary transformations can be performed on right-handed $U$ and $D$ quarks,
as long as corresponding redefinitions of the mass matrix are made.  Moreover,
the simultaneous transformations $L_U^{\dag} \to L_U^{\dag} A$ and $L_D \to
A^{\dag} L_D$, with $A A^{\dag} = 1$, also leave $V$ unchanged.  Thus there is
considerable freedom in the choice of basis for mass matrices, which has led
to a large folklore.  We shall barely scratch the surface.

An early observation \cite{GatOak,WbgWZ} was the relation between the Cabibbo
angle and the ratio of $d$ and $s$ masses: $\sin \theta_c \simeq
(m_d/m_s)^{1/2}$.  This result can be obtained with a correction term
$-(m_u/m_c)^{1/2}$ if one postulates a zero in the upper-left corner of the
mass matrices for two families of quarks: 
\beq
{\cal M}_Q = \left[ \begin{array}{c c} 0 & a \\ a & b \end{array} \right]~~~,
\eeq
where $a \ll b$.  Since the determinant of this matrix is $-a^2$ while its
trace is $b$, its eigenvalues are approximately $-a^2/b$ and $b$, corresponding
to the eigenvectors $[1,-a/b]$ and $[a/b,1]$, respectively.  We then identify
$$
a_U = \sqrt{-m_u m_c}~~,~~~b_U = m_c~~~,
$$
\beq
a_D = \sqrt{-m_d m_s}~~,~~~b_D = m_s~~~.
\eeq
The CKM matrix may be computed from the scalar product of the appropriate
eigenvectors $U_i$ and $D_j$ via \cite{RW} $V_{ij} = U_i^{\dag} \cdot D_j$. 
For example,
\beq \label{eqn:vus}
V_{us} = \left[ 1, - \sqrt{\frac{-m_u}{m_c}} \right] \cdot
\left[ \begin{array}{c} \sqrt{-m_d/m_s} \\ 1 \end{array} \right]
= \sqrt{\frac{-m_d}{m_s}} - \sqrt{\frac{-m_u}{m_c}}~~~.
\eeq
The negative sign of a quark mass is no cause for concern; it can always be
changed by redefining $q \to \gamma_5 q$.  The relative phases of the two terms
on the right-hand side of (\ref{eqn:vus}) can be adjusted by an {\it ansatz}
regarding the phases of terms in the mass matrices.  For $m_s/m_d \simeq 20$
\cite{Leut}, the observed value $V_{us} \simeq 0.22$ is obtained for a relative
phase of the two terms close to 90$^{\circ}$.

The crucial role of zeroes in mass matrices for three-family models was first
exploited by Fritzsch \cite{Fritzsch}, explored in some detail by Ramond and
his group \cite{Ramond}, and has been reviewed recently by Branco {\it et al.}
\cite{Branco}.  These zeroes go under the rubric of ``texture zeroes.'' A
``democratic'' basis has also been proposed \cite{dem} , which explains the
hierarchies $m_u,m_c \ll m_t$ and $m_d,m_s \ll m_b$ by perturbing around
matrices of the form 
\beq
{\cal M} = \frac{1}{3} \left[ \begin{array}{c c c} 1 & 1 & 1 \\
1 & 1 & 1 \\ 1 & 1 & 1  \end{array} \right]~~~
\eeq
with eigenvectors
\beq
\frac{1}{\s} \left[ \begin{array}{r} 1 \\ -1 \\ 0 \end{array} \right]~~,~~~
\frac{1}{\sx} \left[ \begin{array}{r} 1 \\ 1 \\ -2 \end{array} \right]~~,~~~
\frac{1}{\st} \left[ \begin{array}{r} 1 \\ 1 \\ 1 \end{array} \right]~~,~~~
\eeq
corresponding to eigenvalues 0, 0, 1.  By means of a unitary transformation
one can end up with a matrix ${\cal M}$ consisting of all zeroes except for
a 1 in the lower right-hand corner.

A curious basis choice seems to be encountered in several quite different
approaches \cite{RW,Matu}:
\beq \label{eqn:curious}
{\cal M} = \left[ \begin{array}{c c c} 0 & \alpha p & \alpha \\
\alpha p & \beta & \beta q \\
\alpha & \beta q & \gamma \end{array} \right]~~~,
\eeq
where $p$ and $q$ are numbers of order 1.  For $\alpha \ll \beta \ll \gamma$
the eigenvectors and eigenvalues of this matrix are easy to compute.  The CKM
elements and the quark mass ratios which contribute to them (with phases which
must be adjusted by hand) are summarized in Table 11.  The magnitudes of the
mass ratios appear to be appropriate for the sizes of the CKM elements.  No
insight is present in these models (nor in most others) regarding the choice
of relative phases, however.

\begin{table}
\caption{Quark mass ratios contributing to CKM matrix elements in one
basis for quark mass matrices.  Here the symbol $q$ stands for $m_q$.}
\begin{center}
\begin{tabular}{c c c} \hline
CKM element      & Quark mass ratios & Magnitudes of ratios \\ \hline
$V_{us},~V_{cd}$ & $(d/s)^{1/2},~(u/c)^{1/2}$ & 0.22, 0.06 \\
$V_{cb},~V_{ts}$ &        $s/b,~c/t$        & 0.03, 0.007 \\
     $V_{ub}$    & $(ds)^{1/2}/b,~(s/b)(u/c)^{1/2},~(uc)^{1/2}/t$ &
0.007, 0.002, 0.0004 \\
     $V_{td}$    & $(ds)^{1/2}/b,~(c/t)(d/s)^{1/2},(uc)^{1/2}/t$ &
0.007, 0.002, 0.0004 \\ \hline
\end{tabular}
\end{center}
\end{table}
\bigskip

\leftline{\bf B.  A two-family fable}
\bigskip

Perhaps one of the least popular ideas these days is the notion that quarks
are somehow composite objects.  Nonetheless, an exercise (perhaps no more
than a fable, as applied to quarks) shows the potential insight a composite
model can provide into the interplay between masses and mixing \cite{RW,PTP}.

The exercise is based on states for which there are experimental candidates:
the charmed-strange baryons known as $\Xi_c$ illustrated in Fig.~5.  These
are interesting because no two quarks in them have the same mass.  They
consist of $csq$, where $q = u$ or $d$.

The lowest charmed-strange baryon, now called $\Xi_c$ \cite{PDG} and denoted in
Fig.~5 by $\Xi_c^a$ (originally $A$ in Ref.~\cite{GLR}), has its light quarks
$s$ and $q$ in a state primarily of zero spin and flavor SU(3) antisymmetry.
The total spin of the $\Xi_c^a$ is 1/2. 

A candidate for an excited version of the $\Xi_c$, which we label as $\Xi_c^s$
in Fig.~5 (originally $S$ in Ref.~\cite{GLR}), has been claimed by the WA89
Collaboration at CERN \cite{WA89}. It lies $95~\M/c^2$ above the $\Xi_c^a$
and decays to it by photon emission.  It, too, has total spin equal to 1/2,
but its light quarks are coupled up primarily to a state of spin 1 and
flavor SU(3) symmetry.  This state is on shaky ground as it has only been
reported in conference proceedings and is not confirmed by experiments at
the CLEO detector at Cornell.

Finally, the Cornell experiments just mentioned {\it do} see a candidate for a
$\Xi_c^*$ state at higher mass \cite{Xicstar}, which fits well with the
assignment of a state of total spin 3/2 (originally $S^*$ in Ref.~\cite{GLR}). 
This state decays to $\Xi_c^a + \pi$.  The light quarks, of course, have to be
in a pure spin-1 state here.

The masses of many hadrons can be described in terms of sums of
constituent-quark masses and hyperfine-interaction terms inversely proportional
to quark masses \cite{massforms}.  Thus, for example, for a $\Xi_c$, one will
have
\beq
M \simeq m_c + m_s + m_q + a \left(
\frac{\sigma_q \cdot \sigma_s}{m_q m_s} +
\frac{\sigma_q \cdot \sigma_c}{m_q m_c} +
\frac{\sigma_s \cdot \sigma_c}{m_s m_c} \right)~~~
\eeq
We note that $m_q < m_s \ll m_c$.  Now, when $m_c \to \infty$, this
mass operator is diagonal in the basis labeled by $S_{sq}$:
\beq
\left[ \begin{array}{c} S_{sq} = 0 \\ S_{sq} = 1 \end{array} \right]
\leftrightarrow
\left[ \begin{array}{c} \Xi_c^a \\ \Xi_c^s \end{array} \right]~~~
{\rm for}~J = 1/2~{\rm states}~~~.
\eeq
When $m_c \ne \infty$, however, this is only an approximate basis:  The mass
eigenstates are mixed.  Moreover, {\it changing $m_q$ changes the mixing
angle.}  We suggest this as a paradigm for the CKM rotation.  In an attempt to
construct a realistic two-family model \cite{RW,PTP}, specific relations among
parameters are necessary in order to ensure that the $J = 3/2$ state lies much
higher in mass than the $J = 1/2$ ones, and in order to obtain the appropriate
``texture'' zero in the upper left-hand corner of the $2 \times 2$ mass matrix
for the $J = 1/2$ states.  The three-family model constructed in Ref.~\cite{RW}
from similar arguments leads to the mass matrix (\ref{eqn:curious}) with $p = q
= \s$. 
\bigskip

\leftline{\bf C.  Is the top quark special?}
\bigskip

The top quark is heavier, {\it so far}, than any other known fermion.  We
do not know if this is merely because it is near the limit of what one can
produce via the strong interactions at present energies (this is certainly
true), or something more fundamental.  Many suggestions have been made about
the crucial role played by its large mass.  We have mentioned already the
contribution of $m_t$ to loop diagrams in particle-antiparticle mixing (Sec.~2)
and in precise electroweak physics (Sec.~3).  Here we mention some roles of a
more fundamental nature which a heavy top might play. 

Veltman \cite{Vt} proposed that the top quark plays a crucial role in the
cancellation of quadratic divergences in the effective Higgs potential,
leading to the mass relation $12 m_t^2 = 3 M_H^2 + 3 m_Z^2 + 6 M_W^2$
in the limit that other quark masses are neglected.  This implies that the
Higgs boson lies slightly below $2 m_t$ in mass, as if it were a $t \bar t$
bound state.  In the limit in which the electroweak couplings $g$ and $g'$
vanish, $M_H \to 2 m_t$ and the binding energy vanishes.

Nambu \cite{Nambu}, drawing in part on his pioneering work with Jona-Lasinio
\cite{NJL}, notes that in many dynamical symmetry breaking schemes, in which a
condensate of fermion pairs $\langle f \bar f \rangle \ne 0$ leads to chiral
symmetry breaking, there exist in the spectrum not only a zero-mass
pseudoscalar boson, but the fermion $f$ and a scalar excitation ``$\sigma$'' at
$m_\sigma = 2 m_f$.  The analogies suggested by this observation are shown in
Table 12. 

\begin{table}
\caption{Analogy between states in Nambu-Jona-Lasinio (NJL) model and dynamic
symmetry breaking (DSB) in electroweak interactions}
\begin{center}
\begin{tabular}{c l c l} \hline
State in  &      Role      &    State in   & Role \\
NJL model &                &   DSB scheme  & \\ \hline
Pions & Nambu-Goldstone & Higgs bosons $\phi^\pm$ & Provide longitudinal \\
     & bosons of spont. & and $(\phi^0 - \bar \phi^0)/\s$ & components of \\
     & broken chiral symm. &              & $W$ and $Z$ bosons \\ \hline
Nucleon  & Constituent of pion & Top quark & Constituent of \\
         &                     &           & Higgs boson? \\
\hline 
$\sigma$ & Weakly bound        & Higgs boson & Preserves unitarity \\
  & nucleon-antinucleon & $(\phi^0 + \bar \phi^0)/\s$ & in scatt.~of longit.\\
         & state               &             & $W$'s and $Z$'s \\
\hline
\end{tabular}
\end{center}
\end{table}

Bardeen, Hill, and Lindner \cite{BHL} emulate the Nambu-Jona-Lasinio model
with a new interaction to generate the Higgs boson out of $t \bar t$.  A
nonzero condensate $\langle t \bar t \rangle \ne 0$ develops, leading to
electroweak symmetry breaking.  The new interactions are needed both to
bind $t \bar t$ and to communicate the effects of the nonzero condensate
to the other fermions.

In contrast to the above approach, G\'erard and Weyers \cite{GW}
introduce no new interaction.  The Higgs field $\phi$ initially has $m_\phi^2 >
0$; its coupling to $t \bar t$ turns an effective Higgs potential with upward
curvature at $\phi = 0$ into one with negative curvature, so that a non-zero
value of $\langle \phi \rangle$ develops, triggering electroweak symmetry
breaking.  This scheme only is self-consistent for a narrow range of Higgs
boson masses around $85~\G/c^2$, and thus makes a very specific prediction. 

Some versions of supersymmetry, described in more detail in Ref.~\cite{LH}
and by Graham Ross at this Institute \cite{Ross}, have two Higgs doublets with
$m^2$ values which are the same at some very large (unification) mass scale
$\mu$.  One doublet ($\phi_1$) interacts with down-type quarks and charged
leptons, while the other ($\phi_2$) interacts with up-type quarks (including
the top quark) and perhaps neutrinos. As a result of the strong coupling of
$\phi_2$ with the top quark, $m^2(H_2)$ (where $H_2$ is the neutral scalar
object in the complex doublet $\phi_2$) evolves toward a negative value as
$\mu$ is lowered, triggering electroweak symmetry breaking. 

A scheme known as ``reduction of coupling constants'' \cite{OZ} (see also
\cite{WMGYU}) or ``gauge -- Yukawa unification'' \cite{GYU} notes the presence
of special relations for the renormalization-group behavior of coupling
constants when gauge couplings $g$ and Yukawa couplings $g_Y$ are related.
In contrast to supersymmetry, these schemes do not necessarily entail any
extended symmetries.  They posit a special role for finiteness in
perturbation theory to each order.
\bigskip

\leftline{\bf D.  Technicolor}
\bigskip

The electroweak dynamical symmetry breaking scheme described in Table 12 makes
use of the top quark to form a fermion condensate which breaks the symmetry. 
One can also postulate some {\it new} fermion which binds with its antiparticle
under some new superstrong force to form the pseudo-Nambu-Goldstone bosons
which are needed to make the $W$ and $Z$ massive.  This is the essence of the
scheme \cite{LSTC,SWTC} which has come to be known as ``technicolor.'' 

Suppose the $u$ and $d$ quark were massless.  The pion would be a massless
quark-antiquark bound state in QCD.  Its coupling to the $W$ via the divergence
of the axial current would give the $W$ a mass $M_W = g F_\pi/2 \simeq
30~\M/c^2$, where we use the normalization $F_\pi = 93$ MeV here.  In order
to generate a $W$ with mass $M_W = gv/2 = 80.34~\G/c^2$, noting that
$g^2/4 \pi \simeq 1/30$, we need instead some fermion-antifermion pair with a
decay constant $v = 246$ GeV.  Thus, one needs an interaction which becomes
strong at a mass scale roughly 246/0.093 = 2650 times as high as the QCD
scale.  This is the postulated technicolor interaction.  The fermions
responding to this interaction are known as {\it technifermions}.

The $Z-\gamma-\gamma$ vertex described by a triangle diagram like that in
Fig.~2 will be anomaly-free if the sum over technifermions $F$ of $Q^2_F
I_{3L}^F$ vanishes.  The simplest way in which this can be realized is with
a single doublet with $Q_F = \pm 1/2$.  The doublet may come in $N_T$
different ``technicolors.''  This is ``minimal technicolor.''  It works
fine for the masses of the $W$ and $Z$.

An extension of technicolor to the description of quark and lepton masses
immediately runs into difficulty.  A new interaction (``extended
technicolor'' \cite{ETC}) is needed to communicate the effects of $\langle \bar
F F \rangle \ne 0$ to the ordinary quarks and leptons.  This interaction has
the potential for introducing flavor-changing neutral currents and steps must
be taken to prevent them.  The biggest mass (top) naturally poses the biggest
problem for this theory. 
\bigskip

\leftline{\bf E.  Quark-lepton pattern and new physics}
\bigskip

Returning again to Fig.~1, we see that the up-type quarks are less ``dense''
than the down-type quarks or charged leptons on a logarithmic plot.  Moreover,
their center of gravity is higher.  This suggests that the down-type quarks and
charged leptons are being pushed down in mass by mixing with heavier extra
levels with $Q = \pm 1/3$ and $Q = \pm 1$. 

The situation with the neutrinos is even more extreme.  They appear to be very
light if not massless.  (Oscillation results, which would require non-zero
masses, are not yet conclusive.)  A popular way to understand such light masses
is the ``seesaw'' mechanism \cite{seesaw} involving mixing with other very
heavy neutrinos.  The neutrino mass matrix may be written in terms of
left-handed states in a basis $(\nu_L,\nu_L^c)$ as
\beq
{\cal M}_\nu = \left[ \begin{array}{c c} 0 & m_D \\ m_D & M
\end{array} \right]~~~,
\eeq
where $m_D$ stands for a Dirac mass, and $M$ is a large Majorana mass.  The
eigenvalues of this matrix are approximately $-m_D^2/M$ and $M$,
corresponding to the eigenvectors $[1,-m_D/M]$ and $[m_D/M,1]$.

Thus, the large top quark mass could be telling us that at higher masses
one expects more matter which is {\it not} like the top quark:  quarks
of charge $-1/3$ or new leptons.  What can serve as the source for these
new forms of matter?  We seek the answer in several schemes for quark-lepton
unification.
\bigskip

\leftline{\bf F.  Quark-lepton unity}
\bigskip

The ``grand unification'' of strong and electroweak interactions in a larger
symmetry, and the identification of quarks and leptons as objects related to
one another under this symmetry, involves such groups as SU(5) \cite{GG},
SO(10) \cite{SO}, and \es~\cite{E6}.

One forms a $5 \times 5$ matrix of SU(5) by simply putting the $3 \times 3$
matrices of color SU(3) in the upper left-hand corner and the $2 \times 2$
matrices of SU(2)$_L$ in the lower right.  The miracle of SU(5) is that it
also possesses a U(1) $\sim$ diag(2, 2, 2, $-3,-3$) which is just right for
weak hypercharge!

Within SU(5) a specific choice of representations (${\bf 5}^* + {\bf 10}$) is
required for the left-handed fermions in order to accommodate the known states
and to eliminate anomalies. This choice is automatic if left-handed fermions
are assigned to the ${\bf 16}$-dimensional spinor multiplet of SO(10); the
additional state is a right-handed neutrino.  Thus, we have the decomposition
\beq
[{\rm SO(10)}]~~~{\bf 16}_L = {\bf 5}^*_L + {\bf 10}_L + {\bf 1}_L~~~[{\rm
SU(5)}]~~~.
\eeq
The right-handed neutrino can acquire a large Majorana mass without violating
the SU(3)$_c \times$ SU(2)$_L \times$ U(1) symmetry of the standard model.
Anomalies are not present in SO(10), as long as matter belongs to complete
multiplets. 

The group \es~contains SO(10).  Its lowest-dimensional representation (${\bf
27}$) contains the ${\bf 16}$ of SO(10), as well as ${\bf 10}$- and ${\bf
1}$-dimensional (``exotic'') representations of SO(10).  These, in turn, have
the following SU(5) decompositions: 
\beq
[{\rm SO(10)}]~~~{\bf 10}_L = {\bf 5}_L + {\bf 5}^*_L~~~[{\rm SU(5)}]~~~;~~~
[{\rm SO(10)}]~~~{\bf 1}_L = {\bf 1}_L~~~[{\rm SU(5)}]~~~.
\eeq
There has been some interest in \es~as a result of its appearance in certain
versions of superstring theories \cite{E6SS,rev}.
\bigskip

\leftline{\bf G.  The \es~zoo}
\bigskip

\begin{figure}
\centerline{\epsfysize = 2in \epsffile {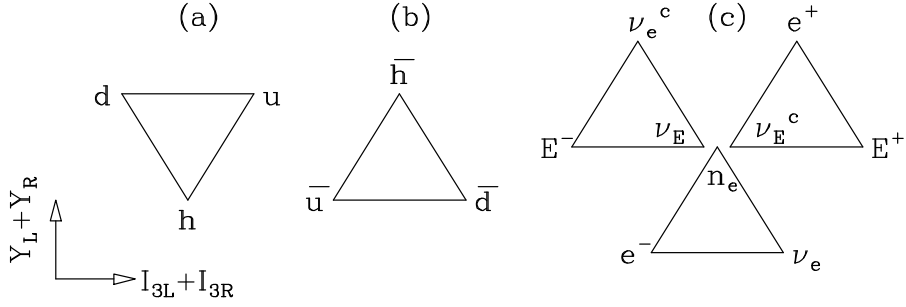}}
\caption{Left-handed fermions belonging to the {\bf 27}-plet of E$_6$.
(a) Quarks [triangle denotes {\bf 3} of SU(3)$_L$]; (b) antiquarks [triangle
denotes {\bf 3}$^*$ of SU(3)$_R$]; (c) leptons [triangles denote {\bf 3}$^*$'s
of SU(3)$_L$; they are arranged in a {\bf 3} of SU(3)$_R$].}
\end{figure}

One way to illustrate the particle content of the {\bf 27}-plet of \es~is via
the decomposition \cite{RS} \es$ \to$ SU(3)$_c \times$ SU(3)$_L \times$
SU(3)$_R$, whereby 
\beq
{\bf 27}_L = ({\bf 3}_c,  {\bf 3}_L,  {\bf 1}_R)
           + ({\bf 3}^*_c,{\bf 1}_L,  {\bf 3}^*_R)
           + ({\bf 1}_c,  {\bf 3}^*_L,{\bf 3}_R)~~~.
\eeq
The left-handed fermions can then be plotted in a space (Fig.~23) in which the
$x$-axis is $I_{3L} + I_{3R}$, while the $y$-axis is $Y_L + Y_R$.  The electric
charge of particles is $Q_{\rm em} = I_{3L} + I_{3R} + (Y_L + Y_R) /2$.  (I
thank Graham Ross for reminding me of the utility of this picture.)  The new
particles [beyond those in the {\bf 16} of SO(10)] consist of $h$ (an
isosinglet $Q = -1/3$ quark), $E^\pm$ (a weak isodoublet charged lepton),
$\nu_E$ and $\nu_E^c$ (a weak isodoublet Dirac neutrino, or else two Majorana
neutrinos each belonging to a weak isodoublet), and an ``inert'' isosinglet
Majorana neutrino $n_e$.  By ``weak isodoublet'' or ``weak isosinglet'' we
mean behavior under SU(2)$_L$.

Another way to visualize the additional states in \es~is to note that the
standard-model fermions belonging to $({\bf 16},{\bf 5}^*)$ of (SO(10),~SU(5))
are a column vector consisting of $(\bar d_1, \bar d_2, \bar d_3, e^-, \nu_e)$.
Then since SU(5) contains all the standard-model generators, equal SU(5)
representations have equal standard-model transformation properties.  (We are
not considering ``flipped'' SU(5) models in which the electric charge lies
outside SU(5) \cite{flip}.)  Thus, in $({\bf 10},{\bf 5}^*) = (\bar h_1, \bar
h_2, \bar h_3, E^-, \nu_E)$, one identifies the $\bar h_i$ states as a
color-triplet of isosinglet $Q = 1/3$ antiquarks, and $(E^-,\nu_E)$ as a weak
isodoublet of leptons.  In the charge-conjugate representation $({\bf 10},{\bf
5}) = (h_1,h_2,h_3, E^+, \nu_E^c)$, the $Q = -1/3$ quarks $h$ are then
weak {\it isosinglets}, while $E^+,\nu_E^c)$ are a weak {\it isodoublet}.
Such fermions are not encountered in the standard model.  Neither is the (${\bf
1}, {\bf 1}$) consisting of the single state $n_e$, which is a weak isosinglet. 

One can calculate the cross sections for $\eep \to (\gamma^*,Z^*) \to
f \bar f$ for the new fermions, in the manner leading to Table 10. In computing
the values of $g_V$ and $g_A$ for $E^-$ and a Dirac neutrino $\nu_E$ both
left-handed and right-handed states have the same value of $I_{3L}$: $-1/2$ for
$E^-$ and $+1/2$ for $\nu_E$.  The results are shown in Table 13. 

\begin{table}
\caption{Cross sections $\sigma$ [in units of $\sigma_0 \equiv \sigma(e^+ e^-
\to \gamma^* \to \mu^+ \mu^-$)] for $e^+ e^-$ production of pairs of exotic
fermions $f \bar f$ via virtual photons and $Z$'s in the direct channel. The
$\nu_E$ is assumed to be a Dirac neutrino. Values of $g_V^f$ are quoted for $x
= 0.2315$.  QCD corrections to $h$ quark production have been neglected.} 
\begin{center}
\begin{tabular}{c c c c c c} \hline
Fermion & $Q_f$  & $g_V^f$   & $g_A^f$ &$\sso$ far &$\sso$ far \\
  $f$   &        &           &         & below $Z$ & above $Z$ \\ \hline
  $h$   & $-1/3$ & ~~0.0772  &    0    &   1/3     &   0.35    \\
  $E^-$ &  $-1$  & $-0.2685$ &    0    &    1      &   1.20    \\
$\nu_E$ &    0   &    1/2    &    0    &    0      &   0.50    \\ \hline
\end{tabular}
\end{center}
\end{table}

All of these states are produced exclusively via the vector current, and hence
their maximum cross section is expected to occur very close to threshold.  If
one is sufficiently far from the $Z$ pole, this maximum occurs at $E_{\rm c.m.}
\simeq 1.18 \eth$.  Closer to the $Z$ pole, the maximum occurs even lower
\cite{eeggevt}. 

The new fermions in \es~are marginally consistent with the constraints on the
parameters $S$ and $T$ mentioned in Sec.~3.  The parameter $T$ describes the
effects of mass difference between $I_{3L} = +1/2$ and $I_{3L} = -1/2$
fermions.  Recalling $\rho/\rho^{\rm nominal} = 1 + \alpha T$, a more precise
expression for $\Delta \rho$ than that given in terms of $m_t$ is \cite{PLJE}
\beq
\Delta \rho = \frac{N_c G_F}{8 \sqrt{2} \pi^2} \left( m_1^2 + m_2^2 -
\frac{4 m_1^2 m_2^2}{m_1^2 - m_2^2} \ln \frac{m_1}{m_2} \right)~~~.
\eeq
The \es~fermions thus make a significant contribution to $\Delta \rho$
only when $m(E^-) \ne m(\nu_E)$.  The exotic quarks do not contribute
because they are isosinglets.

The parameter $S$ arises as a result of $W^0 - B$ mixing \cite{PT}.
Consequently, it receives no contribution in \es~from the isosinglet $h$
quarks.  There are two doublets $(E^-,\nu_E)$ and $(E^+,\nu_E^c)$ of leptons
per family, leading to a contribution $\Delta S = 2(1/6 \pi) \simeq 0.1$ per
family.  One can just barely tolerate three families. 
\bigskip

\leftline{\bf H.  Extended gauge structure}
\bigskip

At any given mass scale, there is an interplay between the fermion spectrum and
the subgroup of any grand unified theory (GUT) which remains unbroken at that
mass scale.  If a fermion is heavy, no GUT subgroup which relies on that
fermion for anomaly cancellation should be unbroken far below that fermion's
mass.  Let us recapitulate how this works in several of the unified gauge
groups we have mentioned.

1.  In SU(5) we need a {\bf 5}$^*$ and a {\bf 10} to avoid anomalies in
$I_{3L} Q^2$.

2.  In SO(10) $\to$ SU(5) $\times$ U(1)$_\chi$ (the notation is that of
Refs.~\cite{rev} and \cite{RR}), the U(1)$_\chi$ symmetry is anomalous (the sum
of $Q_\chi I_{3L}^2$ is non-zero) unless the $\nu_L^c$ is light. 

3.  In SO(10) $\to$ SO(6) $\times$ SO(4) (which is the same as SU(4) $\times$
SU(2)$_L \times$ SU(2)$_R$), the subsequent symmetry breakdowns are SU(4)
$\to$ SU(3)$c \times$ U(1)$_{B-L}$ and SU(2)$_R \to$ U(1)$_R$.  Each of these
U(1)'s separately is anomalous unless the $\nu_L^c$ is light.  However, one
can write the electric charge as
\beq
Q = I_{3L} + I_{3R} + \frac{B-L}{2} = I_{3L} + \frac{Y_W}{2}
\eeq
Since $I_{3L}(\nu_L^c) = -1/2$ while $(-L/2)(\nu_L^c) = +1/2$, one has
$(Y_W/2)(\nu_L^c) = 0$, and the $Y_W$ anomaly is insensitive to $\nu_L^c$.

4.  In \es $\to$ SO(10) $\times$ U(1)$_\psi$ (see, again, \cite{rev,RR})
the cancellation of the U(1)$_\psi$ anomaly requires the entire {\bf 27}-plet
to be light.

One may conjecture that the converse is also true:  If an extended set of
fermions is found to be light, there may exist an extended gauge structure for
which this set of fermions provides the required anomaly cancellation. Thus,
the search for new fermions may have bearing on extended gauge structures. One
can also directly search for new gauge bosons (see, e.g., Ref.~\cite{CDFZp}) in
lepton pair production, identifying them via characteristic forward-backward
asymmetries \cite{CDFas,FBA}. 
\bigskip

\leftline{\bf I.  An unusual event}
\bigskip

The CDF Collaboration \cite{eegg} has reported one event with an
electron-positron pair, two photons, and missing energy (\evt), produced in
proton-antiproton collisions at $E_{\rm c.m.} = 1.8$ TeV.  Popular
interpretations of this event have appeared within the context of
supersymmetry \cite {sseegg}, involving the production of a pair of selectrons
($\tilde e^+ \tilde e^-$) or a pair of charginos ($\chi^+ \chi^-$) which
then decay to $\eep$ and radiatively-decaying neutral superparners.  An
interpretation has also appeared in one non-supersymmetric model \cite{BM}.

The \es~scenario mentioned above can interpret this event in the following way
\cite{eeggevt}:
\beq
\bar p p \to Z_I + \ldots \to E^+ E^- + \ldots
\eeq
followed by the chain
\beq
E^- \to e^- W_I^{(*)} \to e^- \bar N_e \bar n_e \to e^- \gamma n_e \bar n_e
\eeq
and its charge-conjguate for $E^+$ decay.  The $n_e$ state is allowed to be
stable as long as its mass satisfies cosmological bounds (typically less than a
few tens of eV).  The $Z_I$ is a neutral gauge boson with mass greater than
present limits \cite{CDFZp} of a few hundred GeV.  The $W_I$ belongs to a
subgroup \sui~of \es~along with the $Z_I$ \cite{LR}; it occurs in the
decomposition \es$ \to$ \sui$ \times$ SU(6), where the SU(6) breaks
subsequently to the standard SU(5) GUT.  The $W_I$ is probably virtual, as
indicated by the asterisk in parentheses. The neutral nature of all three
bosons in \sui~is a key feature permitting the flavor of $E^-$ to be passed on
to the electron. 

How does one tell the difference between these scenarios?  Supersymmetry
has superpartners of the gauge bosons known as ``gauginos,'' and very
specific relations between couplings involving superpartners and ordinary
couplings.  \es~has exotic fermions but they are not superpartners of Higgs
bosons [despite a superficial resemblance; both belong to {\bf 10}'s of
SO(10).]  A typical reaction expected in \es~ is $\eep \to \nu_E \bar \nu_E$,
followed by $\nu_E \to \nu_e \bar N_1 N_2$, with one of the exotic neutrinos
$\bar N_1$ decaying radiatively to $\bar N_2 + \gamma$.

A feature common to both scenarios is that in addition to the \evt~event, one
should also see in $\bar p p$ collisions some diphoton events with missing
transverse energy and {\it without} charged lepton pairs.  Searches for such
events have been performed in the CDF data \cite{Toback}.  Of all events with
two photons and missing transverse energy, the event \evt~has the highest
$\met$.  Thus, there are no indications yet of non-standard behavior in
the diphoton events with $\met$ but without $\eep$.

There is still a need for extensive discussions of standard-model backgrounds
to this event, such as multiple interactions, radiative production of $W$
pairs, effects of cracks in the detector, and so on.  One cannot conclude
anything from a single event, but it can point the way to searches for related
phenomena. 
\bigskip

\leftline{\bf 7.  CONCLUSIONS}
\bigskip

We have discussed the top quark mass more from the standpoint of what it
does than where it comes from.  It participates in a crucial way in loop
diagrams, leading to mixing of neutral kaons and $B$ mesons, thereby accounting
for CP violation in the kaon system and predicting appreciable CP-ciolating
effects for $B$'s.  It also affects electroweak observables, to the extent that
its mass could be anticipated to within a few tens of $\G/c^2$ before it was
discovered.

The origin of the top quark mass, on the other hand, is related to the origin
of {\it all} the quark masses, which also is linked to the curious pattern
of weak charge-changing couplings expressed by the Cabibbo-Kobayashi-Maskawa
(CKM) matrix.  We have presented a {\it potpourri} of ideas about these
questions while not finding any of them particularly conclusive.  The best
we can do is to note some overall patterns suggested by the quark and lepton
masses in Fig.~1, which may indicate that we are not yet in possession of
the complete experimental picture.  At the very least, this is certainly
the case for neutrinos, where we don't even know if they have masses.  But
there could be more.  I leave you with the following exercise in pattern
recognition:  What familiar pattern do you see in Fig.~24?

\begin{figure}
\centerline{\epsfysize = 1.6in \epsffile {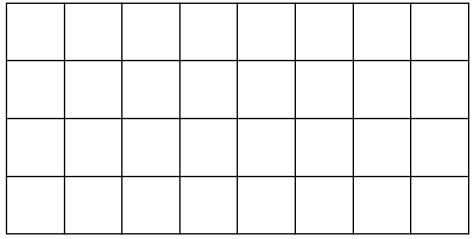}}
\caption{Part of a familiar pattern.}
\end{figure}

One can re-express the pattern as shown in Fig.~25; perhaps it suggests
something at this point.

\begin{figure}
\centerline{\epsfysize = 1.6in \epsffile {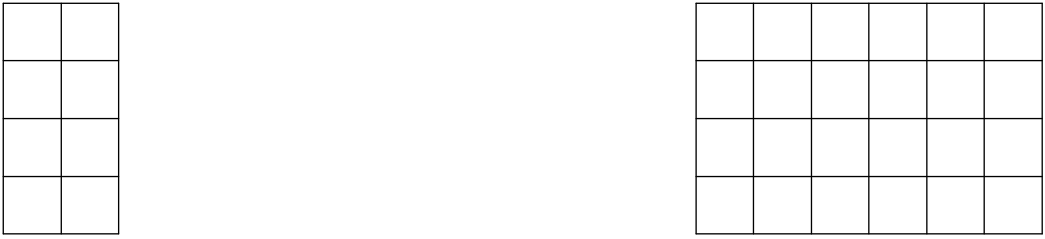}}
\caption{Part of a familiar pattern, expressed differently.}
\end{figure}

Finally, when one adds variety to the pattern, it becomes recognizable as
the periodic table of the elements (Fig.~26).

\begin{figure}
\centerline{\epsfysize = 1.6in \epsffile {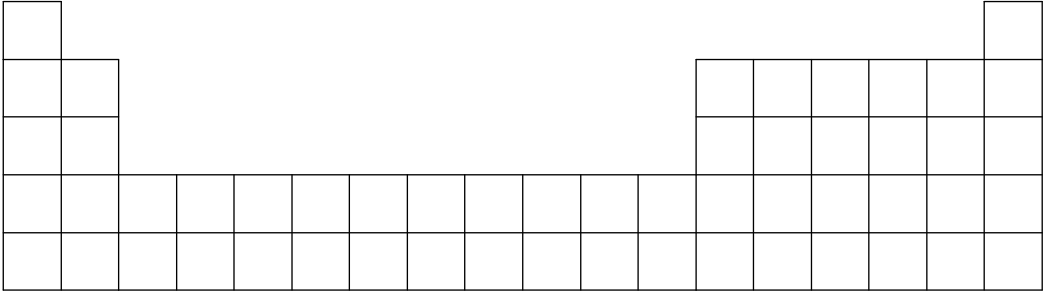}}
\caption{A larger part of a familiar pattern.}
\end{figure}

The variety of the pattern of the elementary particles laid the foundations of
the quark model and our understanding of the fundamental strong interactions.
Will there be a similar advance for quarks and leptons? The pattern of quarks
and leptons has been quite regular up to now, just as if the periodic table of
the elements consisted only of rows of equal length and were missing hydrogen,
helium, the transition metals, the lanthanides, and the actinides.  Whether one
discovers superpartners of the known states, or variety such as predicted in
extended gauge structures, the new states could help us to make sense of the
pattern of the masses of the more familiar ones. In this sense the top quark
mass may not be the end of a story, but just the beginning. 
\bigskip

\leftline{\bf ACKNOWLEDGMENTS}
\bigskip

I would like to thank Jim Amundson, Isi Dunietz, Aaron Grant, Michael Gronau,
Oscar Hern\'andez,  Nahmin Horowitz, Mike Kelly, Harry Lipkin, David London,
Bill Marciano, Alex Nippe, Sheldon Stone, Tatsu Takeuchi, and Mihir Worah for
enjoyable collaborations on some of the topics mentioned in these lectures. In
addition I am grateful to Gideon Alexander, Guido Altarelli, Alain Blondel,
Keith Ellis, Paul Frampton, Henry Frisch, Michaelangelo Mangano, Paolo Nason,
Matthias Neubert, Steve Pollock, Graham Ross, Chris Sachrajda, Bob Sachs, Pat
Sandars, Michael Schmitt, Jack Steinberger, Sam Ting, Tini Veltman, Petr Vogel,
Brian Webber, Larry Wilets, Bruce Winstein, and Lincoln Wolfenstein for
fruitful discussions, and to the CERN Theory Group for its hospitality during
the preparation of this report. Finally, I wish to thank the organizers of the
Summer Institute and all of my other hosts in Carg\`ese for making this first
visit such an enjoyable one. Parts of the research described here were
performed at the Aspen Center for Physics, the Fermilab Theory Group, the
Institute for Nuclear Theory at the University of Washington, and the Technion.
This work was supported in part by the United States Department of Energy under
Grant No.~DE FG02 90ER40560. 
\newpage

\def \ajp#1#2#3{{\it Am.~J.~Phys.} #1:#2 (#3)}
\def \ap#1#2#3{{\it Ann.~Phys.~(N.Y.)} #1:#2 (#3)}
\def \apny#1#2#3{{\it Ann.~Phys.~(N.Y.)} #1:#2 (#3)}
\def \app#1#2#3{{\it Acta Physica Polonica} #1:#2 (#3)}
\def \arnps#1#2#3{{\it Ann.~Rev.~Nucl.~Part.~Sci.} #1:#2 (#3)}
\def \arns#1#2#3{{\it Ann.~Rev.~Nucl.~Sci.} #1:#2 (#3)}
\def \art{and references therein}
\def \ba88{{\it Particles and Fields 3} (Proceedings of the 1988 Banff Summer
Institute on Particles and Fields), edited by A. N. Kamal and F. C. Khanna
(World Scientific, Singapore, 1989)}
\def \baps#1#2#3{{\it Bull.~Am.~Phys.~Soc.} #1:#2 (#3)}
\def \be87{{\it Proceedings of the Workshop on High Sensitivity Beauty
Physics at Fermilab,} Fermilab, Nov. 11--14, 1987, edited by A. J. Slaughter,
N. Lockyer, and M. Schmidt (Fermilab, Batavia, IL, 1988)} 
\def \cmts#1#2#3{{\it Comments on Nucl.~and Part.~Phys.} #1:#2 (#3)}
\def \cn{Collaboration}
\def \corn{{\it Lepton and Photon Interactions:  XVI International Symposium,
Ithaca, NY 1993,} edited by P. Drell and D. Rubin (AIP, New York, 1994)}
\def \cp89{{\it CP Violation,} edited by C. Jarlskog (World Scientific,
Singapore, 1989)} 
\def \dpfa{{\it The Albuquerque Meeting:  DPF 94} (Division of Particles and
Fields Meeting, American Physical Society, Albuquerque, NM, August 2--6,
1994), ed. by S. Seidel (World Scientific, River Edge, NJ, 1995)}
\def \dpff{{\it The Fermilab Meeting -- DPF 92} (Division of Particles and
Fields Meeting, American Physical Society, Fermilab, 10--14 November, 1992),
ed. by C. H. Albright \ite~(World Scientific, Singapore, 1993)} 
\def \dpfm{{\it The Minneapolis Meeting:  DPF 96} (Division of Particles and
Fields Meeting, American Physical Society, Minneapolis, MN, 10--15 August,
1996), to be published}
\def \dpfv{{\it The Vancouver Meeting - Particles and Fields '91}
(Division of Particles and Fields Meeting, American Physical Society,
Vancouver, Canada, Aug.~18--22, 1991), ed. by D. Axen, D. Bryman, and M. Comyn
(World Scientific, Singapore, 1992)} 
\def \efi{Enrico Fermi Institute Report No.~}
\def \fermlg{{\it Proc.~Int.~Symp.~on Lepton and Photon Interactions at High
Energies (Fermilab, August 23--29, 1979)}, T. B. W. Kirk and H. D. I.
Abarbanel, eds., Fermilab, Batavia, IL (1979)}
\def \hb87{{\it Proceeding of the 1987 International Symposium on Lepton and
Photon Interactions at High Energies,} Hamburg, 1987, ed. by W. Bartel
and R. R\"uckl (Nucl.~Phys.~B, Proc. Suppl., vol. 3) (North-Holland,
Amsterdam, 1988)}
\def \ib{{\it ibid.}}
\def \ibj#1#2#3{{\it ibid.} #1:#2 (#3)}
\def \ijmpa#1#2#3{{\it Int.~J. Mod.~Phys.}~A #1:#2 (#3)}
\def \jpb#1#2#3{{\it J. Phys.} B #1:#2 (#3)}
\def \jpg#1#2#3{{\it J. Phys.} G #1:#2 (#3)}
\def \kdvs#1#2#3{{\it Kong.~Danske Vid.~Selsk., Matt-fys.~Medd.} #1:No.~#2
(#3)}
\def \ky85{{\it Proceedings of the International Symposium on Lepton and
Photon Interactions at High Energy,} Kyoto, Aug.~19-24, 1985, edited by M.
Konuma and K. Takahashi (Kyoto Univ., Kyoto, 1985)} 
\def \latm{{\it Lattice 1995} (Proceedings of the International Symposium on
Lattice Field Theory, Melbourne, Australia, 11--15 July 1995, T. D. Kieu,
B. H. J. McKellar, and A. Guttman, eds., North-Holland, Amsterdam, 1996}
\def \lgb{{\it LP95: Proceedings of the International Symposium on Lepton and
Photon Interactions (IHEP)}, 10--15 August 1995, Beijing, People's Republic of
China, Z.-P. Zheng and H.-S. Chen, eds., World Scientific, Singapore, 1996} 
\def \lgg{International Symposium on Lepton and Photon Interactions, Geneva,
Switzerland, July, 1991}
\def \lkl87{{\it Selected Topics in Electroweak Interactions} (Proceedings of 
the Second Lake Louise Institute on New Frontiers in Particle Physics, 15--21
February, 1987), edited by J. M. Cameron \ite~(World Scientific, Singapore,
1987)}
\def \lti{lectures at this Institute}
\def \mpla #1#2#3{{\it Mod.~Phys.~Lett.} A #1:#2 (#3)}
\def \nc#1#2#3{{\it Nuovo Cim.} #1:#2 (#3)}
\def \np#1#2#3{{\it Nucl.~Phys.} #1:#2 (#3)}
\def \npbps#1#2#3{{\it Nucl.~Phys.~B (Proc.~Suppl.)} #1:#2 (#3)}
\def \oxf65{{\it Proceedings of the Oxford International Conference on
Elementary Particles} 19/25 Sept.~1965, ed.~by T. R. Walsh (Chilton, Rutherford
High Energy Laboratory, 1966)}
\def \pascos{{\it PASCOS 94} (Proceedings of the Fourth International
Symposium on Particles, Strings, and Cosmology, Syracuse University, 19--24
May 1994), ed.~by K. C. Wali (World Scientific, Singapore, 1995)}
\def \pisma#1#2#3#4{{\it Pis'ma Zh. Eksp. Teor. Fiz.} #1:#2 (#3) [{\it
JETP Lett.} #1:#4 (#3)]} 
\def \pl#1#2#3{{\it Phys.~Lett.} #1:#2 (#3)}
\def \pla#1#2#3{{\it Phys.~Lett.} A #1:#2 (#3)}
\def \plb#1#2#3{{\it Phys.~Lett.} B #1:#2 (#3)}
\def \ppmsj#1#2#3{{\it Proc.~Phys.~Math.~Soc.~Jap.} #1:#2 (#3)}
\def \pnpp#1#2#3{{\it Prog.~Nucl.~Part.~Phys.} #1:#2 (#3)}
\def \pr#1#2#3{{\it Phys.~Rev.} #1:#2 (#3)}
\def \prd#1#2#3{{\it Phys.~Rev.} D #1:#2 (#3)}
\def \prl#1#2#3{{\it Phys.~Rev.~Lett.} #1:#2 (#3)}
\def \prp#1#2#3{{\it Phys.~Rep.} #1:#2 (#3)}
\def \ptp#1#2#3{{\it Prog.~Theor.~Phys.} #1:#2 (#3)}
\def \ptwaw{Plenary talk, XXVIII International Conference on High Energy
Physics, Warsaw, July 25--31, 1996}
\def \rmp#1#2#3{{\it Rev.~Mod.~Phys.} #1:#2 (#3)}
\def \rp#1{~~~~~\ldots\ldots{\rm rp~}{#1}~~~~~}
\def \si90{25th International Conference on High Energy Physics, Singapore,
Aug. 2-8, 1990, Proceedings edited by K. K. Phua and Y. Yamaguchi (World
Scientific, Teaneck, N. J., 1991)}
\def \slaclg{{\it Proceedings of the 1975 International Symposium on
Lepton and Photon Interactions at High Energies,} Stanford University,
Aug.~21--27, 1975, W. T. Kirk, ed., SLAC, Stanford, CA, (1975)} 
\def \slc87{{\it Proceedings of the Salt Lake City Meeting} (Division of
Particles and Fields, American Physical Society, Salt Lake City, Utah, 1987),
ed. by C. DeTar and J. S. Ball (World Scientific, Singapore, 1987)}
\def \smass82{{\it Proceedings of the 1982 DPF Summer Study on Elementary
Particle Physics and Future Facilities}, Snowmass, Colorado, edited by R.
Donaldson, R. Gustafson, and F. Paige (World Scientific, Singapore, 1982)}
\def \smass90{{\it Research Directions for the Decade} (Proceedings of the
1990 DPF Snowmass Workshop), edited by E. L. Berger (World Scientific,
Singapore, 1991)}
\def \smassb{{\it Proceedings of the Workshop on $B$ Physics at Hadron
Accelerators}, Snowmass, Colorado, 21 June--2 July 1994, ed.~by P. McBride
and C. S. Mishra, Fermilab report FERMILAB-CONF-93/267 (Fermilab, Batavia, IL,
1993)} 
\def \stone{{\it B Decays}, edited by S. Stone (World Scientific, Singapore,
1994)}
\def \tasi90{{\it Testing the Standard Model} (Proceedings of the 1990
Theoretical Advanced Study Institute in Elementary Particle Physics),
edited by M. Cveti\v{c} and P. Langacker (World Scientific, Singapore, 1991)}
\def \waw{XXVIII International Conference on High Energy
Physics, Warsaw, July 25--31, 1996}
\def \yaf#1#2#3#4{{\it Yad.~Fiz.} #1:#2 (#3) [Sov.~J.~Nucl.~Phys.~#1:#4 (#3)]}
\def \zhetf#1#2#3#4#5#6{{\it Zh.~Eksp.~Teor.~Fiz.} #1:#2 (#3) [Sov.~Phys.~--
JETP #4:#5 (#6)]}
\def \zhetfl#1#2#3#4{{\it Pis'ma Zh.~Eksp.~Teor.~Fiz.} #1:#2 (#3) [JETP
Letters #1:#4 (#3)]}
\def \zp#1#2#3{{\it Zeit.~Phys.} #1:#2 (#3)}
\def \zpc#1#2#3{{\it Zeit.~Phys.} C #1:#2 (#3)}


\begin{thebibliography}{249}

\bibitem{Ross} G. Ross, \lti.

\bibitem{LeCompte} T. J. LeCompte, Top decay physics at CDF and measurement of
the CKM element $V_{tb}$, Fermilab report FERMILAB-CONF-96-021-E, Jan., 1996,
presented at 2nd Rencontres du Vietnam, {\it Physics at the Frontiers of
the Standard Model}, Ho Chi Minh City, Vietnam, 21 -- 28 Oct., 1995.

\bibitem{Cab} N. Cabibbo, \prl{10}{531}{1963}.

\bibitem{KM} M. Kobayashi and T. Maskawa, \ptp{49}{652}{1973}.

\bibitem{JRCP} J. L. Rosner, Present and future aspects of CP violation,
\efi 95-36 [hep-ph/9506364], lectures given at the VIII J. A. Swieca Summer
School, {\it Particles and Fields}, Rio de Janeiro, February, 1995, proceedings
to be published by World Scientific. 

\bibitem{GWS} S. L. Glashow, \np{22}{579}{1961}; S. Weinberg,
\prl{19}{1264}{1967}; A. Salam, in {\it Proceedings of the Eighth Nobel
Symposium}, edited by N. Svartholm (Almqvist and Wiksell, Stockholm; Wiley, New
York, 1978), p. 367. 

\bibitem{LEPEWWG} LEP Electroweak Working Group, CERN report LEPEWWG/96-02,
July 30, 1996, presented at \waw.

\bibitem{HMOBG} Y. Hara, \pr{134}{B701}{1964}; Z. Maki and Y. Ohnuki,
\ptp{32}{144}{1964}; B. J. Bjorken and S. L. Glashow, \pl{11}{255}{1964}.

\bibitem{GL} M. Gell-Mann and M. L\'evy, \nc{19}{705}{1960}.

\bibitem{GIM} S. L. Glashow, J. Iliopoulos, and L. Maiani, \prd{2}{1285}{1970}.

\bibitem{BIM} C. Bouchiat, J. Iliopoulos, and Ph. Meyer, \pl{38B}{519}{1972}.
See also H. Georgi and S. L. Glashow, \prd{6}{429}{1972}; D. J. Gross and
R. Jackiw, \ibj{6}{477}{1972}.

\bibitem{hist} For more details see, e.g., V. L. Fitch and J. L. Rosner,
Elementary particle physics in the second half of the twentieth century, in
{\it Twentieth Century Physics}, edited by L. M. Brown, A. Pais, and B. Pippard
(AIP/IOP, New York and Bristol, 1995), ch.~9.

\bibitem{Pauli} W. Pauli, Zur \"alteren und neueren Geschichte des Neutrinos,
in {\it Collected Scientific Papers}, R. Kronig and V. F. Weisskopf, eds.,
Interscience, New York, 1964, v.~2, p.~1313. 

\bibitem{isospin} See, e.g., L. M. Brown, in {\it Twentieth Century Physics},
L. M. Brown, A. Pais, and B. Pippard, eds., AIP/IOP, New York and Bristol,
1995, ch.~5.

\bibitem{NA} S. H. Neddermeyer and C. D. Anderson, \pr{51}{884}{1937}.  See
also J. C. Street and E. C. Stevenson, \pr{51}{1005}{1937}; Y. Nishina, M.
Takeuchi, and T. Ichimaya, \pr{52}{1198}{1937}.

\bibitem{twonu} G. Danby \ite, \prl{9}{36}{1962}.

\bibitem{Ting} J. J. Aubert \ite, \prl{33}{1404}{1974}.

\bibitem{Richter} J.-E. Augustin \ite, \prl{33}{1406}{1974}.

\bibitem{charm} E. G. Cazzoli \ite, \prl{34}{1125}{1975}; G. Goldhaber \ite,
\prl{37}{255}{1976}; I. Peruzzi \ite, \prl{37}{569}{1976}.

\bibitem{Perl} M. Perl \ite, \prl{35}{1489}{1975}; \pl{63B}{466}{1976};
\pl{70B}{487}{1977}.

\bibitem{HH} H. Harari, in \slaclg, p.~317.

\bibitem{upsilons} S. W. Herb \ite, \prl{39}{252}{1977}; W. R. Innes \ite,
\prl{39}{1240, 1640(E)}{1977}.

\bibitem{beauty} S. Behrends \ite~(CLEO \cn), \prl{50}{881}{1983}.

\bibitem{top} F. Abe \ite~(CDF \cn), \prl{73}{225}{1994}; \prd{50}{2966}
{1995}; \prl{74}{2626}{1995}; S. Abachi \ite~(D0 \cn), \prl{74}{2632} {1995}. 

\bibitem{nutau} Fermilab E-872, R. Ramieka, spokesperson.  See B. Lundberg
\ite, Measurement of $\tau$ lepton production from the process $\nu_\tau +
N \to \tau$, Fermilab proposal P-872, Jan., 1994.

\bibitem{Watop} P. Tipton, \ptwaw;  CDF \cn, presented by R. Hughes, reports
PA-05-052, PA05-053, and PA08-108; D0 \cn, presented by A. Zieminski,
reports PA05-027, PA05-028, and PA05-029.

\bibitem{JKT} M. Je$\dot{\rm z}$abek, J. H. K\"uhn, and T. Teubner,
\zpc{56}{653}{1992}. 

\bibitem{etacJP} Mark III \cn, R. M. Baltrusaitis \ite, \prl{52}{2126}{1984}.

\bibitem{EDT} See, e.g., W. Kwong and J. L. Rosner, \prd{38}{279}{1988};
W. Buchm\"uller and S. Cooper, Upsilon spectroscopy, in {\it High-Energy
Electron-Positron Physics}, A. Ali and P. S\"oding, eds., World Scientific,
Singapore (1988), p.~412.

\bibitem{QR} C. Quigg and J. L. Rosner, \pl{71B}{153}{1977}; \cmts{8}{11}
{1978}; \prp{56}{167}{1979}.

\bibitem{Martin} A. Martin, \pl{93B}{338}{1980}; \ibj{100B}{511}{1981}.

\bibitem{Xicstar} CLEO \cn, P. Avery \ite, \prl{75}{4364}{1995};
L. Gibbons \ite, \prl{77}{810}{1996}.

\bibitem{Sigmacstar} CLEO \cn, report PA01-078, presented by R. Kutschke at
\waw.  We thank J. Yelton for communicating this result to us. 

\bibitem{MN} M. Neubert, \lti. See also M. Neubert, CERN report CERN-TH/96-55
[hep-ph/9604412], to be published in Int.~J. Mod.~Phys.~A. 

\bibitem{RGtop} N. Gray, D. J. Broadhurst, W. Grafe, and K. Schilcher,
\zpc{48}{673}{1990}, \art.

\bibitem{AL} A. Ali and D. London, DESY report DESY 96-140 [hep-ph/9607392],
presented at High Energy Physics International Euroconference on QCD (QCD 96),
Montpellier, France, July 4 -- 12, 1996. 

\bibitem{QP} L.-L. Chau and W.-Y. Keung, \prl{53}{1802}{1984}; H. Harari and M.
Leurer, \plb{181}{123}{1986}; J. D. Bjorken and I. Dunietz,
\prd{36}{2109}{1987}. 

\bibitem{WP} L. Wolfenstein, \prl{51}{1945}{1983}.

\bibitem{strange} See, e.g., M. Bourquin \ite, \zpc{21}{27}{1983}; H. Leutwyler
and M. Roos, \zpc{25}{91}{1984}; J. F. Donoghue, B. R. Holstein, and S. W.
Klimt, \prd{35}{934}{1987}; F. J. Gilman, K. Kleinknecht, and B. Renk, in
Review of particle properties, Particle Data Group, \prd{54}{94}{1996}.

\bibitem{BH} T. Browder and K. Honscheid, \pnpp{35}{81}{1995}.

\bibitem{Gib} L. Gibbons, \ptwaw.

\bibitem{UT} L.-L. Chau and W.-Y. Keung, \prl{53}{1802}{1984}; M. Gronau and J.
Schechter, \ibj{54}{385}{1985}; M. Gronau, R. Johnson, and J. Schechter,
\prd{32}{3062}{1985}; C. Jarlskog, in {\it Physics at LEAR with Low Energy
Antiprotons,} proceedings of the workshop, Villars-sur-Ollon, Switzerland,
1987, edited by C. Amsler \ite ~(Harwood, Chur, Switzerland, 1988), p. 571; J.
D. Bjorken and I. Dunietz, \prd{36}{2109}{1987}. 

\bibitem{NQ} Y. Nir and H. Quinn, \arnps{42}{211}{1992}.

\bibitem{CL} T. P. Cheng and L. F. Li, {\it Gauge Theory of Elementary
Particles} (Oxford University Press, 1984).

\bibitem{TASI} J. L. Rosner, in \tasi90, p.~91.

\bibitem{IL} T. Inami and C. S. Lim, \ptp{65}{297}{1981};
\ibj{65}{1772(E)}{1981}. 

\bibitem{PDG} Particle Data Group, \prd{54}{1}{1996}.

\bibitem{latBB} A. Abada \ite, \np{B376}{172}{1992}.

\bibitem{QCDB} A. Buras, M. Jamin, and P. H. Weisz, \np{B347}{491}{1990}.

\bibitem{UA1mix}  UA1 \cn, C. Albajar \ite, \plb{186}{237, 247}{1987};
\ibj{197}{565(E)}{1987}.

\bibitem{ARmix} ARGUS \cn, H. Albrecht \ite, \plb{192}{245}{1987}.

\bibitem{LEPmix} ALEPH \cn, D. Buskulic \ite, \plb{313}{498}{1993}; DELPHI
\cn, P. Abreu \ite, \plb{338}{409}{1994}; OPAL \cn, R. Akers \ite,
\plb{327}{411}{1994}; \ibj{336}{585}{1994}.  For a compilation of the many
new LEP results presented in 1996 see Ref.~\cite{Gib}.

\bibitem{CDFmix} CDF \cn, F. Abe \ite, Fermilab report FERMILAB-CONF-95-231-E,
July, 1995, contributed to \lgb; report PA08-032, presented by P. Sphicas at
\waw. 

\bibitem{SLDmix} SLD \cn, reports PA08-026A, PA08-026B, and PA08-027/028,
presented by D. Su at \waw.

\bibitem{Dmavg} C. Zeitnitz, invited talk at the 4th International Workshop on
$B$-Physics at Hadron Machines (BEAUTY 96), Rome, June 17--21, 1996,
proceedings to be published in Nucl.~Instr.~Meth.  A slightly different average
was presented in Ref.~\cite{Gib}: $\Delta m_d = 0.464 \pm 0.017~{\rm ps}^{-1}$.

\bibitem{Dsmeas} CERN WA75 \cn, S. Aoki \ite, \ptp{89}{131}{1993}; CLEO \cn, D.
Acosta \ite, \prd{49}{5690}{1994}; F. Muheim and S. Stone,
\prd{49}{3767}{1994}; BES \cn, J. Z. Bai \ite, \prl{74}{4599}{1995}; F. Muheim,
parallel session on $B$ physics, in \dpfa; Fermilab E653 \cn, K. Kodama
\ite, \plb{382}{299}{1996}.

\bibitem{Dsfact} ARGUS \cn, H. Albrecht \ite, \plb{219}{121}{1989}; D.
Bortoletto, Ph.~D. Thesis, Syracuse University, 1989; D. Bortoletto and S. L.
Stone, \prl{65}{2951}{1990}; CLEO \cn, D. Bortoletto \ite,
\prd{45}{2212}{1992}; J. L. Rosner, \prd{42}{3732}{1990}; in \smass90, p.~255;
M. Paulini \ite, in {\it Proceedings of the Joint International Symposium and
Europhysics Conference on High Energy Physics,} S. Hegarty, K. Potter, and E.
Quercigh, eds., World Scientific, Singapore (1992), p.~592;  CLEO \cn, A. Bean
\ite, \prl{70}{2681}{1993}. 

\bibitem{Mart} G. Martinelli, \ptwaw.

\bibitem{MkIII} Mark III \cn, J. Adler \ite, \prl{60}{1375}{1988};
\ibj{63}{1658(E)}{1989}.

\bibitem{BEPC} J. Z. Bai \ite~(BEPC \cn), SLAC report SLAC-PUB-7147, 1996.

\bibitem{FBQ} J. Amundson \ite, \prd{47}{3059}{1993}; J. L. Rosner,
\dpff, p.~658.

\bibitem{FBL} See, e.g., A. Duncan \ite, \prd{51}{5101}{1995}; UKQCD \cn, C. R.
Allton \ite, \np{B437}{641}{1995}; MILC \cn, C. Bernard \ite, report
WASH-U-HEP-96-31, June, 1996 [hep-lat/9608092], presented at Lattice 96:  14th
International Symposium on Lattice Field Theory, St. Louis, MO, 4--8 June 1996;
G. Martinelli, Ref.~\cite{Mart}.

\bibitem{Grin} B. Grinstein, \prl{71}{3067}{1993}.

\bibitem{CCFT} J. H. Christenson, J. W. Cronin, V. L. Fitch, and R. Turlay,
\prl{13}{138}{1964}.

\bibitem{CJ} \cp89.

\bibitem{GP} M. Gell-Mann and A. Pais, \pr{97}{1387}{1955}.

\bibitem{KCP} T. D. Lee, R. Oehme and C. N. Yang, \pr{106}{340}{1957}; 
B. L. Ioffe, L. B. Okun' and A. P. Rudik, \zhetf{32}{396}{1957}{5}{328}{1957}.

\bibitem{KL} K. Lande, E. T. Booth, J. Impeduglia, and L. M. Lederman,
\pr{103}{1901}{1956}.

\bibitem{sw} L. Wolfenstein, \prl{13}{562}{1964}.

\bibitem{QCDK} A. J. Buras, M. Jamin, and P. H. Weisz, \np{B347}{491}{1990}; S.
Herrlich and U. Nierste, \np{B419}{292}{1994}; \prd{52}{6505}{1995}. 

\bibitem{HRS} C. Hamzaoui, J. L. Rosner, and A. I. Sanda, in \be87, p. 97; J.
L. Rosner, A. I. Sanda, and M. P. Schmidt, \ib, p. 165.

\bibitem{BKlat} See, e.g., the following sources quoted by \cite{AL}: S.
Sharpe, \npbps{34}{403}{1994}; J. Bijnens and J. Prades, \np{B444}{523}{1995};
A. Soni, Brookhaven National Laboratory report BNL-62284 [hep-lat/9510036];
JLQCD \cn, S. Aoki \ite, \npbps{47}{465}{1996}; APE \cn, M. Crisafulli \ite,
\plb{369}{325}{1996}.  Our quoted error reflect a somewhat more generous spread
in values obtained by different means. 

\bibitem{BCP} For discussions with references to earlier literature see, e.g.,
I. I. Bigi and A. I. Sanda, \np{B281}{41}{1987}; I. Dunietz,
\ap{184}{350}{1988}; M. B. Wise, in \ba88, p.~124; I. I. Bigi, V. A. Khoze, N.
G. Uraltsev, and A. I. Sanda, in Ref.~\cite{CJ}, p.~175; J. L. Rosner, in
\tasi90, p.~91; Y. Nir and H. R. Quinn, \arnps{42}{211}{1992}; B. Winstein and
L. Wolfenstein, \rmp{65}{1113}{1993}. 

\bibitem{Isi} I. Dunietz and J. L. Rosner, \prd{34}{1404}{1986}; I. Dunietz,
Ref.~\cite{BCP}. 

\bibitem{PP} D. London and R. Peccei, \plb{223}{257}{1989}; M. Gronau,
\prl{63}{1451}{1989}; B. Grinstein, \plb{229}{280}{1989}; M. Gronau,
\plb{300}{163}{1993}.

\bibitem{pipi} M. Gronau and D. London, \prl{65}{3381}{1990}.

\bibitem{dms} ALEPH and DELPHI \cn s, presented at \waw; L. Gibbons,
Ref.~\cite{Gib}.

\bibitem{Blifes} I. I. Bigi \ite, in \stone, p.~132.

\bibitem{IsiBs} I. Dunietz, \prd{52}{3048}{1995}.

\bibitem{BP} T. E. Browder and S. Pakvasa, \prd{52}{3123}{1995}. 

\bibitem{CPeven} R. Aleksan, A. Le Yaouanc, L. Oliver, O. P\`ene, and J. C.
Raynal, \plb{316}{567}{1993}.

\bibitem{Beneke} M. Beneke, G. Buchalla, and I. Dunietz, \prd{54}{4419}{1996};
M. Beneke, SLAC report SLAC-PUB-96-7258, presented at \waw.

\bibitem{DDLR} A. S. Dighe, I. Dunietz, H. J. Lipkin, and J. L. Rosner,
\plb{369}{144}{1996}.  See also R. Fleischer and I. Dunietz, Univ.~of Karlsruhe
report TTP 96-07 [hep-ph/9605220] (unpublished); A. Dighe, I. Dunietz, and
R. Fleischer, in preparation.

\bibitem{Tr} H. J. Lipkin, in {\it Proceedings of the SLAC Workshop on Physics
and Detector Issues for a High-Luminosity Asymmetric B Factory}, edited by
David Hitlin, published as SLAC, LBL and Caltech reports SLAC-373, LBL-30097
and CALT-68-1697, 1990, p. 49; I. Dunietz \ite, \prd{43}{2193}{1991}; I.
Dunietz, Appendix in \stone, p.~550. 

\bibitem{CLEOTr} CLEO \cn, CLEO-CONF-96-22, report PA05-074, presented by
R. Kutschke at \waw.

\bibitem{FMJR} J. L. Rosner, \prd{42}{3732}{1990}.

\bibitem{Yang} L. D. Landau, {\it Dokl.~Akad.~Nauk SSSR} 60:207 (1948);
C. N. Yang, \pr{77}{242,722}{1950}.  See also N. P. Chang and C. A. Nelson,
Jr., \prl{40}{1617}{1978}; \prd{20}{2923}{1979}.

\bibitem{pen} J. Ellis, M. K. Gaillard, and D. V. Nanopoulos, \np{B100}{313}
{1975}; \ibj{B109}{213}{1976}; A. I. Va\u{\i}nshte\u{\i}n, V. I. Zakharov, and
M. A. Shifman, \pisma{22}{123}{1975}{55};
\zhetf{72}{1275}{1977}{45}{670}{1977}; J. Ellis, M. K. Gaillard, D. V.
Nanopoulos, and S. Rudaz, \np {B131}{285}{1977}; \ibj{B132}{541(E)}{1978}. 

\bibitem{epspth} M. B. Wise, in \ba88, p.~124; J. M. Flynn and L. Randall,
\plb{224}{221}{1989}; G. Buchalla, A. J. Buras, and M. Harlander,
\np{B337}{313}{1990}; A. J. Buras, M. Jamin, and M. E. Lautenbacher,
\np{B408}{209}{1993}; M. Ciuchini, E. Franco, G. Martinelli, and L. Reina,
\plb{301}{263}{1993}; S. Bertolini, talk given at Workshop on $K$ Physics,
Orsay, France, 30 May -- 4 June 1996, INFN (Trieste) report SISSA-110-96-EP
[hep-ph/9607312]; A. J. Buras, \ptwaw.

\bibitem{E731} Fermilab E731 \cn, L. K. Gibbons \ite, \prl{70}{1203}{1993}.

\bibitem{NA31} CERN NA31 \cn, G. D. Barr \ite, \plb{317}{233}{1993}.

\bibitem{BuFl} A. J. Buras and R. Fleischer, \plb{341}{379}{1995}.

\bibitem{Battle} CLEO \cn, M. Battle \ite, \prl{71}{3922}{1993}.

\bibitem{Wurt} CLEO \cn, D. M. Asner \ite, \prd{53}{1039}{1996}.

\bibitem{CLEOGlas} CLEO \cn,  P. Gaidarev, \baps{40}{923}{1995}.

\bibitem{BPP} M. Gronau, J. L. Rosner, and D. London, \prl{73}{21}{1994}; O. F.
Hernandez, D. London, M. Gronau, and J. L. Rosner, \plb{333}{500}{1994}; M.
Gronau, O. F. Hern\'andez, D. London, and J. L. Rosner, \prd{50}{4529} {1994}.

\bibitem{GReta} M. Gronau and J. L. Rosner, \prd{53}{2516}{1996}.

\bibitem{DGR} M. Gronau and J. L. Rosner, \prl{76}{1200}{1996}; A. S. Dighe, M.
Gronau, and J. L. Rosner, \prd{54}{3309}{1996}; A. S. Dighe and J. L. Rosner,
\prd{54}{4677}{1996}.

\bibitem{GRP} M. Gronau and J. L. Rosner, \plb{376}{205}{1996}.

\bibitem{SilWo} J. Silva and L. Wolfenstein, \prd{49}{R1151}{1994};
F. DeJongh and P. Sphicas, \prd{53}{4930}{1996}.

\bibitem{BB} G. Buchalla and A. J. Buras, hep-ph/9607447.
See also G. Buchalla and A. J. Buras, \np{B412}{106}{1994}. 

\bibitem{Dib} C. O. Dib, Ph.D. Thesis, Stanford University, 1990, SLAC Report
SLAC-364, April, 1990 (unpublished).

\bibitem{RVW} B. Winstein and L. Wolfenstein, \rmp{65}{1113}{1993}.

\bibitem{E787} Brookhaven E787 \cn, S. Adler \ite, \prl{76}{1421}{1996}.

\bibitem{GaL} M. K. Gaillard and B. W. Lee, \prd{10}{897}{1974}.

\bibitem{pinunu} Fermilab E799 \cn, M. Weaver \ite, \prl{72}{3758}{1994}.

\bibitem{APV} J. L. Rosner, \prd{53}{2724}{1996}.

\bibitem{Wref} The earliest such proposals were made by H. Yukawa,
\ppmsj{17}{48}{1935}, and O. Klein, in {\it Les Nouvelles Th\'eories de la
Physique}, Paris, Inst.~de Co\"operation Intellectuelle (1939), p.~81.  For
others (in the 1950's) see Ref.~\cite{hist}.

\bibitem{VA} R. P. Feynman and M. Gell-Mann, \pr{109}{193}{1958};
E. C. G. Sudarshan and R. E. Marshak, \pr{109}{1860}{1958}.

\bibitem{alpha} S. Eidelman and F. Jegerlehner, Paul Scherrer Institute report
PSI-PR-95-1, January, 1995; H. Burkhardt and B. Pietrzyk, \plb{356}{398}{1995}.
Both sets of authors quote $\alpha^{-1}(M_Z) = 128.89 \pm 0.09$. Values
differing only slightly from this are obtained by A. D. Martin and D.
Zeppenfeld, \plb{345}{558}{1995} [$\alpha^{-1}(M_Z) = 128.99 \pm 0.06$] and
by M. Swartz, SLAC report SLAC-PUB-95-7001, Sept.~1995, submitted to
Phys.~Rev.~D [$\alpha^{-1}(M_Z) = 128.96 \pm 0.06$]. 

\bibitem{Sirlinalpha} A. Sirlin, \prl{72}{1786}{1994}.

\bibitem{Pich} A. Pich, \lti.

\bibitem{Treille} D. Treille, \lti.

\bibitem{Tini} M. Veltman, \np{B123}{89}{1977}.

\bibitem{PT} M. Peskin and T. Takeuchi, \prl{65}{964}{1990}; \prd{46}{381}
{1992}, \art.

\bibitem{DKS} G. DeGrassi, B. A. Kniehl, and A. Sirlin, \prd{48}{3963}{1993}.

\bibitem{GS} P. Gambino and A. Sirlin, \prd{49}{1160}{1994}.

\bibitem{MVS} M. Veltman, \plb{91}{95}{1980}.

\bibitem{KenL} D. C. Kennedy and P. G. Langacker, \prl{65}{2967}{1990};
\ibj{66} {395(E)}{1991}. 

\bibitem{Hasert} F. J. Hasert \ite, \plb{46}{138}{1973}; \np{B73}{1}{1974};
A. Benvenuti \ite, \prl{32}{800}{1974}; B. Aubert \ite, \prl{32}{1454}{1974}.

\bibitem{Lls} C. H. Llewellyn Smith, \np{B228}{205}{1983}.

\bibitem{CCFR} CCFR \cn, FERMILAB-CONF-96/227-E, presented by K. S. McFarland
at XXXI Rencontres de Moriond, March, 1996.

\bibitem{CDHS} CDHS \cn, H. Abramowicz \ite, \prl{57}{298}{1986};
A. Blondel \ite, \zpc{45}{361}{1990}.

\bibitem{CHARM} CHARM \cn, J. V. Allaby \ite, \plb{177}{446}{1986};
\zpc{36}{611}{1987}.

\bibitem{Blondel} A. Blondel, \ptwaw.

\bibitem{CDFW} CDF \cn, F. Abe \ite, \prl{75}{11}{1995}; \prd{52}{4784}{1995}.

\bibitem{D0W} D0 \cn, report PA07-038, presented by A. Zieminski at \waw.

\bibitem{LEPW} ALEPH, DELPHI, L3, and OPAL \cn s, presented at \waw. For a
compilation of the results see Ref.~\cite{Blondel}. 

\bibitem{UA2W} UA2 \cn, J. Alitti \ite, \plb{276}{354}{1992}.

\bibitem{Wwidth} J. L. Rosner, M. P. Worah, and T. Takeuchi, 
\prd{49}{1363}{1994}.

\bibitem{CDFWwidth} CDF \cn, F. Abe \ite, \prl{74}{341}{1995}.

\bibitem{SLC} SLD \cn, report PA07-063, presented by D. Su at \waw.

\bibitem{CW} M. C. Noecker, B. P. Masterson, and C. E. Wieman, \prl{61}{310}
{1988}.

\bibitem{Csth} V. A. Dzuba, V. V. Flambaum, and O. P. Sushkov, \pla{141}{147}
{1989}; S. A. Blundell, W. R. Johnson, and J. Sapirstein, \prl{65}{1411}{1990};
\prd{45}{1602}{1992}.

\bibitem{MR} W. Marciano and J. L. Rosner, \prl{65}{2963}{1990}; \ibj{68}
{898(E)}{1992}.

\bibitem{JRRC} J. L. Rosner, \prd{42}{3107}{1990}.

\bibitem{PS} P. G. H. Sandars, \jpb{23}{L655}{1990}.

\bibitem{TlS} P. A. Vetter, D. M. Meekhof, P. K. Majumder, S. K. Lamoreaux,
and E. N. Fortson, \prl{74}{2658}{1995}.

\bibitem{TlO} N. H. Edwards, S. J. Phipp, P. E. G. Baird, and S. Nakayama,
\prl{74}{2654}{1995}.

\bibitem{PSBL} P. G. H. Sandars and B. W. Lynn, \jpb{27}{1469}{1994}.

\bibitem{Tlth} V. A. Dzuba, V. V. Flambaum, P. G. Silvestrov, and O. P.
Sushkov, \jpb{20}{3297}{1987}.

\bibitem{Tlthr} A. C. Hartley, E. Lindroth, and A. M. M{\aa}rtensson-Pendrill,
\jpb{23}{3417}{1990}; A. C. Hartley and P. G. H. Sandars, \jpb{23}{4197}{1990}.

\bibitem{AZbb} ALEPH \cn, reports PA10-014 and PA10-015, presented by J. Carr
at \waw. 

\bibitem{LLM} P. Langacker, M. Luo, and A. K. Mann, \rmp{64}{87}{1992}.

\bibitem{EFL} J. Ellis, G. Fogli, and E. Lisi, CERN report CERN-TH/96-216
[hep-ph/9608239] (unpublished).

\bibitem{deB} W. de Boer, A. Dabelstein, W. Hollik, W. M\"osle, and
U. Schwickerath, Karlsruhe Univ.~report IEKP-KA/96-08 [hep-ph/9609209]
(unpublished).

\bibitem{Schmelling} M. Schmelling, \ptwaw.

\bibitem{Zbb} See, e.g., R. S. Chivukula, in \dpfa, p.~273; T. Takeuchi, A. K.
Grant, and J. L. Rosner, {\it ibid.}, p.~1231; A. K. Grant,
\prd{51}{207}{1995}. 

\bibitem{heavy} ALEPH, DELPHI, L3, and OPAL \cn s,
LEP Heavy Flavor Working Group, D. Abbaneo \ite, CERN report CERN-PPE/96-017,
Feb.~1996, submitted to {\it Nucl.~Inst.~Meth.}

\bibitem{JS} J. Steinberger, seminar at CERN, August 27, 1996; ALEPH \cn,
A. O. Bazarko \ite~[hep-ex/9609005], presented at \dpfm.

\bibitem{Zbbcorrs} J. Bernabeu, A. Pich, and A. Santamaria,
\plb{200}{569}{1988}; B. A. Kniehl and J. H. K\"uhn, \plb{224}{229}{1989};
K. G. Chetyrkin and J. H. K\"uhn, \plb{248}{359}{1990};
G. Altarelli and R. Barbieri, \plb{253}{161}{1991};
K. G. Chetyrkin, J. H. K\"uhn, and A. Kwiatkowski, \plb{282}{221}{1992}.
For a recent discussion see A. K. Grant, \prd{51}{207}{1995}.

\bibitem{SLDRb} SLD \cn, Measurement of $R_b$ at SLD, report PA10-23, presented
by D. Su at \waw. 

\bibitem{EHLQ} E. Eichten, I. Hinchliffe, K. Lane, and C. Quigg,
\rmp{56}{579}{1984}; \ibj{58}{1065(E)}{1986}.

\bibitem{JRtop} J. L. Rosner, \pl{146B}{108}{1984}.

\bibitem{Feltesse} J. Feltesse, \lti.

\bibitem{CDFMor} CDF \cn, Fermilab reports FERMILAB-CONF-96-118-E,
March, 1996 (presented by P. Azzi), and FERMILAB-CONF-96-146-E, June, 1996
(presented by A. Barbaro-Galtieri), at XXXI Rencontre de Moriond, {\it
QCD and High-Energy Hadronic Interactions}, Les Arcs, France, 23--30
March 1996.

\bibitem{Laenen} E. Laenen \ite, \plb{321}{254}{1994}.

\bibitem{ESW} R. K. Ellis, W. J. Stirling, and B. Webber, to be published
by Cambridge University Press (1996).  I thank R. K. Ellis for updating
the plot of Fig.~21.

\bibitem{CMNT} S. Catani, M. L. Mangano, P. Nason, and L. Trentadue,
\plb{378}{329}{1996}.

\bibitem{Kwong} W. Kwong, \prd{43}{1488}{1991}.

\bibitem{PeS} M. E. Peskin and M. J. Strassler, \prd{43}{1500}{1991}.

\bibitem{GatOak} R. Gatto, G. Sartori, and M. Tonin, \pl{28B}{128}{1968};
N. Cabibbo and L. Maiani, \ibj{28B}{131}{1968}; R. J. Oakes, \ibj{29B}
{683}{1969}.

\bibitem{WbgWZ} S. Weinberg, in {\it A Festschrift for I. I. Rabi},
{\it Trans.~N. Y. Acad.~Sci.~Ser.~II}~38 (1977);
F. Wilczek and A. Zee, \pl{70B}{418}{1977}; \ibj{72B}{504(E)}{1978};
H. Fritzsch, \ibj{70B}{436}{1977}.

\bibitem{RW} J. L. Rosner and M. P. Worah, \prd{46}{1131}{1992}.

\bibitem{Leut} H. Leutwyler, \lti.

\bibitem{Fritzsch} H. Fritzsch, \pl{85B}{81}{1979}; \np{B155}{189}{1979}.

\bibitem{Ramond} H. Arason \ite, \prd{46}{3945}{1992}; H. Arason, D. J.
Castano, E. J. Piard, and P. Ramond, \prd{47}{232}{1992}; P. Ramond, R. G.
Roberts, and G. G. Ross, \np{B406}{19}{1993}; D. J. Castano, E. J. Piard, and
P. Ramond, \prd{49}{4882}{1994}; T. Blazek, M. Carena, S. Raby, and C. Wagner,
Ohio State Univ.~report OHSTPY-HEP-T-96-014 [hep-ph/9608273], \art. 

\bibitem{Branco} G. C. Branco, W. Grimus, and L. Lavoura, \plb{380}{119}{1996};
G. C. Branco, D. Emmanuel-Costa, and J. I. Silva-Marcos, report
[hep-ph/9608477] (unpublished).

\bibitem{dem} H. Harari, H. Haut and J. Weyers, \pl{78B}{459}{1978}; Y. Nambu,
in {\it New Theories in Physics} (23--27 May, 1988, Kazimierz, Poland), Z.
Ajduk, S. Pokorski, and A. Trautman, eds., World Scientific, Singapore, 1989,
p.~1; in {\it Strong Coupling Gauge Theories and Beyond,} Proceedings of the
Second International Workshop, Nagoya, Japan, 1990, T. Muta and K. Yamawaki,
eds., World Scientific, Singapore (1991), p.~3; \efi 92-03, in {\it Proceedings
of the Workshop on Electroweak Symmetry Breaking}, Hiroshima, 1991, p.~1; \efi
92-37, Spontaneous symmetry breaking and the origin of mass, invited talk at
Int.~Conf.~on Fluid Mech.~and Theor.~Phys. in honor of Professor Chou
Pei-Yuan's 90th Birthday, 1992; P. Kaus and S. Meshkov, \mpla{3}{1251}{1988};
\ibj{4}{603(E)}{1989}; \prd{42}{1863}{1990}.  See also the second of
Refs.~\cite{Branco}. 

\bibitem{Matu} K. Matumoto and T. Matsuoka, \ptp{83}{373}{1990};
\ibj{84}{53}{1990}; K. Matumoto, \ibj{84}{185,787}{1990}; \ibj{85}{1149}
{1991}; second of Refs.~\cite{Branco}.

\bibitem{PTP} J. L. Rosner, \ptp{66}{1421}{1981}.

\bibitem{GLR} M. K. Gaillard, B. W. Lee, and J. L. Rosner, \rmp{47}{277}
{1975}.

\bibitem{WA89} WA89 \cn, S. Paul \ite, \np{A585}{183c}{1995}.  See,
however, CLEO \cn, report PA01-079, presented by R. Kutschke at \waw.

\bibitem{massforms} A. De R\'ujula, H. Georgi, and S. L. Glashow, 
\prd{12}{147}{1975}; S. Gasiorowicz and J. L. Rosner, \ajp{49}{954}{1981}.

\bibitem{Vt} M. Veltman, in \fermlg, p.~529; \app{B12}{437}{1981}.

\bibitem{Nambu} Y. Nambu, Ref.~\cite{dem}.

\bibitem{NJL} Y. Nambu and G. Jona-Lasinio, \pr{122}{345}{1961}; \ibj{124}
{246}{1961}.

\bibitem{BHL} W. A. Bardeen, C. T. Hill, and M. Lindner, \prd{41}{1647}{1990}.

\bibitem{GW} J.-M. G\'erard and J. Weyers, \pl{146B}{411}{1984}; J. P. Fatelo,
J.-M. G\'erard, T. Hambye, and J. Weyers, \prl{74}{492}{1995}; M. Chaichian, P.
Chiappetta, J.-M. G\'erard, R. Gonzalez Felipe, and J. Weyers,
\plb{365}{141}{1996}, \art. 

\bibitem{LH} L. Hall, The heavy top quark and supersymmetry, lectures given at
the 1995 SLAC Summer Institute, July 10--21, 1995, report LBL-38110
[hep-ph/9605258]. 

\bibitem{OZ} R. Oehme, \efi 95-47 [hep-th/9511006], \art.

\bibitem{WMGYU} W. Marciano, \prl{62}{2793}{1989}; \prd{41}{219}{1990}.

\bibitem{GYU} J. Kubo \ite, Max-Planck-Institut report MPI-PHT-95-132
[hep-ph/9512400], \art.

\bibitem{LSTC} L. Susskind, \prd{20}{2619}{1979}.

\bibitem{SWTC} S. Weinberg, \prd{13}{974}{1976}; \ibj{19}{1277}{1979}.

\bibitem{ETC} For a review see E. Farhi and L. Susskind, \prp{74}{277}{1981}.

\bibitem{seesaw} M. Gell-Mann, P. Ramond, and R. Slansky, in {\it
Supergravity}, edited by P. van Nieuwenhuizen and D. Z. Freedman
(North-Holland, Amsterdam, 1979), p.~315; T. Yanagida, in {\it Proc.~Workshop
on Unified Theory and Baryon Number in the Universe}, edited by O. Sawada and
A. Sugamoto (KEK Report No.~79-18, Tsukuba, Japan, 1979).

\bibitem{GG} H. Georgi and S. L. Glashow, \prl{32}{438}{1974}.

\bibitem{SO} H. Georgi in {\it Proceedings of the 1974 Williamsburg DPF
Meeting}, ed. by C. E. Carlson  (New York, AIP, 1975) p.~575; H. Fritzsch and
P. Minkowski, \apny{93}{193}{1975}. 

\bibitem{E6} F. G\"ursey, P. Ramond, and P. Sikivie, \pl{60B}{177}{1976}.

\bibitem{E6SS} E. Witten, \np{B258}{75}{1985}; E. Cohen, J. Ellis, K. Enqvist,
and D. V. Nanopoulos, \pl{165B}{76}{1985}; J. L. Rosner, \cmts{15}{195}{1986}.

\bibitem{rev} For a review of \es~phenomenology in the context of superstring
theories see J. L. Hewett and T. G. Rizzo, \prp{183}{193}{1989}.  More recent
discussions of extended gauge structures motivated by \es~include those by
P. Langacker and M. Luo, \prd{45}{278}{1992};
P. Langacker, in {\it Precision Tests of the Standard Model}, edited by
P. Langacker (World Scientific, Singapore, 1995), p.~883;
M. Cveti\v{c} and S. Godfrey, in {\it Electro-weak Symmetry Breaking and
Beyond the Standard Model}, edited by T. Barklow, S. Dawson, H. Haber, and
J. Siegrist (World Scientific, Singapore, 1995), \art;
M. Cveti\v{c} and P. Langacker, \prd{54}{3570}{1996}.

\bibitem{RS} R. Slansky, \prp{79}{1}{1981}.

\bibitem{flip} See, e.g., S. M. Barr, \pl{112B}{219}{1982}; J. Ellis, J. L.
Lopez, and D. V. Nanopoulos, \plb{371}{65}{1996}; J. L. Lopez, D. V.
Nanopoulos, and A. Zichichi, \prd{53}{5253}{1996}; \art. 

\bibitem{eeggevt} J. L. Rosner, CERN report CERN-TH/96-209, hep-ph/9607467,
submitted to Phys.~Rev.~D.

\bibitem{PLJE} P. Langacker and J. Erler, in Ref.~\cite{PDG}, p.~103.

\bibitem{RR} R. W. Robinett and J. L. Rosner, \prd{26}{2396}{1982};
P. G. Langacker, R. W. Robinett, and J. L. Rosner, \prd{30}{1470}{1984}.

\bibitem{CDFZp} CDF \cn, F. Abe \ite, \prd{51}{949}{1995}; T. Kamon, Fermilab
report FERMILAB-CONF-96-106-E [hep-ex/9605006], presented at XXXI Rencontre de
Moriond:  QCD and High-energy Hadronic Interactions, 23 -- 30 March, 1996;
CDF \cn, M. K. Pillai, E Hayashi, K. Maeshima, C. Grosso-Pilcher,
P. de Barbaro, A. Bodek, B. Kim, and W. Sakumoto, report hep-ex/9608006,
August, 1996, presented at \dpfm.

\bibitem{CDFas} CDF \cn, F. Abe \ite, \prl{77}{2616}{1996}.

\bibitem{FBA} J. L. Rosner, \prd{54}{1078}{1996}.

\bibitem{eegg} CDF \cn, F. Abe \ite, presented by S. Park, in {\it Proceedings
of the 10th Topical Workshop on Proton-Antiproton Collider Physics,} Fermilab,
May 9--13, 1995, AIP Conference Proceedings 357, edited by R. Raja and J. Yoh,
(AIP, Woodbury, NY, 1996), p.~62.

\bibitem{sseegg} S. Ambrosanio, G. L. Kane, G. D. Kribs, S. P. Martin, and
S. Mrenna, \prl{76}{3498}{1996}; Univ.~of Michigan reports hep-ph/9605398 and
9607414 (unpublished); S. Dimopoulos, M. Dine, S. Raby, and S. Thomas,
\prl{76}{3494}{1996}; S. Dimopoulos, S. Thomas, and J. D. Wells, \prd{54}
{3283}{1996}; S. Dimopoulos, M. Dine, S. Raby, S. Thomas, and J. D. Wells,
SLAC-PUB-7236 [hep-ph/9607450] (unpublished); J. L. Lopez and D. V. Nanopoulos,
Rice University reports DOE/ER/40717-29 [hep-ph/9607220] and DOE/ER/40717-32
[hep-ph/9608275] (unpublished); K. S. Babu, C. Kolda, and F. Wilczek, Institute
for Advanced Study report IASSNS-HEP-96/55 [hep-ph/9605408]; J. Hisano, K.
Tobe, and T. Yanagida, Univ.~of Tokyo report UT-754 [hep-ph/9607234]
(unpublished). 

\bibitem{BM} G. Bhattacharyya and R. N. Mohapatra, \prd{54}{4204}{1996}.

\bibitem{LR} D. London and J. L. Rosner, \prd{34}{1530}{1986}.

\bibitem{Toback} D. Toback, Fermilab report FERMILAB-CONF-96/240-E, August
1996, presented for the CDF \cn~at \dpfm. 
\end{thebibliography}
\end{document}